\documentclass[%
aps,
prx,
reprint,
superscriptaddress,
nofootinbib,
amsmath,
amssymb
]{revtex4-2}
\usepackage{
    physics,
    graphicx,
    multirow,
    tabularx,
    mathtools, 
    placeins, 
    nicefrac, 
    hyperref, 
    threeparttable, 
    siunitx,
    enumitem,
    makecell
}

\usepackage[capitalize]{cleveref}
\Crefname{section}{Sec.}{Secs.}

\usepackage[dvipsnames]{xcolor}
\hypersetup{
    colorlinks,
    linkcolor={blue!50!black},
    citecolor={blue!50!black},
    urlcolor={blue!80!black}
}

\usepackage[caption=false]{subfig} 
\newcommand{\supop}[1]{\ensuremath{\overbracket[0.1ex][0.3ex]{#1}}}
\DeclarePairedDelimiter\pbra{\langle\!\langle}{\rvert}
\DeclarePairedDelimiter\pket{\lvert}{\rangle\!\rangle}
\DeclarePairedDelimiterX\pbraket[2]{\langle\!\langle}{\rangle\!\rangle}{#1 \delimsize\vert #2}

\newtheorem{theorem}{Theorem}

\newtheorem{corollary}{Corollary}

\begin{document}


\title{Virtual Channel Purification}

\newcommand{\qmaddress}{\affiliation{Quantum Motion, 9 Sterling Way, London N7 9HJ, United Kingdom}}
\newcommand{\oxddress}{\affiliation{Department of Materials, University of Oxford, Parks Road, Oxford OX1 3PH, United Kingdom}}

\newcommand{\thuddress}{\affiliation{Center for Quantum Information, Institute for Interdisciplinary Information Sciences, Tsinghua University, Beijing 100084, China}}

\author{Zhenhuan Liu}
\thuddress
\author{Xingjian Zhang}
\affiliation{QICI Quantum Information and Computation Initiative, Department of Computer Science, The University of Hong Kong, Pokfulam Road, Hong Kong}
\author{Yue-Yang Fei}
\affiliation{Hefei National Research Center for Physical Sciences at the Microscale and School of Physical Sciences, University of Science and Technology of China, Hefei 230026, China}
\affiliation{CAS Center for Excellence in Quantum Information and Quantum Physics, University of Science and Technology of China, Hefei 230026, China}
\author{Zhenyu Cai}
\email{cai.zhenyu.physics@gmail.com}
\oxddress
\qmaddress

\date{\today}

\begin{abstract}
Quantum error mitigation is a key approach for extracting target state properties on state-of-the-art noisy machines and early fault-tolerant devices. 
Using the ideas from flag fault tolerance and virtual state purification, we develop the virtual channel purification (VCP) protocol, which consumes similar qubit and gate resources as virtual state purification but offers stronger error suppression with increased system size and more noisy operation copies.
The application of VCP does not require specific knowledge about the target quantum state, the target problem and the gate noise model in the target circuit, and can still offer rigorous performance guarantees for practical noise regimes as long as the noise is incoherent.
Further connections are made between VCP and quantum error correction to produce the virtual error correction (VEC) protocol, one of the first protocols that combine quantum error correction (QEC) and quantum error mitigation beyond directly applying error mitigation protocols on top of logical qubits. Assuming perfect syndrome extraction, VEC can virtually remove all correctable noise in the channel while paying only the same sampling cost as low-order purification. It can achieve QEC-level protection on an unencoded register when transmitting it through a noisy channel, removing the associated encoding qubit overhead. 
Another variant of VEC can mimic the error suppression power of the surface code by inputting only a bit-flip and a phase-flip code.
Our protocol can also be adapted to key tasks in quantum networks like channel capacity activation and entanglement distribution.
\end{abstract}

\maketitle

\section{Introduction}\label{sec:intro}
Despite the rapid advances in quantum hardware in recent years, there is still an extended stretch of time ahead before we can successfully implement quantum error correction (QEC) to achieve fully fault-tolerant computation. 
In order to reach practical quantum advantages during this time, it is essential to develop noise suppression techniques compatible with existing quantum technologies. 
One key method is quantum error mitigation (QEM)~\cite{caiQuantumErrorMitigation2023}, which can extract target information from noisy quantum circuits without physically recovering the noiseless quantum state. Due to the low hardware requirement, QEM has become a prevailing tool in many of the state-of-the-art quantum computation experiments~\cite{googlequantumaiandcollaboratorsHartreeFockSuperconductingQubit2020,googlequantumaiandcollaboratorsFormationRobustBound2022,kimEvidenceUtilityQuantum2023} and is expected to play a key role also in the early fault-tolerant era.

Over the years, various QEM protocols have been proposed, but each comes with its own sets of assumptions:
\begin{itemize}[leftmargin=1.5em]
    \item Probabilistic error cancellation~\cite{temmeErrorMitigationShortDepth2017,endoPracticalQuantumError2018} relies on detailed knowledge about the noise models and the assumption that the noise remains the same across different times and different qubits.
    \item Zero-noise extrapolation~\cite{temmeErrorMitigationShortDepth2017,liEfficientVariationalQuantum2017} requires the ability to tune hardware noise without significantly modifying the noise model and it can only offer rigorous performance guarantee at small noise.
    \item Virtual state purification~\cite{hugginsVirtualDistillationQuantum2021,koczorExponentialErrorSuppression2021} requires the ideal input and output state to be pure states and also the noiseless component to be the dominant component of the noisy output state.
    \item Symmetry verification~\cite{mcardleErrorMitigatedDigitalQuantum2019,bonet-monroigLowcostErrorMitigation2018} and subspace expansion~\cite{mccleanHybridQuantumclassicalHierarchy2017} require problem-specific knowledge about the symmetry or energy constraints on the output state.
\end{itemize}
In this article, we are going to introduce a QEM technique called \emph{virtual channel purification} that removes all of the assumptions above, i.e. it imposes no requirements on specific knowledge about the incoming and output states, the problem we try to solve and the gate error models of the target circuit; does not require additional hardware capability beyond gate-model computation; while still offers rigorous performance guarantee for the most practical noise regime. 
Our protocol does assume the noise in the ideal unitary operation is incoherent, which is the case for most practical scenarios~\cite{koczorDominantEigenvectorNoisy2021,dalzellRandomQuantumCircuits2021}, especially with the help of Pauli twirling~\cite{bennettPurificationNoisyEntanglement1996,bennettMixedstateEntanglementQuantum1996,caiConstructingSmallerPauli2019,wallmanNoiseTailoringScalable2016}. In addition, its full error suppression power can only be achieved when the noise introduced by the additional \textsc{cswap} gates is small compared to the noise in the main circuit, or when the noise model of the \textsc{cswap}s is known. 
Do note that our protocols would prefer quantum hardware that is equipped with transversal entangling gates between 2D qubit arrays, while many other techniques above have no such preference.

Virtual channel purification (VCP) is obtained by combining ideas from virtual state purification (VSP), Choi–Jamiołkowski isomorphism, and the circuit for flag fault-tolerance~\cite{chaoQuantumErrorCorrection2018}. 
Just like how VSP uses $M$ copies of a noisy state to virtually prepare a purified output state whose infidelity falls exponentially with $M$, VCP is able to use $M$  copies of a noisy \emph{channel} to virtually implement a purified channel whose infidelity falls exponentially with $M$. 
The circuit implementation costs of VSP and VCP are similar, but VCP removes many assumptions of VSP as mentioned and provides much stronger error suppression power, extending the applicability of such methods into deeper and noisier circuits. 

Due to the key roles we expect QEC and QEM to both play in practical quantum computation, there is always a strong desire to develop a framework that combines the two. 
So far the attempts are mostly about concatenating QEM on top of QEC~\cite{suzukiQuantumErrorMitigation2022,lostaglioErrorMitigationQuantumAssisted2021,piveteauErrorMitigationUniversal2021}, i.e. directly applying QEM on top of logical qubits to mitigate logical errors. 
Through further generalisation of VCP and the use of the Knill-Laflamme condition~\cite{knillTheoryQuantumErrorcorrecting1997}, we are able to obtain one of the first integrated protocols that combine QEM and QEC beyond concatenation called virtual error correction (VEC). In this protocol, by paying the same sampling cost as channel purification using two copies, we are able to \emph{remove all correctable noise} in the noisy channel, reaching beyond the standard bias-variance trade-off limit in pure QEM~\cite{caiQuantumErrorMitigation2023}. We can also use VEC circuits to combine the error suppression power of different codes, greatly reducing the qubit overhead compared to conventional QEC codes. The discussion of VEC here assumes noiseless syndrome extraction, thus further explorations are needed for its practical implementation. In addition, we have also shown how QEM can be applied before (underneath) QEC without affecting the performance of the QEC.
A modification of VCP can also be used to physically (instead of virtually) suppress the channel errors and we will discuss its role in applications like channel capacity activation and entanglement distribution in quantum networks. 

This paper is organised as follows. In \cref{sec:vcp}, we introduce the basic idea of VCP and the circuit for its implementation. 
In \cref{sec:vcp_performance}, we discuss the performance of VCP, including its error suppression power, flexibility, practical implementation, and sample complexity, compared to VSP. 
We also use numerical simulations to validate our results.
In \cref{sec:comb_qec}, we discuss further generalisation of VCP and its combination with QEC to achieve higher error suppression at lower costs.
In \cref{sec:other_app}, VCP is further modified to use post-selection on its measurement results and we discuss the corresponding applications, especially in quantum networks. 
We make our conclusion in \cref{sec:conclusion} by summarising our results and discussing possible future directions.

\section{Virtual Channel Purification}\label{sec:vcp}
\begin{figure}
\centering
\includegraphics[width=0.48\textwidth]{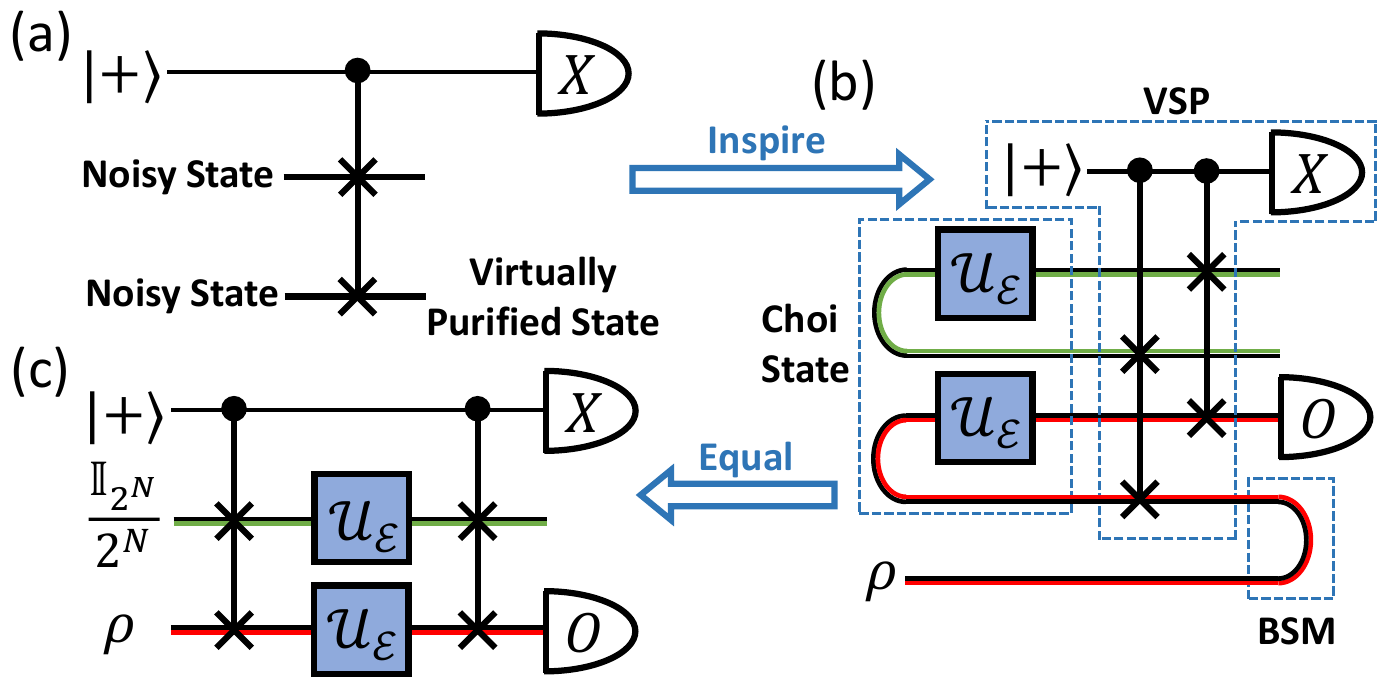}
\caption{The logic for constructing VCP circuit, taking $M=2$ without loss of generality. (a) The circuit for VSP, consists of two copies of the noisy input state, a single control qubit initialised as $\ket{+}$, and a \textsc{cswap} gate. By measuring the control qubit in Pauli-$X$ basis, one can virtually prepare the purified state on the second register. (b) One possible circuit implementation of VCP, obtained from performing the VSP circuit in (a) on two copies of Choi states of the noisy channel. Each curved line at the input end stands for a $2N$-qubit maximally entangled state $\ket{\Phi^+}=\frac{1}{\sqrt{2^N}}\sum_{i=0}^{2^N-1}\ket{ii}$, which is used to prepare the Choi state of the noisy channel $\mathcal{U}_{\mathcal{E}}$. Then, the VSP circuit acts on two noisy Choi states to virtually obtain a purified Choi state. The curved line at the output end stands for Bell state measurement (BSM) post-selected on the all-zero result, which is for implementing the purified channel using the purified Choi state. (c) A circuit implementation of VCP, obtained from straightening the curved lines in green and red in (b). By measuring the control qubit in Pauli-$X$ basis, one can virtually implement the purified channel on the second register. This circuit is easily generalised to a larger value of $M$ by employing $M-1$ copies of maximally mixed state and changing \textsc{swap} to the $M$th order permutation as shown in  \cref{fig:higher_order}(a).}
\label{fig:circuit}
\end{figure}

We consider the case where our ideal operation is a unitary channel $\mathcal{U}$, but in practice we can only implement the noisy channel $\mathcal{U}_{\mathcal{E}}=\mathcal{E}\mathcal{U}$, which is the ideal operation $\mathcal{U}$ followed by the noise channel 
\begin{equation}\label{eqn:id_noise_channel}
\mathcal{E}=p_0\mathcal{I}+\sum_{i=1}^{4^N-1}p_i{\supop{E}}_i.
\end{equation}
Here $N$ is the number of qubits, $\mathcal{I}$ stands for the identity channel, and ${\supop{E}}_i(\rho)=E_i\rho E_i^\dagger$ stands for the channel for a given error component. In line with the assumptions made in VSP~\cite{koczorExponentialErrorSuppression2021,hugginsVirtualDistillationQuantum2021}, we will assume $\mathcal{E}$ is Pauli noise with $E_i$ being Pauli operators and also $p_0 > p_i$ for all $i$ (i.e. identity is the leading component). More general noise can be transformed into Pauli noise via Pauli twirling~\cite{wallmanNoiseTailoringScalable2016,caiConstructingSmallerPauli2019} and the applicability of our method beyond Pauli noise is also discussed in \cref{sec:deviation_from_Pauli}. 
Our target for VCP is consuming $M$ copies of $\mathcal{U}_{\mathcal{E}}$ to implement $\mathcal{U}_{\mathcal{E}^{(M)}}=\mathcal{E}^{(M)}\mathcal{U}$ with the purified noise channel $\mathcal{E}^{(M)}$ being
\begin{equation}\label{eqn:purified_channel}
\mathcal{E}^{(M)}=\frac{1}{\sum_{i=0}^{4^N-1}p_i^M}\left(p_0^M\mathcal{I}+\sum_{i=1}^{4^N-1}p_i^M{\supop{E}}_i\right).
\end{equation}
As $p_0$ is the leading term, the noise rate of $\mathcal{E}^{(M)}$ decays exponentially with $M$ as expected. 

Based on the Choi–Jamiołkowski isomorphism, the Pauli noise channel $\mathcal{E}$ have a one-to-one correspondence to the Choi state $\rho_{\mathcal{E}}$ whose dominant eigenvector corresponds to the identity channel. 
It is easy to show that under Pauli noise, the Choi state of $\mathcal{E}^{(M)}$ is proportional to the $M$th power of the Choi state of $\mathcal{E}$: $\rho_{\mathcal{E}^{(M)}}=\nicefrac{\rho_{\mathcal{E}}^M}{\Tr[\rho_{\mathcal{E}}^M]}$, which can be obtained by performing state purification on $M$ copies of $\rho_{\mathcal{E}}$~\cite{hugginsVirtualDistillationQuantum2021,koczorExponentialErrorSuppression2021}. Similar arguments also apply to the Choi states of $\mathcal{U}_{\mathcal{E}^{(M)}}$ and $\mathcal{U}_{\mathcal{E}}$. 
Hence, the most straightforward way to implement $\mathcal{U}_{\mathcal{E}^{(M)}}$ is first using $M$ copies of $\mathcal{U}_{\mathcal{E}}$ to prepare $M$ copies of their Choi state; next purifying the Choi state of $\mathcal{U}_{\mathcal{E}^{(M)}}$ from them; and finally using this purified Choi state to implement the purified channel $\mathcal{U}_{\mathcal{E}^{(M)}}$.

The circuit of state purification is shown in \cref{fig:circuit}(a), for which the control-qubit measurement outcome probability and the corresponding output state of the second register can be denoted as $p_\pm$ and $\rho_\pm$, respectively. As shown in Ref.~\cite{koczorExponentialErrorSuppression2021,hugginsVirtualDistillationQuantum2021}, the purified state is given as $p_+\rho_+ - p_-\rho_-$, which can be virtually prepared for information extraction by simultaneously measuring the controlled qubit and the second register plus post-processing. As shown in \cref{fig:circuit}(b), using the exact same circuit with the input noisy states being the Choi states of $\mathcal{U}_{\mathcal{E}}$, prepared by acting $\mathcal{U}_{\mathcal{E}}$ on one side of the maximally entangled states $\ket{\Phi^+}$, we can virtually obtain the Choi state of the purified channel $\mathcal{U}_{\mathcal{E}^{(M)}}$. This Choi state can be used to apply the purified channel $\mathcal{U}_{\mathcal{E}^{(M)}}$ on some input state $\rho$ using the technique of quantum teleportation. 
Specifically, we need to perform Bell state measurements (BSM) on input state $\rho$ and one side of the purified Choi state, post-select based on the measurement result, and measure the other side to extract the value of $\Tr[O\mathcal{U}_{\mathcal{E}^{(M)}}(\rho)]$.

Though conceptually simple, this method requires significant quantum and classical resources. It needs more than two times qubit overhead compared to VSP for preparing the Choi states. Furthermore, to transform a state into a channel through teleportation, one would require the $2N$-qubit Bell state measurement result to be all $0$, whose success probability decays exponentially with the number of qubits. 
As we have drawn the whole procedure using the circuit in \cref{fig:circuit}(b) with tensor network notations, we notice that the circuit has some internal structures that allow us to further simplify it.

This circuit utilises $\ket{\Phi^+}$ multiple times including the preparation of Choi states and the BSM.
In tensor network notation~\cite{bridgemanHandwavingInterpretiveDance2017}, $\ket{\Phi^+}$ and the BSM projector are represented by bending a straight line as shown in \cref{fig:circuit}(b). In this way, by simply straightening the wires using tensor network rules, we can obtain a simplified circuit shown in \cref{fig:circuit}(c), in which the first \textsc{cswap} gate goes before the action of two channels and the green open wire on the input is mapped to the maximally mixed state based on tensor network rules. 
This is a much more efficient circuit with the same qubit overhead as VSP with no post-selection needed. We have further proved algebraically that the circuit in \cref{fig:circuit}(c) can achieve VCP and generalise it to $M$th order purification in \cref{sec:pur_circ_output}, giving rise to the first key results in VCP.
\begin{theorem}\label{thrm:purified_channel}
Using a single input state $\rho$, a single control qubit initialised as $\ket{+}$, $M-1$ copies of the maximally mixed state $\nicefrac{\mathbb{I}_{2^N}}{2^N}$, $M$ copies of the noisy channel $\mathcal{U}_{\mathcal{E}}=\mathcal{E}\mathcal{U}$, and two controlled permutation operations, the measurement result of \cref{fig:higher_order}(a) ($M$th order generalisation of \cref{fig:circuit}(c)) gives
\begin{equation}
\frac{\langle X\otimes O\rangle}{\langle X\otimes \mathbb{I}_{2^N}\rangle}=\Tr\left[O\mathcal{E}^{(M)}\mathcal{U}(\rho)\right],
\end{equation}
where $\mathcal{E}$ and $\mathcal{E}^{(M)}$ follow the relation of \cref{eqn:id_noise_channel,eqn:purified_channel}.
\end{theorem}

As discussed in \cref{sec:pur_circ_output}, even when the Pauli noise channels acting on the different copies are different, as long as identity is the dominant component for all of them, we can still achieve a similar level of error suppression using VCP 

In the context of quantum error correction (QEC), the main part of the circuit in \cref{fig:circuit}(c) shares a similar structure with flag fault-tolerance~\cite{chaoQuantumErrorCorrection2018,chaoFlagFaultTolerantError2020}, which detects circuit faults between the two probe operations from the control qubit. It can also be viewed as space-time checks (or detector)~\cite{delfosseSpacetimeCodesClifford2023,mcewenRelaxingHardwareRequirements2023} on the noisy channels using permutation symmetry. This is not surprising since the original VSP is partly inspired by the circuit for performing permutation symmetry checks on states~\cite{hugginsEfficientNoiseResilient2021,barencoStabilizationQuantumComputations1997}. Such a pair of \textsc{cswap}s is also used in the circuit for quantum switches~\cite{chiribellaQuantumComputationsDefinite2013}, error filtration~\cite{leeErrorSuppressionArbitrarySize2023}, and superposed quantum error mitigation \cite{miguel-ramiroSuperposedQuantumError2023}. These connections are further discussed in \cref{sec:generalisation}.

\section{Performance} \label{sec:vcp_performance}

\subsection{Advantages Compared with VSP}\label{sec:stronger_suppresion}

Just like VSP, the VCP protocol can suppress noise exponentially with more copies of the noisy channel, but there are also additional advantages arising from directly mitigating errors in quantum channels instead of states as outlined below.

\subsubsection{Exponentially stronger error suppression for global noise} \label{sec:more_qubit_more_suppress}
VCP offers stronger error suppression compared with VSP due to a lower error rate for each individual error component.
Consider a $N$-qubit system with a globally depolarising channel that has probability $P$ of mapping any input state into the maximally mixed state,
\begin{equation}\label{eq:depolarizing}
\mathcal{E}(\rho)=(1-P)\rho+P\frac{\mathbb{I}_{2^N}}{2^N}.
\end{equation}
It contains approximately $4^N$ Pauli error components
each with an error probability $4^{-N}P$. If the input state is a pure state, the output noisy state has approximately $2^N$ error components each with an error probability $2^{-N}P$.
Therefore, if we perform $M$th order VCP and VSP on the channel and the output state, respectively, which use the same number of qubits and channels, the resultant error rate is suppressed by a factor of $(4^{N}/P)^{M-1}$ for the VCP and $(2^{N}/P)^{M-1}$ for the VSP.
Therefore, \emph{VCP can suppress $2^{N(M-1)}$ times more errors than VSP, which is exponentially larger with the number of qubits and the order of purification.} A more rigorous proof of this can be found in \cref{sec:exp_err_suppr}.

Note that the noise we considered above is not the noise model of the individual gates, but the effective noise model of the whole circuit, obtained by commuting all gate noise to the end of the circuit. Our arguments above assume a globally depolarising channel; thus, the exponential improvement above cannot be fully realised in practice. However, as we increase the depth of the circuit, the local noise will be scrambled into a form closer and closer to the global depolarising channel~\cite{dalzellRandomQuantumCircuits2021,foldagerCanShallowQuantum2023}, taking us closer to the improvement factor mentioned above.
The global noise assumption has also been used in large-scale experiments~\cite{urbanekMitigatingDepolarizingNoise2021,miInformationScramblingQuantum2021} in order to successfully employ the ``rescale and shift'' QEM techniques.

Going beyond the globally depolarising noise model, the advantage of VCP over VSP applies to more general cases. 
Due to the larger dimensionality of channels compared to the corresponding states, different error components in a noise channel can be mapped into the same error component in a noisy state. Thus in general, the error probability of the leading error component in a noise channel is smaller than that of the resultant noisy state, leading to stronger error suppression when we purify the channel compared to the state. 

In \cref{fig:data}(b), we have performed a numerical simulation demonstrating the error suppression power of VSP and VCP increase with the deeper circuit and more qubits, due to the increased number of error components in the circuit. As discussed, the circuit noise model will move closer to the globally depolarising channel as the circuit depth increases, and correspondingly we indeed can see the improvement of VCP over VSP increases as the circuit depth increases.

\begin{figure}[htbp]
\centering
\includegraphics[width=0.5\textwidth]{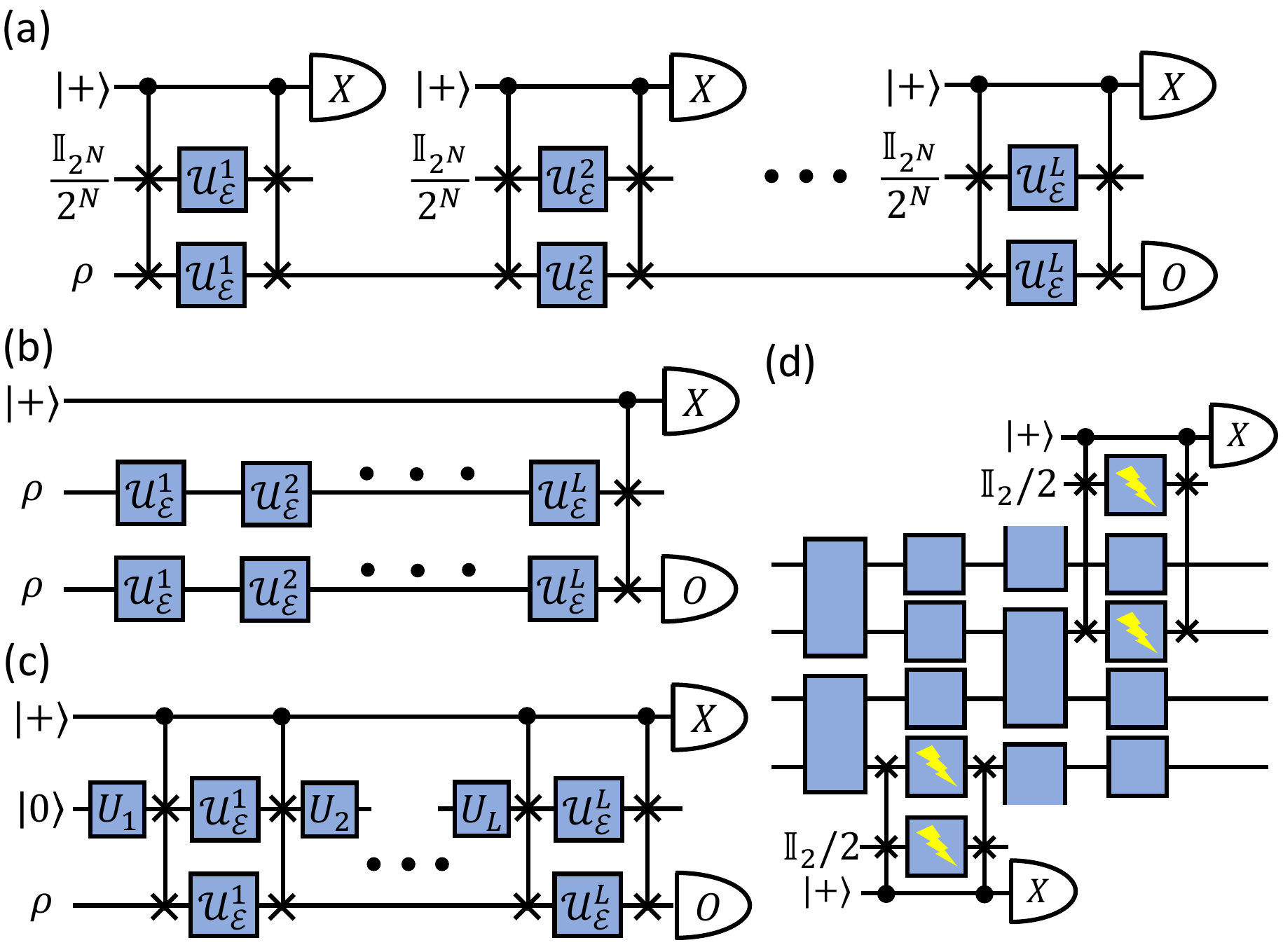}
\caption{Comparison between VSP and VCP with $M=2$.
(a) One way to implement the VCP layer-by-layer, which relies on techniques of state reset and mid-circuit measurement.
(b) VSP protocol can only be performed once at the end of the quantum circuit with one control qubit and two identical copies.
(c) Another implementation of layer-by-layer VCP.
The control qubit can be reused and measured at the end of the circuit.
The ancillary register can be initialised as some pure state and all maximally mixed states are realised using random unitaries labelled by $U_1$, $U_2$, ..., $U_L$.
(d) VCP can target specific noisy gates in the quantum circuit. }
\label{fig:layered_VCD}
\end{figure}

\subsubsection{Layer-wise implementation}

\begin{figure*}
\centering
\includegraphics[width=1\textwidth]{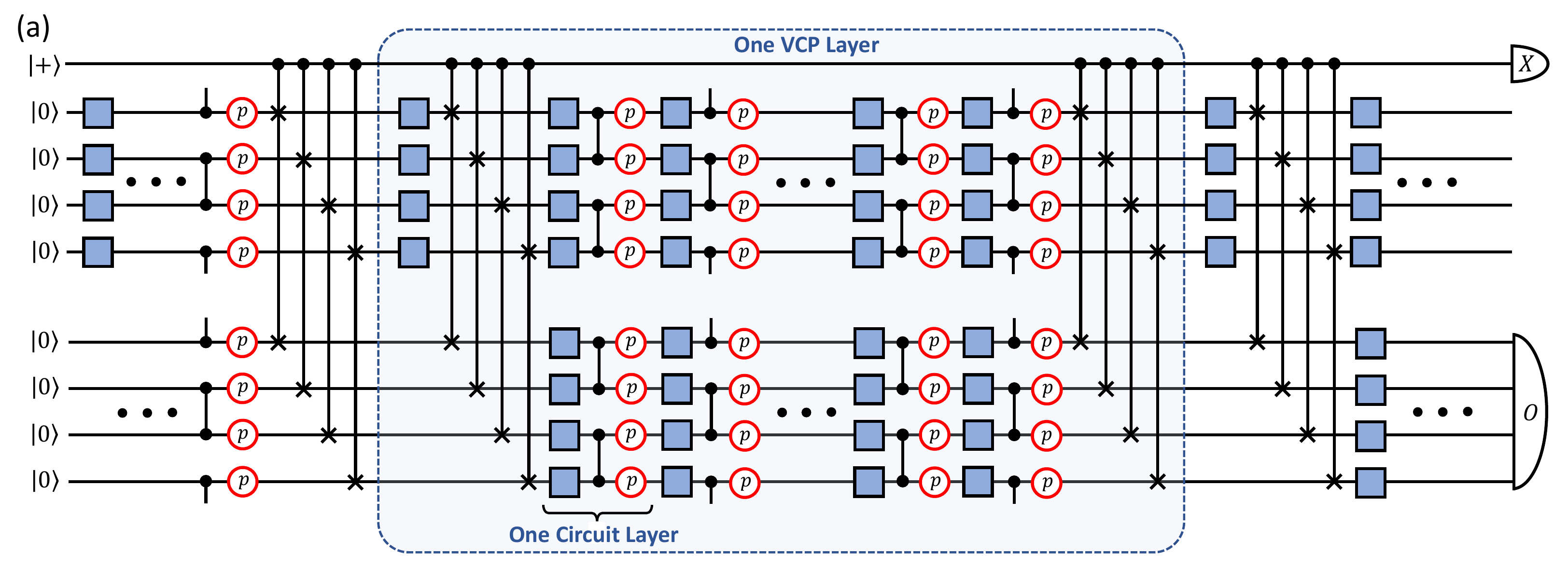}

\includegraphics[width=1\textwidth]{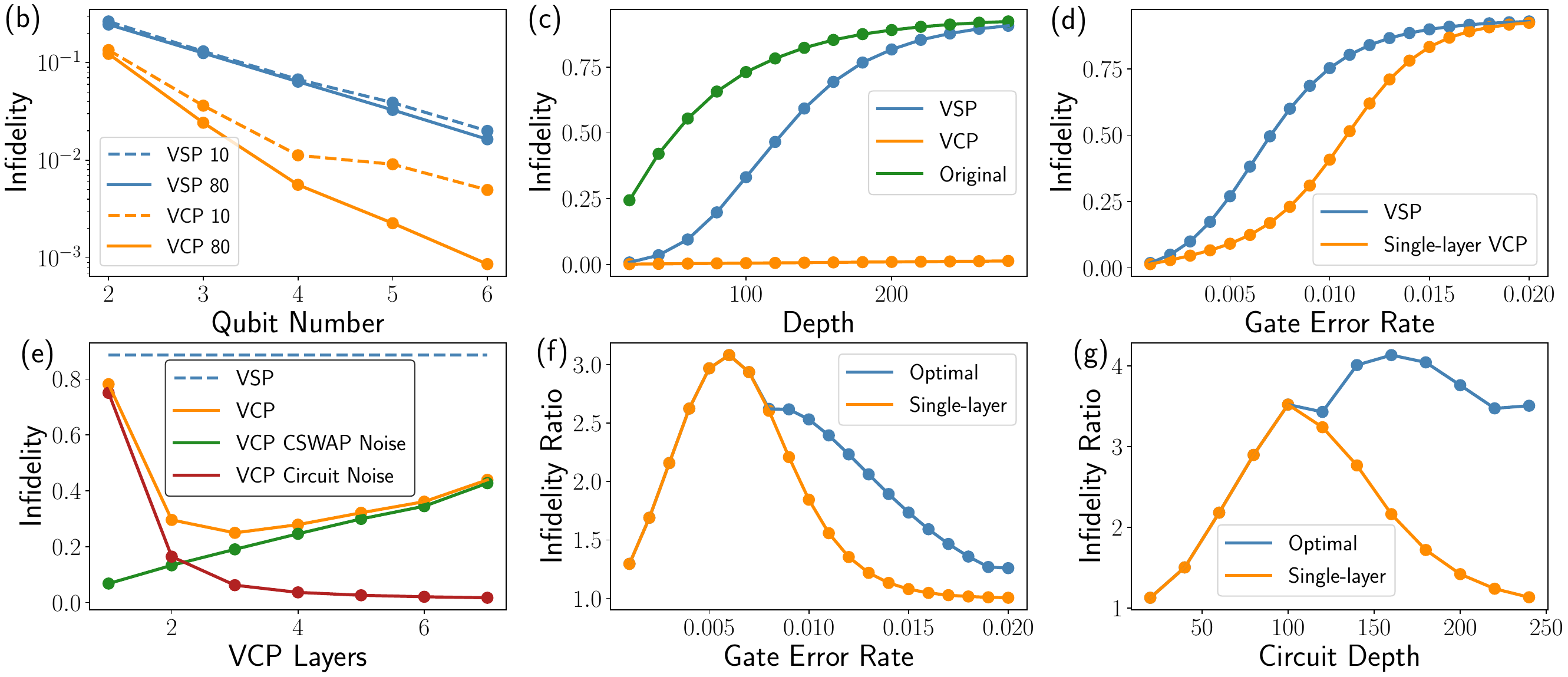}
\caption{
(a) The circuit for our numerical simulation with VCP applied. Except for (b), the main circuit is a four-qubit system initialised in the all-$0$ state evolving through a random circuit consisting of alternating layers of random single-qubit gates (blue boxes) and \textsc{cnot} gates (two dots connected by a vertical line). The \textsc{cnot} gates are affected by two single-qubit depolarising channels each with an error rate $p$ (red circles, thus the two-qubit gate error rate is $2p$). Except for (b) and (c), every \textsc{cswap} gate is followed by three single-qubit depolarising channels each with an error rate $5p$ (the total gate error rate for each \textsc{cswap} is thus $15p$). 
(b) The infidelities after VSP and single-layer VCP with different qubit numbers at the circuit depths of 10 and 80. For each data point, we use similar circuit structures and keep the original infidelity between the unmitigated state and the noiseless state to be $0.5$. 
(c) The infidelities after VSP and VCP for different circuit depths \emph{without any \textsc{cswap} noise}. The gate error rate is $p=0.005$ and the depth of each VCP layer is $20$.
(d) The infidelities after VSP and single-layer VCP at different values of $p$. The circuit depth is set to be $80$. 
(e) The infidelity behaviour after VCP with different numbers of VCP layers for a circuit with a depth $240$ and a gate error rate $p=0.005$.
The green and red lines represent cases without \textsc{cnot} and \textsc{cswap} errors, respectively.
(f) The infidelity ratios between VSP and single-layer VCP, and between VSP and VCP with the optimal number of VCP layers, at different gate noise rates with a circuit depth of $80$.
(g) The infidelity ratios at different circuit depths with a fixed gate error rate of $p=0.005$.
}
\label{fig:data}
\end{figure*}

With the ability to reset the maximally mixed state and the control qubit, VCP can be performed layer-by-layer, as shown in \cref{fig:layered_VCD}(a).
On the other hand, VSP can only be conducted at the end of the whole circuit as shown in \cref{fig:layered_VCD}(b), there are no existing protocols for performing VSP layer-by-layer. Such flexibility further enhances the scalability of VCP compared with VSP. For both VSP and VCP protocols to work, we need the target pure state (unitary) to be the dominant component of the noisy state (channel). Hence, for a given gate error rate, there is a maximum circuit depth beyond which \emph{VSP is not applicable} as the noisy components start to dominate over the noiseless part. On the other hand, the advantages of VCP over the unmitigated circuit can be extended to any depths by simply adding more VCP layers. This is because while the full circuit error rate is high, VCP only cares about the error rate of individual layers that VCP acts on, which can be much smaller and is unchanged if we maintain the same layer depth as the circuit depth increases. As shown in the numerical experiment in \cref{fig:data}(c), the fidelity of VSP drops to nearly zero when the circuit depth approaches the transition point. On the other hand, employing VCP layer by layer is able to maintain the fidelity at close to one over the whole range of circuit depth.

Even in the region where VSP is applicable and its dominant error component has the same error rate as that of VCP, there is still an advantage in applying layer-by-layer VCP. For simplicity, let us consider the example in which both the error channel and the corresponding noisy state have only one error component with the same error rate $P$ (more general cases are considered in \cref{sec:vcp_err}).
This error component can be suppressed to $P^M$ using VSP. 
When applying VCP to the same circuit with the circuit divided into $L$ layers, the error rate of the dominant noise component in each layer is then $\sim P/L$, which can suppressed to $(P/L)^M$ after applying VCP to each layer. Thus the total error rate of the $L$-layer circuit after VCP is $L(P/L)^M$. This is $L^{M-1}$ smaller than VSP, which is exponential in $M$.

It is worth mentioning that, if the noise channel of the circuit layer is a tensor product of local noisy channels, then applying VCP to the individual local channels will have the same effect as applying VCP to the entire circuit layers,
\begin{equation}
\left(\mathcal{E}_1\otimes\mathcal{E}_2\right)^{(M)}=\mathcal{E}_1^{(M)}\otimes\mathcal{E}_2^{(M)}.
\end{equation}
Thus, it is possible to use one single control qubit to purify many independent quantum channels.
As shown in \cref{fig:layered_VCD}(c), to purify a sequence of individual noisy gates, we can also use a single control qubit to remove the requirement for mid-circuit measurements for the VCP circuit.
The maximally mixed ancillary input can be implemented by randomly initialising its constituent qubit to $0$ or $1$. In the cases in which mid-circuit initialisation is challenging, one can simply apply random Pauli gates between two layers of VCP to obtain the maximally mixed input on the ancillary register.
We prove in Appendix~\ref{sec:impl_max_mix} that these implementations of the maximally mixed state do not change the sample complexity of VCP.
Therefore, we can remove all the requirements for mid-circuit measurement and state reset for the layer-wise VCP.

\subsubsection{Wider application scenarios}
As VCP directly mitigates the noise in quantum channels, there is no restriction on the input and output states. This is also the reason behind our ability to implement layer-wise VCP in the last section. Going further beyond layer-wise implementation, we can also apply VCP to a subset of qubits to target the noisiest gate as shown in \cref{fig:layered_VCD}(d), while VSP can only mitigate noise for the entire state as a whole \cite{koczorExponentialErrorSuppression2021,hugginsVirtualDistillationQuantum2021,hakoshimaLocalizedVirtualPurification2023}.
This offers great flexibility on the amount and the type of noise that we can mitigate, and can be particularly useful when the noise is concentrated only on a small number of qubits or some local gates. In such cases, applying VCP may require fewer \textsc{cswap}s and ancillary qubits compared to VSP while offering greater error suppression power.

In the cases where we are only interested in some local observable of a state, we can use VCP to target operations within the casual light-cone drawn from the target observable~\cite{tranLocalityErrorMitigation2023}, while VSP needs to be applied to the entire circuit. In settings like modular/networked quantum computation and quantum communication, we only have access to a part of the system, which means that we can use VCP but not VSP to mitigate the noise in these subsystems. We will further explore its usage in quantum networks in Sec.~\ref{sec:other_app}.

Mixed states are also at the heart of many quantum algorithms like quantum principal component analysis \cite{lloydQuantumPrincipalComponent2014} (which utilises mixed states to encode the data) and dissipative algorithms \cite{harringtonEngineeredDissipationQuantum2022} (which rely on engineered open system dynamics to perform computation). VSP is not applicable in these scenarios since our target state is not pure anymore, but VCP can still be applied to mitigate the noise of the unitaries inside these circuits.

\subsection{Practical Implementation}\label{sec:practical_imple}
So far we have not considered the noise in the \textsc{cswap}s needed for implementing VCP. Compared to VSP, single-layer VCP requires an additional layer of \textsc{cswap}s before the main circuit, which may introduce additional noises. It is worth noting that, any noise happening after these additional \textsc{cswap}s can be absorbed into the main circuit noise and thus can be mitigated by VCP, especially when the noise is uncorrelated among the different qubits that the individual \textsc{cswap} acts on. 
In such cases, the noise effect caused by this additional layer of \textsc{cswap}s would be less serious than expected. This effect is demonstrated in \cref{fig:data}(d), where we see single-layer VCP can offer much stronger error suppression than VSP even in the presence of strong \textsc{cswap} noise for the full range of gate error rate. The \textsc{cswap} error rate is set to $7.5$ times the two-qubit gate error rate, since we need at least five two-qubit gates to construct a \textsc{cswap} gate~\cite{smolinFiveTwobitQuantum1996}.

Moving onto multi-layer VCP discussed in the last section, the error rate after VCP \emph{without \textsc{cswap} noise}, denoted as $P_{\mathrm{c,cir}}(L)$, will decrease with the number of VCP layers $L$ and thus we should have as many VCP layers as possible. 
However, the error rate due to \textsc{cswap}s, denoted as $P_{\mathrm{c,sw}}(L)$, increases with the number of VCP layers $L$. Therefore, instead of increasing the number of layers $L$ indefinitely, we will only increase $L$ until the \emph{approximate} optimal point $L^*$ (see \cref{sec:vcp_err} for the exact optimal point) such that
\begin{align}\label{eqn:optimal_layer_cond}
    P_{\mathrm{c,cir}} (L^*) \sim P_{\mathrm{c,sw}}(L^*).
\end{align}
Increasing VCP layers beyond this point will increase the total errors since the \textsc{cswap} errors $P_{\mathrm{c,sw}}$ start to dominate. 
This is illustrated in \cref{fig:data}(e) where we have plotted how $P_{\mathrm{c,cir}}$, $P_{\mathrm{c,sw}}$ and the total error rate $P_{c,\mathrm{tot}}$ change with the number of VCP layers in a numerical experiment. We can see that the optimal number of VCP layers that minimise $P_{c,\mathrm{tot}}$ indeed occur around $P_{\mathrm{c,cir}} \sim P_{\mathrm{c,sw}}$.

In order to obtain expressions of the optimal number of VCP layers, we can zoom into each individual VCP layer whose depth is denoted as $d$ and the qubit number is denoted as $N$. We will focus on the noise regime $dNp \ll 1$ such that the unmitigated error rate per VCP layer is well approximated by $dNp$, which can be suppressed to $P_{\mathrm{c,cir}}(L)/L \propto (dNp)^M$ using VCP. There are two layers of \textsc{cswap}s in each VCP layer and thus the corresponding error rate is simply proportional to the error rate of individual gate layer $P_{\mathrm{c,sw}}(L)/L \propto Np$. As mentioned above, the optimal number of layers, or equivalently, the optimal depth of each VCP layer $d^*$, can be reached when \cref{eqn:optimal_layer_cond} is satisfied, which implies $P_{\mathrm{c,cir}}(L^*)/L^* \sim P_{\mathrm{c,sw}}(L = L^*)/L^*$ and thus:
\begin{align}
    (d^*Np)^M \propto Np \quad \Rightarrow \quad d^* \propto (Np)^{\frac{1-M}{M}}.
\end{align}
The arguments above are mainly for demonstrating the scaling of $d^*$ w.r.t. $p$. There is further dependence on $N$, through the number of error components, that is not shown here as discussed in \cref{sec:vcp_err}. If we denote the depth of the circuit as $D$, the optimal number of VCP layers is given as
\begin{align}\label{eqn:L_scaling}
   L^* = D / d^* \propto D (Np)^{\frac{M-1}{M}}
\end{align}
For a fixed gate layer error rate $Np$, we have a fixed $d^*$ and thus the number of optimal VCP layers $L^*$ is linearly related to $D$. We use numerical experiments shown in \cref{fig:data_figure_app} to validate our conclusion.

The total error rate of the circuit with the optimal number of layers is thus given as:
\begin{align*}
    P_{c,\mathrm{tot}}(L^*) \sim 2P_{\mathrm{c,sw}}(L^*) = 2\frac{P_{\mathrm{c,sw}}(L^*)}{L^*}L^* \propto D(Np)^{\frac{2M-1}{M}}.
\end{align*}
This error rate at the optimal number of layers is by definition smaller than that of single-layer VCP, which is in turn smaller than that of VSP in most cases as discussed above. 

Now given the ideas of how VCP performs with the optimal number of layers, we can try to compare it against VSP. We will focus on the parameter regime where VSP is applicable and use $P_{\mathrm{s,cir}}$ and $P_{\mathrm{s,tot}}$ to denote the error rate after VCP without and with \textsc{cswap} noise, respectively. The ratio between the error rate achieved by VSP over that of VCP with the optimal layer number as shown in \cref{sec:vsp_vcp_compare} is given as:
\begin{align}\label{eqn:improvement_ratio}
    R = \frac{P_{\mathrm{s,tot}}}{P_{\mathrm{c,tot}}} \approx \frac{1}{2}{L^*}^{M-1}\frac{P_{\mathrm{s,cir}}}{P_{\mathrm{c,cir}}(L=1)}.
\end{align}
where $P_{\mathrm{s,cir}}/P_{\mathrm{c,cir}}(L = 1)$ is simply the ratio between the error rate after VSP and single-layer VCP \emph{without any \textsc{cswap} noise}. When looking at the individual factors in $R$, we see that $P_{\mathrm{s,cir}}/P_{\mathrm{c,cir}}(L = 1)$ can increase exponentially with the number of qubits for global noise as discussed in \cref{sec:stronger_suppresion}; while $L^*$ will increase linearly with the depth of the circuit $D$ and also increase with the gate error rate $p$ following \cref{eqn:L_scaling}. Since there is implicit exponential dependence on $M$ in $P_{\mathrm{s,cir}}/P_{\mathrm{c,cir}}(L = 1)$ (see \cref{sec:vcp_err}), we also see that $R$ will increase exponentially with $M$. 

It is worth emphasising that all the expressions in this section are mainly for gaining intuitions on the scaling behaviours of the various quantities rather than being the exact expressions. More explicit formulas and the relevant assumptions made are outlined in \cref{sec:vcp_err}. There we will also show that in some rare parameter regime, $P_{\mathrm{c,cir}}$ can increase with the increase of $L$ such that single-layer VCP is the best-performing option.

We have conducted a series of numerical experiments to see the exact factor of improvement we can obtain using VCP over VSP. As shown in \cref{fig:data}(f) and (g), the ratio between the resultant infidelities using VSP over VCP is larger than $1$ throughout different gate error rates and different circuit depths, with the possibility of achieving up to $4$ times more error suppression in some region. Note that this is under a \textsc{cswap} noise that is $5$ times stronger than the gate error rate, which is close to the achievable experimental value~\cite{kimHighfidelityThreequbitIToffoli2022,chapmanHighRatioBeamSplitterInteraction2023} for most of the gate error rates we explore.  

Even though we have only performed up to $6$-qubit numerical simulations in this section ($13$-qubit if we include the control qubit and the ancillary qubits), we are performing experiments on a large enough circuit size such that the circuit fault rate (i.e. the average number of faults per circuit run) is of $\lambda \sim \order{1}$ and beyond. The circuit fault rate is the key quantity that determines the performance of QEM techniques~\cite{caiQuantumErrorMitigation2023} and $\lambda \sim \order{1}$ is expected to be the parameter regime that we will implement QEM in practice. Hence, our numerical experiments still provide a good indication of the performance of VCP in practice despite its small system size due to the use of a realistic circuit fault rate.

If all \textsc{cswap}s follow a similar error model, we can actually efficiently characterise the noise in the \textsc{cswap}s and try to mitigate their noise using probabilistic error cancellation~\cite{temmeErrorMitigationShortDepth2017}. In this way, we can remove almost all \textsc{cswap} noises to obtain the full benefit of VCP. Of course, we can also mitigate the \textsc{cswap} noise using zero-noise extrapolation~\cite{temmeErrorMitigationShortDepth2017,liEfficientVariationalQuantum2017} which is shown to be effective in VSP~\cite{koczorExponentialErrorSuppression2021}. 

\begin{figure}[htbp]
    \centering
    \includegraphics[width=0.35\textwidth]{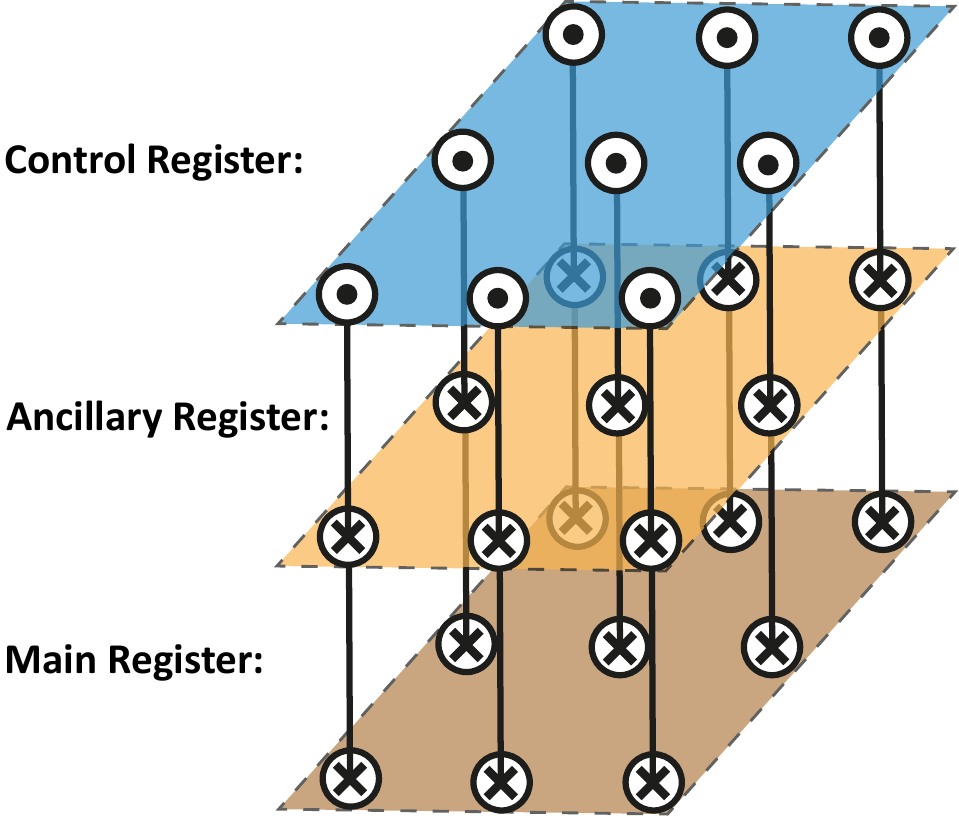}
    \caption{The hardware layout for implementing VCP with transversal \textsc{cswap}s. The control qubits are initialised in $\ket{+}$ states. We will measure their collective $X$ parity at the end.}
    \label{fig:hardware_layout}
\end{figure}

Another challenge in the physical implementation is the connectivity required for the \textsc{cswap}s. This was extensively discussed in VSP~\cite{koczorExponentialErrorSuppression2021}, and the methods there are also applicable to VCP. Furthermore, similar to what is proposed for VSP in Ref.~\cite{caiLoopedPipelinesEnabling2023}, we have shown in \cref{sec:circ_variant} that the control register can be implemented with separate control qubits for different pairs of corresponding qubits in the ancillary and main registers. In this way, the \textsc{cswap}s can be implemented using transversal physical 
\textsc{cswap}s as shown in \cref{fig:hardware_layout}. The connectivity required for such transversal operations has been demonstrated in state-of-the-art trapped ion and neutral atom platforms~\cite{ryan-andersonHighfidelityTeleportationLogical2024,bluvsteinLogicalQuantumProcessor2024}, with architectures proposed for silicon spin qubits~\cite{caiLoopedPipelinesEnabling2023}.

\subsection{Sampling Overhead}
As discussed, VCP is performed by measuring the observables outlined in \cref{thrm:purified_channel}. 
Compared to the unmitigated circuit, additional circuit runs (samples) are needed in order to obtain the VCP result since we are trying to extract useful information out of the noisy data. 
The factor of increase in the number of samples needed compared to the unmitigated circuit is called the sampling overhead. The sampling overhead for VCP can be obtained through the same arguments as the sampling overhead of VSP~\cite{koczorExponentialErrorSuppression2021,hugginsVirtualDistillationQuantum2021}, where the \emph{range} of the effective random variable we used for expectation value estimation is increased by $P_M^{-1}$, with $P_M = \sum_{k=0}^{4^N-1} p_k^{M}$ being normalising constant of the purified channel in \cref{eqn:purified_channel}. Hence, using the Hoeffding's inequality, the sampling overhead is given as
\begin{align*}
    C_{\mathrm{em}} \sim P_M^{-2}.
\end{align*}
This is similar to the sample complexity scaling obtained for VSP. 
A more exact expression is derived in \cref{sec:sample_complexity}.

Analysis of the scaling of $C_{\mathrm{em}}$ follows the same arguments in Ref.~\cite{caiPracticalFrameworkQuantum2021}, which gives a lower bound of $C_{\mathrm{em}} \sim P_M^{-2} \geq \frac{e^{M\lambda}}{1 + (e^{\lambda}-1)^M}$ with $\lambda$ being the average number of errors in each circuit run (circuit fault rate). This bound is the same as VSP~\cite{caiQuantumErrorMitigation2023}. When performing layer-by-layer VCP, the sampling overhead for the whole circuit is simply the product of the sampling overhead for individual layers.

\section{Combination with Quantum Error Correction}\label{sec:comb_qec}
\begin{figure}
\centering
\includegraphics[width=0.5\textwidth]{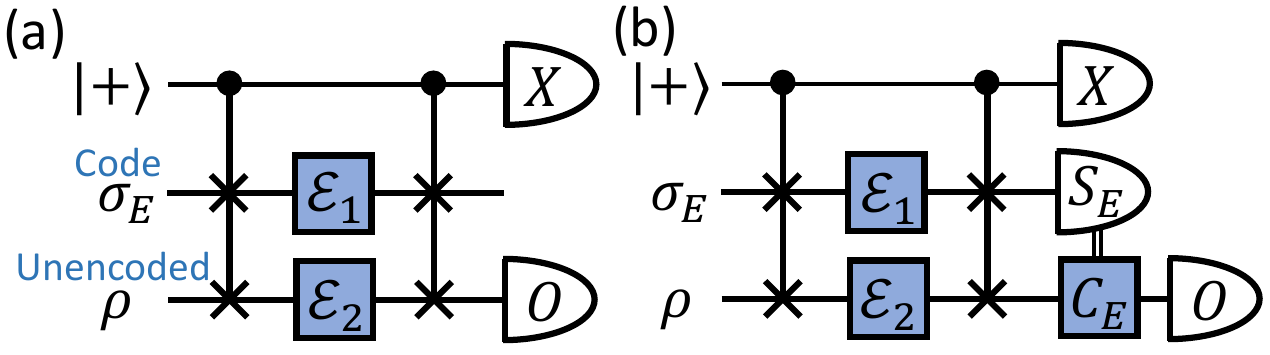}
\caption{
(a) The variant of the VCP circuit with the ancillary input state initialised as the code state.
(b) The circuit for virtual error correction (VEC), in which on top of the circuit in (a), we also measure the error syndrome of the ancillary register and perform the correction on the main register.
}
\label{fig:vec_circuit}
\end{figure}

\subsection{Merging with Quantum Error Corrections}\label{sec:comb_with_qec}
Looking back at the VCP circuit in \cref{fig:circuit}, it is counter-intuitive that the maximally mixed state, which is supposedly the most ``noisy'' state, can be used as the ancillary input for mitigating errors. It is natural to ask whether we can improve the performance of the circuit using some other ancillary input. Let us consider the circuit variant in \cref{fig:vec_circuit}(a) where the ancillary input is some state $\sigma_E$ and the Pauli noise channels acting on the ancillary register and the main register are $\mathcal{E}_1(\rho) = \sum_{i} p_i E_i \rho E_i^\dagger$ and $\mathcal{E}_2(\sigma_E) = \sum_{j} q_j E_j \sigma_E E_j^\dagger$, respectively. We will further assume that these noise elements are all correctable by some \emph{non-degenerate stabiliser code} defined by the code space projector $\Pi_E$. This code space will then need to satisfy the Knill-Laflamme condition~\cite{knillTheoryQuantumErrorcorrecting1997} for our setting:
\begin{align}\label{eqn:kl_ortho}
    \Pi_E E^\dagger_j E_i \Pi_E = \delta_{ij} \Pi_E.
\end{align} 
Being able to correct $\{E_i\}$ also implies that we can correct any linear combination of $\{E_i\}$, thus our arguments here can be extended beyond Pauli noise.

The ancillary input $\sigma_E$ we use in \cref{fig:vec_circuit}(a) will be a code state satisfying $\sigma_E \Pi_E = \Pi_E \sigma_E = \sigma_E$. Through the action of the \textsc{cswap}s and measuring the $X$ observable on the control qubit, the effective post-processed output ``state'' of the two registers before the measurement can be shown to be (see \cref{sec:vir_with_qec}):
\begin{equation}\label{eqn:output_state}
\begin{aligned}
&\sum_{i,j} p_{i}q_{j}\underbrace{\left(E_{j}\sigma_E E_{i}^\dagger\right)}_{\text{ancillary}} \otimes \underbrace{\left(E_{i}\rho E_{j}^\dagger \right)}_{\text{main}}.
\end{aligned}
\end{equation}
Here we see that the noise elements occurring on the code register and the unencoded register become entangled. 

If we directly trace out the ancillary code state at this point as shown in \cref{fig:vec_circuit}(a), the factor of contribution from the ancillary register is
\begin{align}
    \Tr(E_{i}\sigma_E E_{j}^\dagger) = \Tr(\sigma_E \Pi_E E^\dagger_j E_i \Pi_E) &= \delta_{ij}
\end{align}
using the Knill-Laflamme condition in \cref{eqn:kl_ortho}. Substituting back into \cref{eqn:output_state}, we see that all error cross terms with $i \neq j$ are removed, and we can achieve the same performance as VCP just like when we were using maximally mixed ancillary input.

To remove even more noise than before, we can perform measurements on the ancillary code register. By measuring the code projector $\Pi_E$ on the ancillary code state at the end, we have
\begin{align}\label{eqn:purified_cond_qec}
    \Tr(\Pi_E E_{i}\sigma_E E_{j}^\dagger) = \Tr\left(\Pi_E E_i\Pi_E\sigma_E \Pi_EE_j^\dagger\Pi_E\right)=\delta_{i0}\delta_{j0}
\end{align}
using the Knill-Laflamme condition in \cref{eqn:kl_ortho}. Substituting back into \cref{eqn:output_state}, we see that \emph{all error terms are removed}, leaving only $E_0 = \mathbb{I}_{2^N}$. The measurement of the code projector $\Pi_E$ can be carried out through stabiliser measurement and post-selecting on the all-zero syndrome. 
It can also be carried out at a lower hardware cost via single-qubit Pauli measurements and post-processing~\cite{mccleanDecodingQuantumErrors2020}. This additional post-selection of the ancilla stabiliser measurements will further increase the sampling cost of our protocol from $(\sum_{i} p_iq_i)^{-2}$ to $(p_0q_0)^{-2}$ as shown in \cref{sec:circuit_generalisation}, in line with the usual bias-variance tradeoff in QEM~\cite{caiQuantumErrorMitigation2023} which states that more sampling cost is needed for removing more errors. 

We can break this bias-variance trade-off by making use of the error syndrome obtained from the stabiliser measurements and performing active corrections to properly merge QEC into our QEM protocol, which is shown in \cref{fig:vec_circuit}(b). The stabiliser measurements on the ancillary code register will project the errors acting on the code register to some error $E_k$ that corresponds to the syndrome we obtain. Due to the entanglement between the errors on the code register and the unencoded register, the error on the unencoded register will also be projected into $E_k$. Hence, we can perform the corresponding correction $C_E = E_k$ on the \emph{unencoded} register, which will \emph{remove all the noise in the main register using the same sampling cost as 2nd-order VCP} (shown in \cref{sec:mix_unencode_code}). This whole process of virtually entangling the errors between the two registers, performing stabiliser measurements on the code registers and applying the corresponding correction on the unencoded register will be called \emph{virtual error correction} (VEC). Compared to VCP, here we have removed more noise without increasing the sampling cost, thus breaking the bias-variance tradeoff in QEM~\cite{caiQuantumErrorMitigation2023}, demonstrating the power of such a native combination between QEM and QEC. 

So far for simplicity, we have only considered correctable errors acting on the registers. In practice, uncorrectable errors can also occur. Just like in QEC, we need to choose the right code (and the right code size) such that the uncorrectable errors in the noise channels are small enough for our purpose. Other than the requirement on the amount of uncorrectable noise for the given code, there are no other restrictions on the noise channel acting on the two registers in VEC. The two noise channels can be very different and unlike in VCP, they do not need to have the identity as the dominant component.

\subsection{Practical Applications}\label{sec:prac_vec}

In the discussion above, we have not mentioned the noise in the encoding circuit and the stabiliser measurements (note that encoding can also be achieved via stabiliser measurements). Our arguments will still be valid if the noise in encoding and stabiliser checks is much weaker than the noise channels we try to mitigate. This can happen, for example, in the communication setting where the noise in the transmission is dominating over the noise in any local operations. And this will likely happen in the (early) fault-tolerant era in which we can confidently prepare and maintain a code state. 

Even though we can use QEC to tackle noise in the fault-tolerant era, we can protect orders of magnitude more qubits by directly protecting the unencoded register using VEC without any encoding qubit overhead. Let us consider the case where there is a noise channel $\mathcal{E}$ whose noise can be suppressed using a code with $K$ physical qubits per logical qubit, and suppose we want to transmit a $K$-qubit state $\rho$ through this noisy channel $\mathcal{E}$ (for measuring some observable $O$ on the state). We can protect $\rho$ using VEC as shown in \cref{fig:vec_circuit}(b), which will require $2K + 1$ qubits. If we go for the pure QEC approach, then we need to encode our $K$-qubit state $\rho$ into $K^2$ qubits in order to be resilient against $\mathcal{E}$. Hence, we are able to achieve $\order{K}$ qubit overhead saving by using VEC, which often translates into orders of magnitude qubit saving in practice due to the large qubit overhead of many practical QEC codes like surface codes. In \cref{sec:apply_vec}, we also discuss the case where the number of qubits in $\rho$ is $N \neq K$. The same $\order{K}$ qubit saving can still be achieved when $N \gtrsim K$. A factor of $N$ saving is achieved instead when $N \ll K$. Furthermore, the process of encoding a general $\rho$ into the code can be very costly, while in our protocol, $\sigma_E$ can be any code state, which can be chosen to one that is easiest to prepare.  In VEC, there will be additional noise coming from the control qubit and the \textsc{cswap}s. These can be mitigated using other QEM techniques like probabilistic error cancellation and zero-noise extrapolation as mentioned in \cref{sec:practical_imple}.

Besides protecting against phenomenological (environmental) noise channels, our setup can also protect against noise arising from gate operations. Just like VCP, we need to apply the same target operation to both the encoded and unencoded register in VEC and moreover the operation needs to be a logical operation for the code register so that its noise can be corrected. In the cases where part (or all) of the circuits for encoding and stabiliser checks happen to be the operation that we want to perform on our main unencoded register, we can also sandwich these circuits in between the \textsc{cswap}s such that their noise can also be mitigated. Since VEC can use any code state in the given code as the ancillary input, this offers a degree of freedom that we can optimise over in order to choose the most suitable encoding circuit for our purpose. Considering the practical implementation of our techniques and extending its possible application scenarios to more general noisy operations will be a very interesting direction to explore. Here in this paper, with our focus being demonstrating the non-trivial conceptual shifts arising from the combination of QEC and VCP, we will for simplicity continue to focus on the phenomenological noise model in this section. 

\subsection{Combining Two Different Codes}
\begin{figure}
\centering
\includegraphics[width=0.5\textwidth]{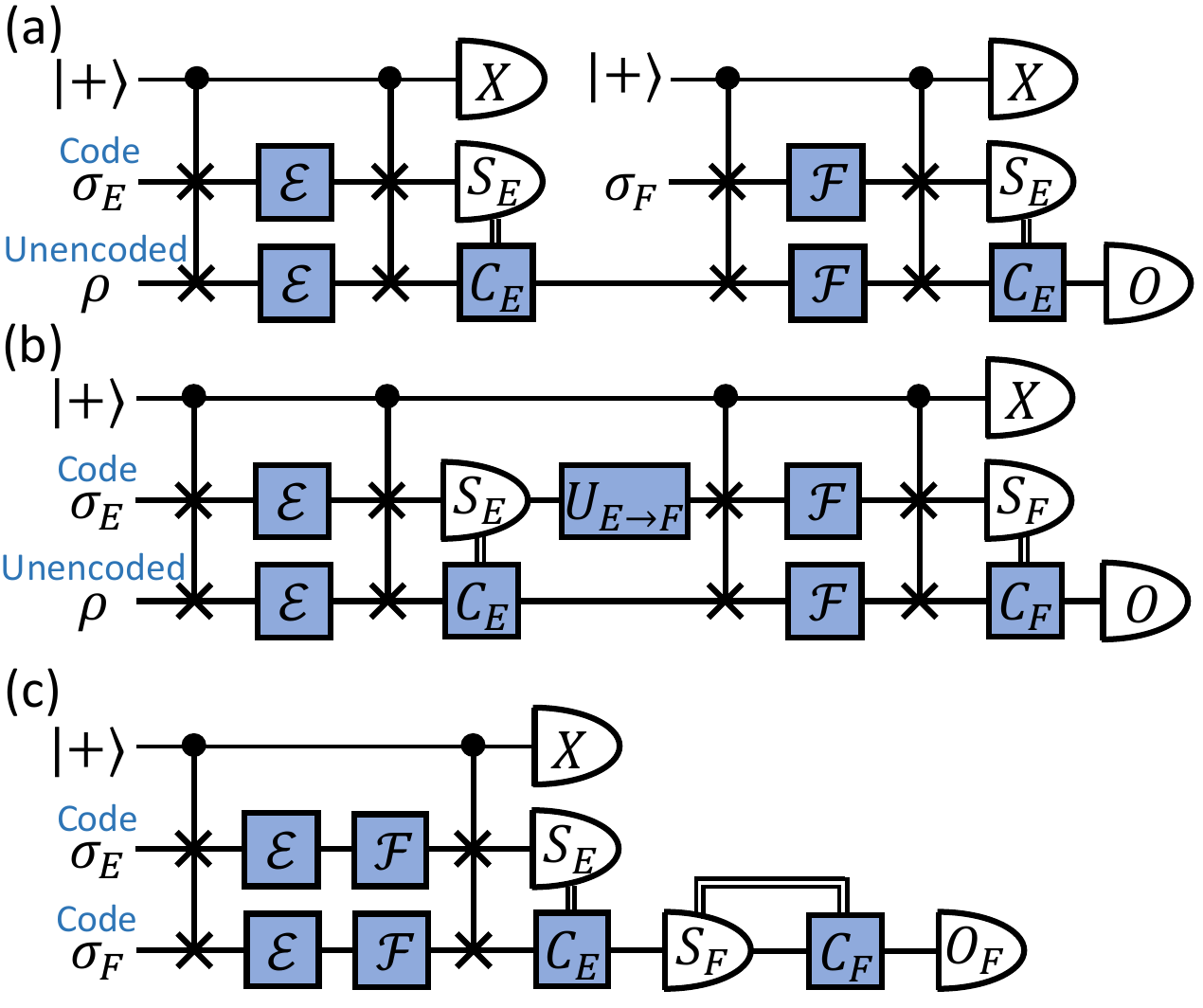}
\caption{
(a)~The VEC circuit for tackling two consecutive error channels separately. The circuit here measures $\Tr(O\rho)$. 
(b)~A variant of (a) with code transformation unitary.
(c)~The VEC circuit for tackling the combination of two error channels. The circuit here measures $\Tr(O_F\sigma_F)$. Note that in (c) $\mathcal{E}$ and $\mathcal{F}$ cannot share error elements, e.g. one is bit-flip and one is phase-flip. Here for simplicity we have shown the same $\mathcal{E}$ and $\mathcal{F}$ acting on the two registers. However, the noise acting on the two registers need not to be the same as discussed at the end of \cref{sec:comb_with_qec}.
}
\label{fig:vec_circuit_gen}
\end{figure}
An interesting scenerio to consider is when we need to transmit our state through consecutive error channels $\mathcal{E}$ and $\mathcal{F}$. To tackle this using QEC, we can build a code state that is resilient against both error channels. On the other hand, using VEC, we can mitigate the $\mathcal{E}$ and $\mathcal{F}$ separately using different codes as shown in \cref{fig:vec_circuit_gen}(a). These codes will typically require much fewer qubits than the code that can correct both errors. For example, if $\mathcal{E}$ is some bit-flip channel and $\mathcal{F}$ is some phase-flip channel, then a code that can detect purely bit-flip or phase-flip errors up to weight $d-1$ will require $d$ qubits, while a surface code that can detect both types of errors will require $d^2$ qubits. Furthermore, the circuits needed to construct the bit-flip and phase-flip codes are much simpler than that of the surface code. Hence, VEC can be easily extended to deal with different error channels in the circuit separately, resulting in lower qubit overhead and simpler circuits. 

In the above setting, another option for using QEC is to perform code transformation from code $E$ to code $F$ in between the two noise channels. Similar methods have been explored in Ref.~\cite{piveteauErrorMitigationUniversal2021} in the context of concatenating QEM on top of QEC, where they tried to harness the powers of multiple codes using code switching and post-processing. For stabiliser codes, this code transformation can be carried out by applying Clifford circuits to transform the stabilisers or via code deformation that gradually changes the set of stabiliser checks we perform. Thus in general it can be more complicated than the VEC circuit. However, in the special case of transforming between bit-flip and phase-flip code, only transversal Hadamard is required. Hence, in this setting, QEC with code transformation is also a good option. However, we need to keep in mind that the lower qubit overhead of VEC discussed in the last section (\cref{sec:prac_vec}) still applies since we are transmitting information using the unencoded state $\rho$ in VEC, i.e. if $\rho$ is a $d$-qubit state, we will require $\order{d}$ qubits using VEC while we need $\order{d^3}$ qubits for QEC. The transformation between the two codes can also be used for a different implementation of VEC as shown in \cref{fig:vec_circuit_gen}(b).

\begin{figure}
\centering
\includegraphics[width=1.0\linewidth]{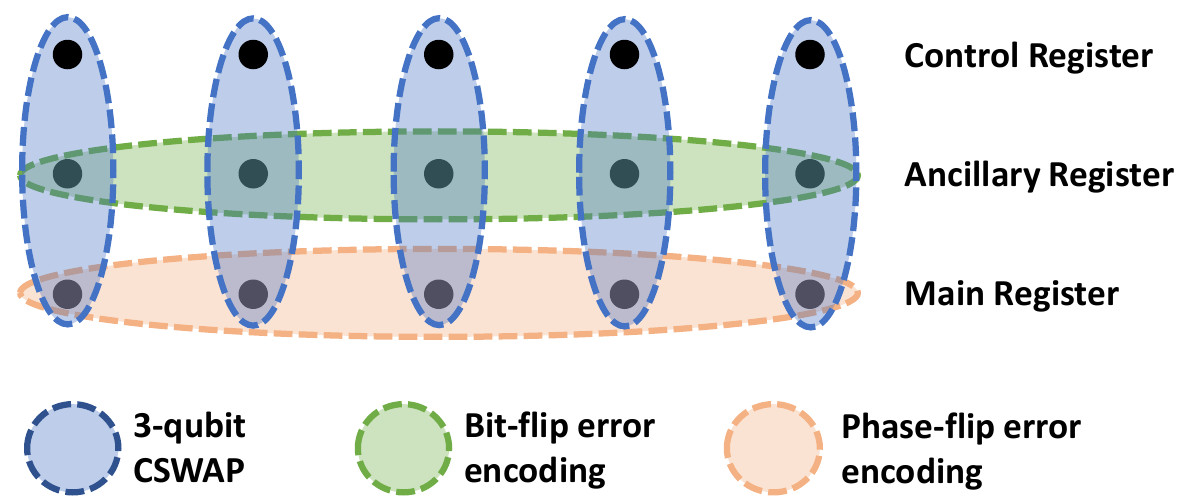}
\caption{Connectivity of VEC. All the qubits in the control register are initialised in the $\ket{+}$ state.}
\label{fig:connectivity}
\end{figure}

Compared to the setting above in which the two error channels happen in stages, a more practical scenario is when the error channels $\mathcal{E}$ and $\mathcal{F}$ occur simultaneously, which means we are not allowed to perform any operations in between the noise channels. In this case, when using QEC, we cannot perform code transformation and can only use a code that can tackle both  $\mathcal{E}$ and $\mathcal{F}$. When using VEC, we can no longer deal with the two noise channels separately. However, we can use a variant of VEC circuit shown in \cref{fig:vec_circuit_gen}(c) that inputs two different code states instead of one code state and one unencoded. The logical information that we are interested in is encoded in the code state $\sigma_F$ here. The two code states here are targeting the two error channels. In the case where $\mathcal{E}$ and $\mathcal{F}$ are bit-flip and phase-flip channels, respectively, we have shown in \cref{sec:mixing_code_state} that by inputting a distance-$d$ bit-flip and phase-flip repetition code at the two registers and performing the respective stabiliser checks, the resultant syndromes will tell us the bit-flip and phase-flip errors occurring to \emph{both} registers since their errors are entangled. By applying the corresponding bit-flip and phase-flip corrections to any one of the registers, we can correct any combination of bit-flip and phase-flip errors up to weight $\lfloor d-1/2 \rfloor$ for that register. Hence, we are able to achieve the same error correction power as a distance-$d$ surface code while reducing the qubit overhead from $\order{d^2}$ to $\order{d}$ and also have simpler code construction circuits. Note that the control qubit can instead be implemented using a tensor product of $\ket{+}$ states plus additional post-processing as outlined in \cref{sec:circ_variant}, which enables the \textsc{cswap}s to be implemented using a tensor product of physical \textsc{cswap}s. In this way, since the repetition codes only require linear connectivity, \cref{fig:vec_circuit_gen}(c) can be implemented using just three lines of qubits corresponding to the two code registers and the control register, respectively, as shown in Fig.~\ref{fig:connectivity}.

We show in \cref{sec:mixing_code_state} that our arguments above will be generally true when the two codes do not share any correctable elements besides the identity, which will be applicable for combining any bit-flip and phase-flip codes built from any classical codes. It will be interesting to extend beyond the repetition code examples and explore new application scenarios.

If we want to tackle noise arising from gate operations, then similar to before, we will need the same operation to be applied to both registers such that the same noise channel will arise, and this operation needs to be the logical operation for both registers. An example of such operations will be transversal $X$ or $Z$ in our example. In this sense, operating on our entangled code state is more restricted than surface codes. 

In all of our discussion above, we can also use the ancillary code register to transmit information by also applying the corresponding correction on them. They will typically carry less information than the main register since the main register is unencoded in many cases above, but this will be very relevant when both registers are encoded.

\subsection{Implementing QEM before QEC}\label{sec:qem_before_qec}
At the end of \cref{sec:practical_imple}, we have mentioned how the errors in the \textsc{cswap}s in VCP can be removed using other QEM techniques like probabilistic error cancellation (PEC), here we will see how this can be extended to VEC. The native application scenario for QEM is on unencoded physical qubits, and it can also be easily implemented on logical qubits of early fault-tolerant machines, where QEM is performed \emph{after} (on top of) QEC to tackle the left-over logical noise~\cite{suzukiQuantumErrorMitigation2022,piveteauErrorMitigationUniversal2021}. On the other hand, to deal with the noise in the \textsc{cswap}s in the VEC circuit, we actually need to perform QEM \emph{before} (underneath) the QEC process. It is less obvious on whether this can be directly implemented since the additional operations introduced by QEM will interact with the QEC process. We will provide a simple proof on why performing QEM before QEC will still give us the same target results. 

Here we will use the process matrix notation where density matrix $\rho$ is written as a vector $\pket{\rho}$ and in line with the notation earlier, we will use ${\supop{U}}\pket{\rho} = \pket{U\rho U^\dagger}$ to denote the super-operator for a given operator $U$. 
In the context of QEC, for a given incoming noisy state $\pket{\rho}$, when the stabiliser measurement outputs the syndrome $s$, it will collapse the incoming state into the output state $\frac{1}{p_s}{\supop{\Pi}}_s \pket{\rho}$ where $\Pi_s$ and $p_s$ are the projector and the probability for the corresponding syndrome. We will then apply the corresponding correction operations $\mathcal{C}_s$, giving the resultant state of $\frac{1}{p_s}\mathcal{C}_s{\supop{\Pi}}_s \pket{\rho}$. Taking the mixture of all the possible outcomes, we have the resultant state: 
\begin{align*}
    \sum_s p_s \frac{1}{p_s}\mathcal{C}_s{\supop{\Pi}}_s \pket{\rho} = \sum_s \mathcal{C}_s{\supop{\Pi}}_s \pket{\rho}.
\end{align*}
i.e. the quantum error correction process can be viewed as a quantum channel of the form:
\begin{align*}
    \mathcal{R}_{\mathrm{qec}} = \sum_s \mathcal{C}_s{\supop{\Pi}}_s.
\end{align*}

Suppose our goal is to measure some observable $O$, then the target expectation value on the error-corrected state is given by
\begin{align*}
    \expval{O}_{\mathrm{qec}} = \pbra{O}\mathcal{R}_{\mathrm{qec}}\pket{\rho}.
\end{align*}
Now if we have some additional noise $\mathcal{E}$ (on top of the existing noise in $\pket{\rho}$) happens before the syndrome measurement, we can try to counteract its effect using probabilistic error cancellation (PEC)~\cite{temmeErrorMitigationShortDepth2017}. There we will implement inverse of the noise channel $\mathcal{E}^{-1} = \sum_{i} \beta_i \mathcal{B}_i$ by breaking it down into a linear combination of implementable basis operations $\{\mathcal{B}_i\}$ with coefficients $\{\beta_i\}$ that can be negative. By running the circuit with different $\mathcal{B}_i$ implemented after the noise channel $\mathcal{E}$ and then combining the results via post-processing, we have
\begin{align*}
    \expval{O}_{\mathrm{mit}} &= \sum_{i} \beta_i \pbra{O}\mathcal{R}_{\mathrm{qec}} \mathcal{B}_i\mathcal{E}\pket{\rho}\\
    &= \pbra{O}\mathcal{R}_{\mathrm{qec}} \mathcal{E}^{-1}\mathcal{E}\pket{\rho}\\
    &=\pbra{O}\mathcal{R}_{\mathrm{qec}} \pket{\rho} = \expval{O}_{\mathrm{qec}}
\end{align*}
i.e. the application of PEC is not affected by QEC channel $\mathcal{R}_{\mathrm{qec}} = \sum_s \mathcal{C}_s{\supop{\Pi}}_s$ behind it at all. In the specific case here, we have tried to use PEC to cancel out all the noise components in $\mathcal{E}$. However, if there are noise components in $\mathcal{E}$ that can directly be tackled by the given QEC code $\mathcal{R}_{\mathrm{qec}}$ (on top of the existing noise in $\pket{\rho}$), then we can leave these correctable components outside PEC to save the sampling cost. In \cref{sec:cswap_noise_vec}, we have explicitly shown how the arguments above can be applied to the \textsc{cswap} noise in VEC.

The key insight here is because the QEC channel, being a valid quantum channel, acts linearly on the incoming quantum states. This enables us to take a quasi-probability mixture of the different circuit configurations to cancel out their noise. In this way, we can see that \emph{QEC is also compatible with other QEM techniques that utilise linear combinations of different circuit configurations}, which includes most of the mainstream QEM techniques~\cite{caiPracticalFrameworkQuantum2021}.

\section{Application beyond computation}\label{sec:other_app}
So far, we have been keeping all output states from the VCP circuit and post-processing them to \emph{virtually} prepare the purified channel for recovering the measurement statistics. If we simply post-select the measurement outcome of the control qubit without further post-processing, we can \emph{physically} obtain a quantum channel with higher purity from copies of noisy ones for tasks beyond expectation value estimation. As shown in \cref{sec:post_select_gen}, by performing a post-selection on the control qubit, we have the following result.

\begin{corollary}\label{coro:channel_purification}
In the VCP circuit given in \cref{fig:higher_order}(a) without the measurement of the observable $O$. Suppose the $M-1$ ancillary registers are initialised in the maximally mixed state, and all of these $M$ registers are acted transversally by $M$ copies of the noisy channel $\mathcal{U}_{\mathcal{E}}$. After applying the quantum circuit, upon measuring $+$ on the control qubit, which occurs with probability $P_+=\frac{1}{2}\left(1+P_M\right)$, and with the ancillary system discarded, the effective channel acting on the main register is 
\begin{equation}
  \mathcal{U}_{\mathcal{E}_{+}}=\frac{1}{1+P_M}\left[\mathcal{U}_{\mathcal{E}}+P_M\mathcal{U}_{\mathcal{E}^{(M)}}\right],
\end{equation}
where $P_M=\sum_{i=0}^{4^N-1}p_i^M$.
\end{corollary}
The protocol effectively mixes in some purified noise components $\mathcal{U}_\mathcal{E}^{(M)}$, improving the purity and the fidelity of the channel. One would notice that performing post-selection on the control qubit in this case shares a lot of similarities to performing quantum error detection using a space-time check as discussed in \cref{sec:detector}.

\begin{figure}[htbp!]
\centering
\includegraphics[width=0.48\textwidth]{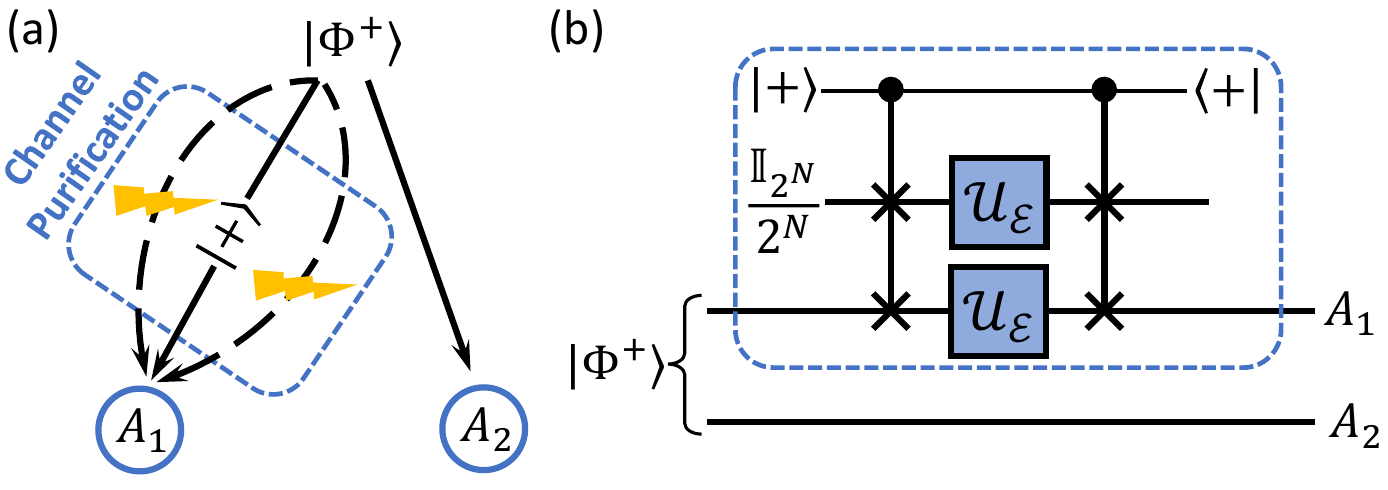}
\caption{(a) Application of channel purification in entanglement distribution. Blue circles represent two users and $\ket{\Phi^+}$ is the entangled state that is distributed. In the case where only some of the quantum channels are noisy, denoted by the dashed lines with flashes, the relevant users can apply channel purification using one clean channel to obtain a purified channel. 
(b) The circuit for channel purification. Compared to the VCP circuit, we post-select on the measurement results of the control qubit and keep only the $\ket{+}$ outcome. The dashed blue box corresponds to the purification process in (a). From the perspective of entanglement-assisted quantum communication, $A_2$ corresponds to the reference system. }
\label{fig:GMEDist}
\end{figure}

The ability to physically purify a quantum channel can benefit applications where high-quality quantum resources are indeed necessary, such as high-fidelity entanglement distribution for quantum networks~\cite{azumaToolsQuantumNetwork2021}. 
This is because the input state $\rho$ of the VCP circuit can be arbitrary states, including the subsystem of a large entangled state.
In \cref{fig:GMEDist}(a), we depict the simplest case with two users. 
Due to factors like platform hardware constraints and network topology, while some channel links are behaving normally, represented by the solid lines, some others may suffer from a high level of noise, represented by the dashed lines. Notably, not all the users in the network may suffer from a highly noisy channel link, such as $A_2$ in the figure.
By applying the channel purification scheme in \cref{coro:channel_purification}, the user $A_1$ can focus on the ill-behaved channel links and apply the channel purification scheme without disturbing the other user, resulting in a high-quality network for entanglement distribution. 
The channel purification scheme uses a single clean channel link to transmit one clean qubit for purifying arbitrarily large quantum channels. 
In practical situations, the clean qubit can be reasonably facilitated, such as by employing the path freedom in a linear optics system~\cite{eckerExperimentalSingleCopyEntanglement2021}.

For high-fidelity entanglement distribution, our channel purification scheme can be advantageous compared to applying entanglement distillation after receiving multiple copies of a state. As stated above, the channel purification scheme is simply constrained to the users suffering from noisy links. 
On the other hand, typical entanglement distillation schemes~\cite{bennettMixedstateEntanglementQuantum1996}, including those not restricted to local operations and classical communication~\cite{childsStreamingQuantumState2023}, involve all the users over various nodes, who must collect copies of the distributed state and perform operations jointly. Moreover, in many applications, the channel links may be easily re-used, like the optical fibres in a communication network. On the contrary, preparing a highly entangled state can be costly. By purifying the channel links in advance, a high-quality entangled state can be shared among the users in every round of entanglement distribution. These features render our protocol to be resource-efficient. 
Moreover, the channel purification protocol permits us to activate some valuable quantum resources, such as entanglement, which is forbidden for typical entanglement distillation schemes.

We take a channel-theoretic viewpoint and quantitatively compare the channel behaviour before and after channel purification. The entanglement distribution scenario can be regarded as an entanglement-assisted quantum communication protocol~\cite{devetakFamilyQuantumProtocols2004,devetakResourceFrameworkQuantum2008}, as shown in \cref{fig:GMEDist}(b). The entanglement source, which we call the sender in this protocol, sends a part of a maximally entangled state into the circuit, depicted as one half of $\ket{\Phi^+}$, and leaves the other part untouched, which we call the reference system. The two users $A_1$ and $A_2$, which we jointly call the receiver, collect the joint state of the main system and the reference system. The final joint state is the Choi state of the channel in the dashed blue box. For simplicity, we consider $\mathcal{U}_{\mathcal{E}}$ to be the depolarising channel characterized by \cref{eq:depolarizing}, which may turn the state into the fully mixed state with probability $P$,
and take $M=2$ copies of the noisy channel in the channel purification scheme. We consider a single-qubit channel ($N=1$) and a two-qubit channel ($N=2$), respectively. 

\begin{figure}[htbp!]
\centering
\includegraphics[width=0.47\textwidth]{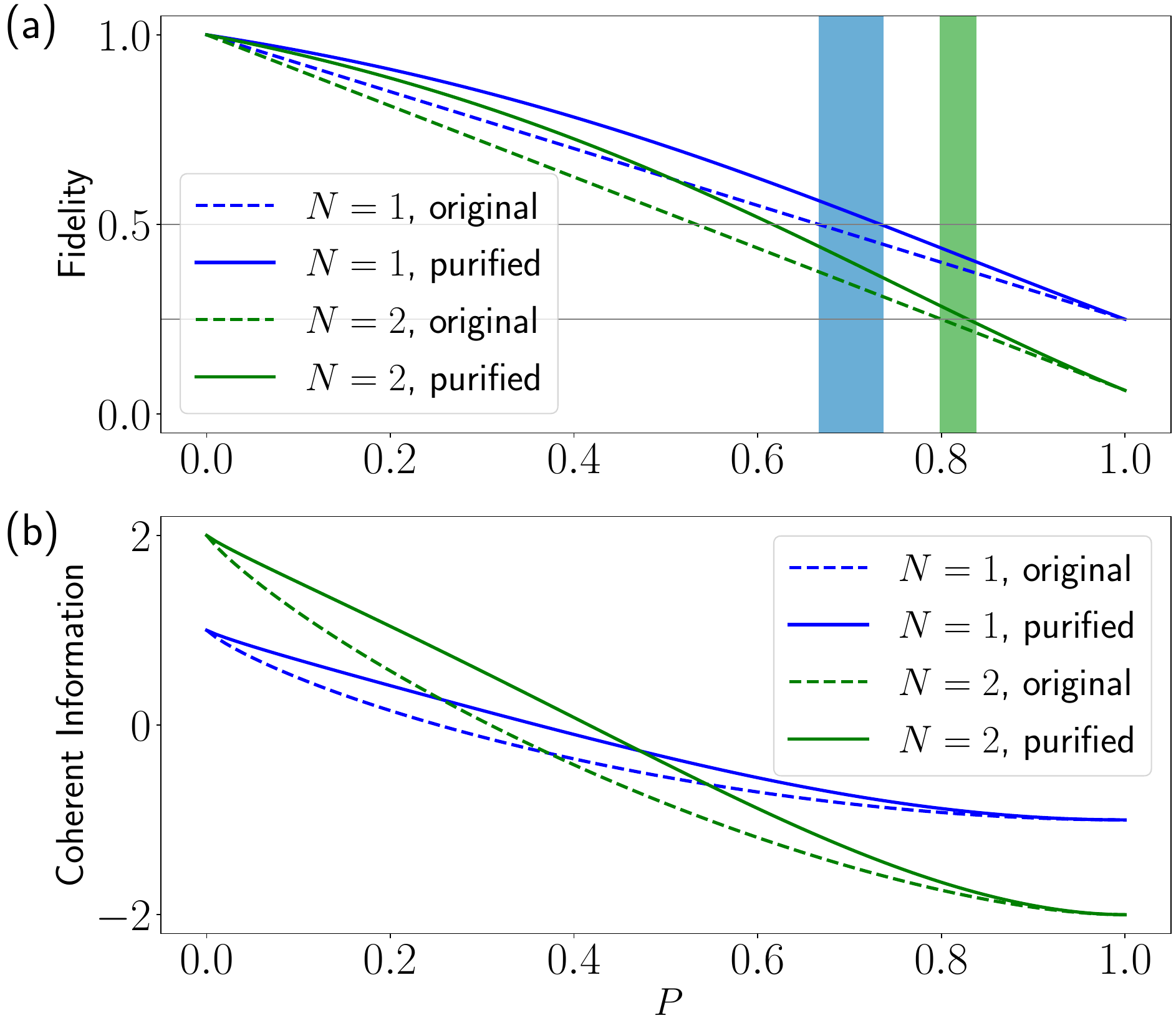}
\caption{Channel performance before (dashed lines) and after (solid lines) channel purification. 
(a) The fidelity of the distributed state with respect to the maximally entangled state for different qubit numbers. Phenomena of entanglement activation are observed in ranges denoted by blue and green shadows.
(b) The channel coherent information with respect to the maximally entangled state for different numbers of qubits.
}
\label{fig:Fidelity}
\end{figure}

We first quantify the state fidelity with respect to the maximal entangled state, $\bra{\Phi^+}\rho\ket{\Phi^+}$, with $\ket{\Phi^+}=\sum_{i=0}^{2^N-1}\ket{ii}/\sqrt{2^N}$ and $\rho$ being the distributed state. In \cref{fig:Fidelity}(a), we depict the final state fidelity with respect to $P$ when it is directly transmitted through the noisy channel and transmitted through the purified channel. 
As can be seen from the curves in blue, after applying the channel purification, the state fidelity is boosted. In particular, in some range of $P$ where a direct state transmission breaks entanglement, the output state through a purified channel can become entangled. That is, the fidelity becomes higher than $0.5$ for $N=1$ and $0.25$ for $N=2$. 

From another perspective, the fact that the Choi state of the channel becomes entangled implies a nontrivial quantum channel capacity, where the Choi states of all entanglement-breaking channels are separated states~\cite{horodeckiEntanglementBreakingChannels2003}. As a figure of merit, we calculate the coherent information of the channels with respect to the maximally entangled state. For a channel $\mathcal{U}_\mathcal{E}$, the measure $I_C(\mathcal{U}_\mathcal{E},\ket{\Phi^+})$ is defined as
\begin{equation}
I_C(\mathcal{U}_\mathcal{E},\ket{\Phi^+})=S_E(\Tr_R(\mathcal{U}_\mathcal{E}\otimes\mathcal{I}[\Phi^+]))-S_E(\mathcal{U}_\mathcal{E}\otimes\mathcal{I}[\Phi^+]),
\end{equation}
where $\mathcal{U}_\mathcal{E}$ and the identity channel $\mathcal{I}$ each acts on a half of the state, $\Phi^+$ represents the density matrix of $\ket{\Phi^+}$, $S_E(\cdot)$ represents the von Neumann entropy, and the subscript $R$ represents the reference system. This measure can be viewed as a capacity measure for entanglement transmission and serves as a lower bound on the quantum channel capacity~\cite{schumacherQuantumDataProcessing1996,devetakPrivateClassicalCapacity2005,minarPhasenoiseMeasurementsLongfiber2008}. As seen from \cref{fig:Fidelity}(b), the coherent information of the purified channel is also boosted after the purification scheme.

There are other schemes like error filtration~\cite{gisinErrorFiltrationEntanglement2005,vijayanRobustWstateEncoding2020,leeErrorSuppressionArbitrarySize2023} and coherent routing~\cite{chiribellaQuantumShannonTheory2019,abbottCommunicationCoherentControl2020,kristjanssonQuantumNetworksCoherent2023} that also use a control system to superpose channels coherently to combat the noise. However, the original error filtration scheme mainly works for communication tasks where the target channel is restricted to the identity channel. The coherent routing schemes require the knowledge of how the channel couples the transmitted state and the environment to maximize its efficacy. 
In contrast, our channel purification scheme works for a general target unitary. Besides, by employing a fully mixed state in the ancillary system, we can simply focus on the subsystem of interest, removing the additional requirement for a benchmark or even tomography of the channel.

\section{Conclusion}\label{sec:conclusion}
In this article, we have introduced virtual channel purification (VCP), which is a way to virtually implement a purified channel using multiple copies of the noisy channels, and the infidelity of the purified channel is suppressed exponentially with more copies of the noisy channels. 
Compared to its state counterpart, virtual state purification (VSP), VCP offers stronger noise suppression due to the larger dimensionality of channels compared to states (exponentially stronger with more qubits in the specific case of globally depolarising noise).
While VSP requires the ideal output state to be a pure state, VCP places no such restrictions as it directly purifies the noisy channel and thus can be applied to any input state. 
Furthermore, while VSP can only be applied to the entire quantum circuit as a whole, VCP can be applied in a layer-by-layer manner or even target specific gates in the circuits. 
This provides flexibility and, more importantly, removes the restriction in VSP that the noiseless component must be the dominant component for the noisy circuit output. 
In this way, VCP is applicable to much deeper and noisier circuits than VSP.

When considering the practical implementation of VCP, the single-layer variant only requires one additional layer of \textsc{cswap}s compared to VSP. 
Furthermore, the noise of this additional layer of \textsc{cswap} is naturally mitigated by VCP itself. 
We have seen in numerics that the errors due to \textsc{cswap}s are essentially the same for both VCP and VSP. 
Hence, single-layer VCP is almost always preferred over VSP due to its stronger noise suppression power at a similar implementation error cost. We can further optimise the number of VCP layers to outperform the single-layer variant. In numerical simulation across a range of circuit depths and gate error rates (with the \textsc{cswap} error rate five times stronger), optimal-layer VCP always outperform VSP and can offer up to 4 times more error suppression. The advantage is expected to be even stronger using more copies of noise channels and more qubits.

Thus far in the standard configuration of VCP, the additional copies of the noisy channels are acting on a maximally mixed input state at the ancillary registers. Using the Knill-Laflamme condition, we show that we can also perform purification using a quantum error correction (QEC) code state as the ancillary input state. If we are able to perform \emph{perfect} stabiliser checks on the ancillary register and apply the correction on the \emph{main register} based on these check results, we can actually remove \emph{all correctable noise} from the main register while paying only the sampling cost of the second-order VCP. This provides one of the first frameworks that seamlessly combines QEM and QEC beyond the concatenation of QEC and QEM~\cite{piveteauErrorMitigationUniversal2021,lostaglioErrorMitigationQuantumAssisted2021,suzukiQuantumErrorMitigation2022}. It allows us to achieve the same level of protection as QEC on an unencoded register when transmitting a state through a noisy channel, removing the associated encoding qubit overhead. Using the same circuit but two different QEC codes as inputs, we are actually able to combine the error suppression power of these two codes under environmental noise, e.g. we can achieve the same error resilience as the surface code using a bit-flip code and a phase-flip code as inputs. The current analysis of VEC does assume perfect syndrome extraction, but it opens up new possibilities for exploring possible practical schemes for combining QEC and QEM. We have also shown that QEM can be applied before QEC without degrading each other's performance, which allows us to use QEM to remove the noise in the \textsc{cswap}s in VEC. There can be many more application scenarios for using QEM before QEC.

Using the same circuit as VCP, but performing post-selection on the controlled qubit measurement results instead of post-processing, we can physically (instead of virtually) obtain a purified channel, which can then be applied to many tasks in quantum networks. 
For example in entanglement distribution, compared to previous works based on entanglement purification, our channel purification protocol does not require multipartite joint operations and multiple identical copies of the distributed states for local quantum noises. Furthermore, with a single clean channel to transmit a single clean qubit, our protocol can purify noisy channels with arbitrarily large dimensions and enable the activation of valuable quantum resources.
The fact that our protocol is applicable to arbitrary incoming states and arbitrary incoherent noise also opens up the door for many other possible applications in communication. 

Due to the presence of the controlled permutation operator at the beginning of the circuit which is coherently connected to the controlled permutation operator at the end, VCP actually lies outside the QEM frameworks presented for the discussion of the fundamental limits of QEM~\cite{takagiUniversalSamplingLower2023,tsubouchiUniversalCostBound2023,quekExponentiallyTighterBounds2024}. Hence, hopefully VCP can inspire a new range of QEM protocols outside these frameworks, like the combination with symmetry verification as we have mentioned in \cref{sec:additional_symmetry}. One can also develop more general frameworks of QEM that incorporate VCP, which may have more desirable properties compared to the previous QEM framework. 

One promising way to do this is by finding deeper connections to QEC. Our current discussion on VEC is still restricted to non-degenerate codes, noiseless syndrome extraction and second-order purification. It will be interesting to see how we can generalise our arguments beyond these restrictions. 
One may also want to search for other QEM methods that can be naturally merged with QEC beyond concatenation, similar to what we have done. This can be the start of a more general error suppression framework that naturally incorporates both QEM and QEC.
Further possible combination between VCP and VEC and other quantum information techniques like classical shadow \cite{seifShadowDistillationQuantum2023,huLogicalShadowTomography2022} will also be an interesting direction to explore.

\section*{Acknowledgements}
The authors would like to thank Simon Benjamin, Jinzhao Sun, Bálint Koczor, Ryuji Takagi, Wenjun Yu, Wentao Chen, Hong-Ye Hu, You Zhou, Yuxuan Yan, Xiangjing Liu, Rui Zhang, and Yuxiang Yang for insightful discussions and suggestions.
ZL acknowledges support from the National Natural Science Foundation of China Grant No.~12174216 and the Innovation Program for Quantum Science and Technology Grant No.~2021ZD0300804 and No.~2021ZD0300702..
XZ acknowledges support from the Hong Kong Research Grant Council through the Senior Research Fellowship Scheme SRFS2021-7S02.
ZC acknowledges support from the EPSRC QCS Hub EP/T001062/1, EPSRC projects Robust and Reliable Quantum Computing (RoaRQ, EP/W032635/1), Software Enabling Early Quantum Advantage (SEEQA, EP/Y004655/1) and the Junior Research Fellowship from St John’s College, Oxford.

\appendix

\section{VCP Protocol and its Generalisation}
\subsection{Validation of VCP Circuit}\label{sec:pur_circ_output}
\begin{figure}[htbp]
\centering
\includegraphics[width=0.5\textwidth]{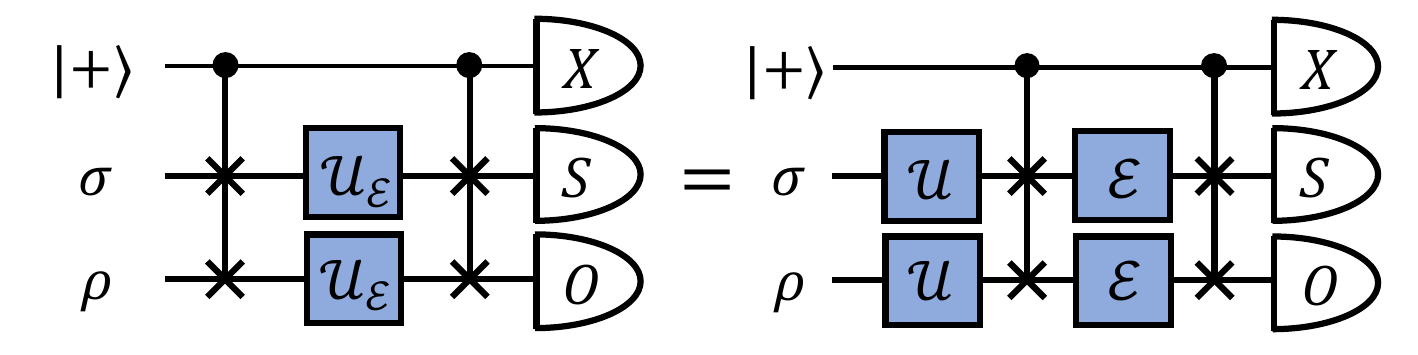}
\caption{Under the assumption of $\mathcal{U}_{\mathcal{E}}=\mathcal{E}\mathcal{U}$, the unitary channel can be moved before the first \textsc{cswap} gate. }
\label{fig:comment_circuit}
\end{figure}

Here, we provide the proof for the validation of the VCP circuit, starting from the $M=2$ case. The channel of the noisy operation is  $\mathcal{U}_{\mathcal{E}}=\mathcal{E}\mathcal{U}$. Since the tensor product of two identical unitaries $\mathcal{U} \otimes \mathcal{U}$ commutes with the controlled register-swap gate, we can perform our analysis of how the noise is mitigated by only analysing $\mathcal{E}$ (instead of $\mathcal{U}_{\mathcal{E}}$) while viewing the input states as $\mathcal{U}(\rho)$ and $\mathcal{U}(\sigma)$ as shown in \cref{fig:comment_circuit}.

Using a Kraus representation of the noise channel $\mathcal{E}(\rho)=\sum_{i=0}^{4^N-1}p_iE_i\rho E_i^\dagger$, the measurement result of \cref{fig:circuit}(c) is
\begin{equation}\label{eq:full_exp_val}
\begin{aligned}
&\expval{X\otimes O}\\=&\frac{1}{2}\sum_{i,j=0}^{4^N-1}\frac{p_ip_j}{2^N}\bigg[\bra{0}X\ket{0}\Tr\left(O E_i\rho E_i^\dagger\right)\Tr\left(E_j\mathbb{I}_{2^N}E_j^\dagger\right)\\
&+\bra{1}X\ket{1}\Tr\left(O E_j\rho E_j^\dagger\right)\Tr\left(E_i\mathbb{I}_{2^N}E_i^\dagger\right)\\
&+\bra{0}X\ket{1}\Tr\left(O E_i\rho E_j^\dagger\right)\Tr\left(E_j\mathbb{I}_{2^N}E_i^\dagger\right)\\
&+\bra{1}X\ket{0}\Tr\left(O E_j\rho E_i^\dagger\right)\Tr\left(E_i\mathbb{I}_{2^N}E_j^\dagger\right)\bigg].
\end{aligned}
\end{equation}
As mentioned in the main text, we will focus on the case where $\mathcal{E}$ is a Pauli channel, thus $\frac{1}{2^N}\Tr\left(E_iE_j^\dagger\right)=\delta_{ij}$. Further using $\bra{0}X\ket{0}=\bra{1}X\ket{1}=0$ and $\bra{0}X\ket{1}=\bra{1}X\ket{0}=1$, the expectation value above becomes
\begin{equation}\label{eq:exp_O}
\expval{X\otimes O}=\sum_{i=0}^{4^N-1}p_i^2\Tr\left(OE_i\rho E_i^\dagger\right)=P_2\Tr\left[O\mathcal{E}^{(2)}(\rho)\right].
\end{equation}
By setting $O=\mathbb{I}_{2^N}$, we have the expectation value of the control qubit measurement as
\begin{equation}\label{eq:exp_X}
\expval{X\otimes\mathbb{I}_{2^N}}=\sum_{i=0}^{4^N-1}p_i^2=P_2.
\end{equation}
Combining \cref{eq:exp_O,eq:exp_X}, we finish the proof of \cref{thrm:purified_channel} by
\begin{equation}
\frac{\expval{X\otimes O}}{\expval{X\otimes\mathbb{I}_{2^N}}}=\frac{\sum_{i=0}^{4^N-1}p_i^2\Tr\left(OE_i\rho E_i^\dagger\right)}{\sum_{i=0}^{4^N-1}p_i^2}=\Tr\left[O\mathcal{E}^{(2)}(\rho)\right].
\end{equation}

\begin{figure}[htbp]
\centering
\includegraphics[width = 0.5\textwidth]{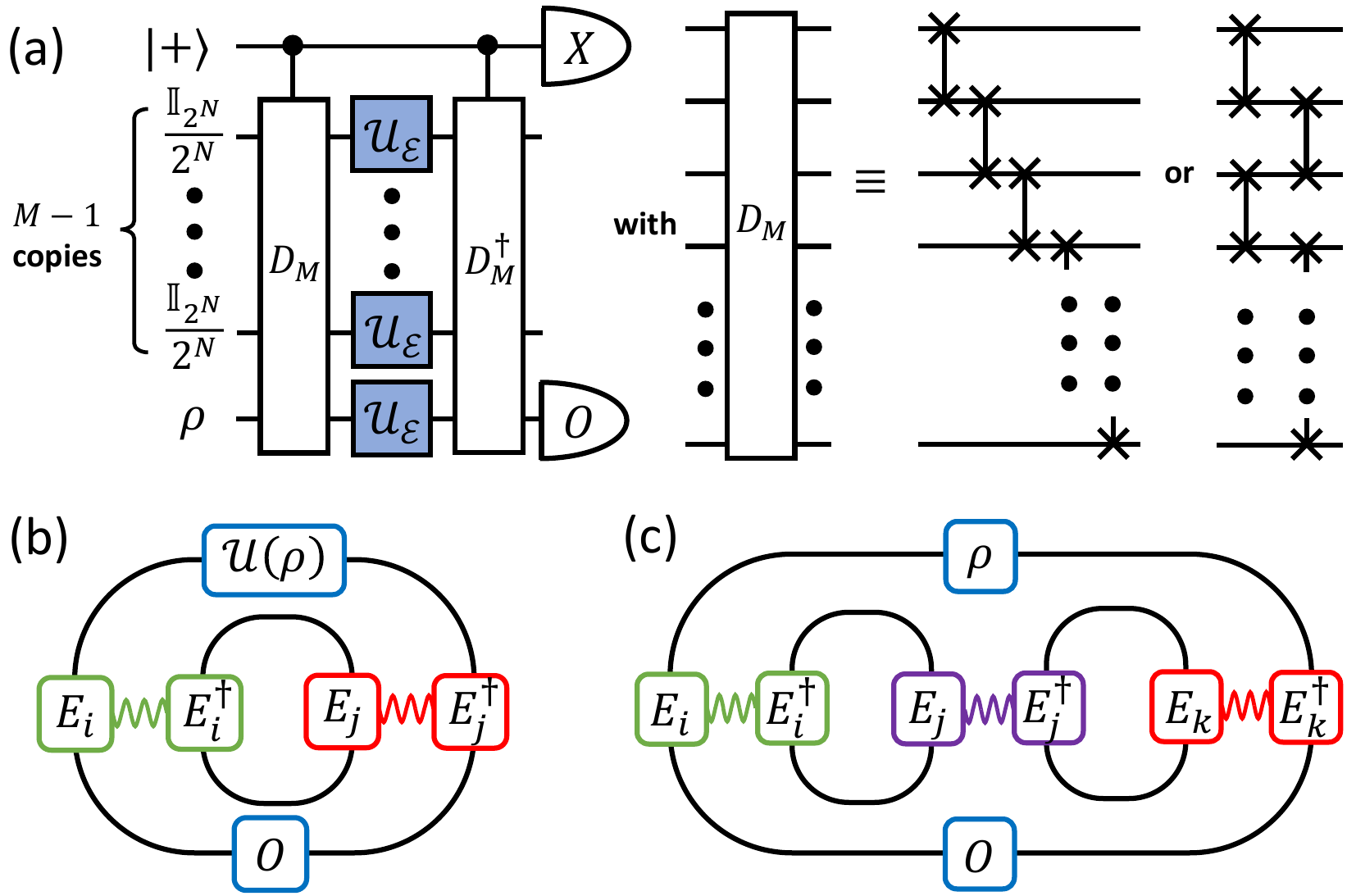}
\caption{(a) The circuit for $M$th order VCP protocol, where $M-1$ copies of maximally mixed states and controlled $M$th order permutation gates, $D_M$, are employed. $D_M$ can be any highest-order permutation gate and thus can be constructed in many different ways. (b) and (c) The core tensor network diagram for the validation of the VCP circuit for $M=2$ and $3$. The solid black lines stand for matrix indices contraction and coloured tilde lines stand for summation over labelled indices. }
\label{fig:higher_order}
\end{figure}

As stated in the main text, by replacing the \textsc{cswap} gates with controlled $M$th order permutation gates and inputting $M-1$ copies of maximally mixed states, the VCP protocol can be generalised to a higher-order version.
Following a similar derivation, one can show that the expectation value in \cref{fig:higher_order}(a) gives
\begin{equation}\label{eq:exp_val_higher_order}
\begin{aligned}
\expval{X\otimes O}=&\sum_{i_1,\cdots,i_M=0}^{4^N-1}\frac{p_{i_1}\cdots p_{i_M}}{2^{N(M-1)}}\Tr\left(E_{i_M}\rho E_{i_1}^\dagger O\right)\\
& \times \prod_{m=1}^{M-1}\Tr\left(E_{i_m}\mathbb{I}_{2^N}E_{i_{m+1}}^\dagger\right)\\
=&\sum_{i=0}^{4^N-1}p_i^M\Tr\left(E_i\rho E_i^\dagger O\right).
\end{aligned}
\end{equation}
Similarly, we have $\expval{X\otimes\mathbb{I}_{2^N}}=\sum_{i=0}^{4^N-1}p_i^M$ and $\nicefrac{\expval{X\otimes O}}{\expval{X\otimes\mathbb{I}_{2^N}}}=\Tr\left[O\mathcal{E}^{(M)}(\rho)\right]$.

Tensor network diagrams are introduced to illustrate the core ideas of our deviation, as shown in \cref{fig:higher_order}(b) and (c). To summarise, the controlled permutation gates together with the measurement of control qubit in Pauli-$X$ basis help to superpose multiple channels and extract terms where different Kraus operators of $\mathcal{E}$ act on different sides of $\rho$, such as $p_ip_jE_i\rho E_j^\dagger$. Then, the maximally mixed state helps to cancel the cross terms with $i\neq j$ and keep the diagonal terms with $i=j$, such that the weight of the target component can be amplified.

Our argument can be extended to the case in which the error channels acting on different copies are different. For the $M = 2$ case, let one of the noise channel be $\mathcal{E}_{1} = \sum_{i}p_i E_i \rho E_i^\dagger$ and the other be $\mathcal{E}_{2}(\rho) = \sum_{i}q_i E_i \rho E_i^\dagger$. Following the same arguments above, all of our results still applied with $p_iq_j$ in place of $p_ip_j$. As a result, the errors in our registers are suppressed from $\sum_{i \neq 0} p_i$ and $\sum_{i \neq 0} q_i$ to around $\sum_{i \neq 0} p_iq_i$ as long as identity is the dominating component in both $\mathcal{E}_{1}$ and $\mathcal{E}_{2}$. The factor of error suppression achieved over one of the error channels is the inverse of the error rate of the other channel. Extending the same argument beyond $M = 2$, VCP will still work when all $M$ of the error channels are identity-dominant Pauli channels. With the $m$th error channel being 
\begin{align*}
\mathcal{E}_{m} = p_{m,0}\mathcal{I}+\sum_{i=1}^{4^N-1}p_{m,i}{\supop{E}}_i.
\end{align*}
The effective noise channel we obtained after VCP will be
\begin{align*}
\mathcal{E}^{(M)} \propto \left(\prod_{m=1}^M p_{m,0}\right)\mathcal{I}+\sum_{i=1}^{4^N-1}\left(\prod_{m=1}^M p_{m,i}\right){\supop{E}}_i.
\end{align*}
with error rate suppressed from $\sum_i p_{m,i}$ to $\sum_i \prod_{m=1}^M p_{m,i}$.

\subsection{Post-selection}\label{sec:post_select_gen}
If we introduce post-selection to the VCP circuit, we can physically, instead of virtually, prepare a better quantum channel, as stated in \cref{coro:channel_purification}. Similar to \cref{eq:full_exp_val}, if instead of measuring $X$ on the control qubit, we post-select only the $+$ result, while still measuring $O$ on the main register in \cref{fig:circuit}, we then have
\begin{equation}
\begin{aligned}\nonumber
\expval{\ketbra{+}\otimes O}=&\frac{1}{4}\sum_{i,j=0}^{4^N-1}\frac{p_ip_j}{2^N}\bigg[\Tr\left(O E_i\rho E_i^\dagger\right)\Tr\left(E_j\mathbb{I}_{2^N}E_j^\dagger\right)\\
&+\Tr\left(O E_i\rho E_j^\dagger\right)\Tr\left(E_j\mathbb{I}_{2^N}E_i^\dagger\right) + c.c. \bigg]
\end{aligned}
\end{equation}
which can be simplified to 
\begin{equation}\label{eq:post_exp_+X}
\begin{aligned}
\expval{\ketbra{+}\otimes O}
=&\frac{1}{2}\sum_{i=0}^{4^N-1}\left[p_i\Tr\left(OE_i\rho E_i^\dagger\right)+p_i^2\Tr\left(OE_i\rho E_i^\dagger\right)\right]\\
=&\Tr\left[O\left(\frac{\mathcal{E}(\rho)+P_2\mathcal{E}^{(2)}(\rho)}{2}\right)\right].
\end{aligned}
\end{equation}
And the probability of measuring $\ket{+}$ on the control qubit is
\begin{equation}\label{eq:post_exp_+}
\begin{aligned}
p_+=&\expval{\ketbra{+}\otimes \mathbb{I}_{2^N}} 
=\frac{1}{2}(1+P_2).
\end{aligned}
\end{equation}
We see that after post-selecting on the control qubit measurement, the main register has collapsed to a physical state
\begin{equation}
\rho_+=\frac{\mathcal{E}(\rho)+P_2\mathcal{E}^{(2)}(\rho)}{1+P_2}.
\end{equation}

This result is easy to be generalized to the VCP circuit with higher order $M$, as shown in \cref{fig:higher_order}(a). 
Thus, with post-selection and tracing out the ancillary systems, we can physically implement the following channel on the main register
\begin{equation}
\mathcal{E}^{(M)}_{+}=\frac{\mathcal{E}+P_M\mathcal{E}^{(M)}}{1+P_M}.
\end{equation}
Noticing that $\mathcal{E}^{(M)}$ has the same Kraus components as $\mathcal{E}$ while with a higher weight of the target component.
Thus, the noise rate of $\mathcal{E}^{+}$ is lower than $\mathcal{E}$, and we have physically acquired a better quantum channel.

\subsection{Sample Complexity}\label{sec:sample_complexity}
Our estimator is constructed by division, $\nicefrac{\expval{X\otimes O}}{\expval{X\otimes\mathbb{I}_{2^N}}}=\Tr[O\mathcal{E}^{(M)}(\rho)]$. 
Notice that observables on the nominator and denominator commute with each other, so that we can measure both observables together in each circuit run.
Suppose these two values are calculated using a number of $K$ circuit runs, then we can use the formula
\begin{equation}\label{eq:var_1}
\mathrm{Var}\left(\frac{x}{y}\right)\approx\frac{\mu_x^2}{\mu_y^2}\left(\frac{\mathrm{Var}(x)}{\mu_x^2}-2\frac{\mathrm{Cov(x,y)}}{\mu_x \mu_y}+\frac{\mathrm{Var}(y)}{\mu_y^2}\right)
\end{equation}
to estimate the variance, where $x$ and $y$ stand for estimators of $\expval{X\otimes O}$ and $\expval{X\otimes\mathbb{I}_{2^N}}$, $\mu$ is the expectation value of the corresponding variable, and $\mathrm{Cov}$ stands for the covariance.

Starting with the simple case $M=2$, we can use the results in the last two sections to calculate $\mathrm{Var}\left(\frac{x}{y}\right)$. 
According to \cref{eq:exp_O,eq:exp_X}, we have
\begin{equation}\label{eq:var_2}
\mu_x = P_2\Tr\left[O\mathcal{E}^{(2)}(\rho)\right] \ , \ \mu_y = P_2.
\end{equation}
According to the standard expressions for variance, we have 
\begin{equation}\label{eq:var_3}
\begin{aligned}
\mathrm{Var}(x)=\frac{1}{K}\left(\expval{\mathbb{I}_2\otimes O^2}-\expval{X\otimes O}^2\right)
\end{aligned}
\end{equation}
Slightly changing the expression of \cref{eq:full_exp_val}, we have
\begin{equation}\label{eq:var_4}
\begin{aligned}
&\expval{\mathbb{I}_2\otimes O^2}\\
=&\frac{1}{2}\sum_{i,j=0}^{4^N-1}\frac{p_ip_j}{2^N}[\bra{0}\mathbb{I}_2\ket{0}\Tr\left(O^2 E_i\rho E_i^\dagger\right)\Tr\left(E_j\mathbb{I}_{2^N}E_j^\dagger\right)\\
&+\bra{1}\mathbb{I}_2\ket{1}\Tr\left(O^2 E_j\rho E_j^\dagger\right)\Tr\left(E_i\mathbb{I}_{2^N}E_i^\dagger\right)\\
&+\bra{0}\mathbb{I}_2\ket{1}\Tr\left(O^2 E_i\rho E_j^\dagger\right)\Tr\left(E_j\mathbb{I}_{2^N}E_i^\dagger\right)\\
&+\bra{1}\mathbb{I}_2\ket{0}\Tr\left(O^2 E_j\rho E_i^\dagger\right)\Tr\left(E_i\mathbb{I}_{2^N}E_j^\dagger\right)]\\
=&\frac{1}{2}\sum_{i,j=0}^{4^N-1}\frac{p_ip_j}{2^N}[\Tr\left(O^2 E_i\rho E_i^\dagger\right)\Tr\left(E_j\mathbb{I}_{2^N}E_j^\dagger\right)\\
&+\Tr\left(O^2 E_j\rho E_j^\dagger\right)\Tr\left(E_i\mathbb{I}_{2^N}E_i^\dagger\right)]\\
=&\Tr\left[O^2\mathcal{E}(\rho)\right].
\end{aligned}
\end{equation}
Substituting \cref{eq:var_2,eq:var_4} into \cref{eq:var_3}, we have
\begin{equation}\label{eq:var_5}
\mathrm{Var}(x)=\frac{1}{K}\left\{\Tr\left[O^2\mathcal{E}(\rho)\right]-P_2^2\Tr\left[O\mathcal{E}^{(2)}(\rho)\right]^2\right\}
\end{equation}
Similarly, we have
\begin{equation}\label{eq:var_6}
\begin{aligned}
\mathrm{Var}(y)=\frac{1}{K}\left(\expval{\mathbb{I}_2\otimes\mathbb{I}_{2^N}}-\expval{X\otimes\mathbb{I}_{2^N}}^2\right)=\frac{1}{K}\left(1-P_2^2\right).
\end{aligned}
\end{equation}

The calculation of covariance is similar to variance,
\begin{equation}\label{eq:var_7}
\mathrm{Cov}(x,y)=\frac{1}{K}\left(\expval{\mathbb{I}_2\otimes O}-\expval{X\otimes O}\expval{X\otimes\mathbb{I}_{2^N}}\right).
\end{equation}
Changing the observable $O^2$ in \cref{eq:var_4} to $O$, we have
\begin{equation}\label{eq:var_8}
\mathrm{Cov}(x,y)=\frac{1}{K}\left\{\Tr\left[O\mathcal{E}(\rho)\right]-P_2^2\Tr\left[O\mathcal{E}^{(2)}(\rho)\right]\right\}.
\end{equation}
Substituting \cref{eq:var_2,eq:var_5,eq:var_6,eq:var_8} into \cref{eq:var_1}, we get the final expression of the variance
\begin{equation}\label{eq:var_9}
\begin{aligned}
\mathrm{Var}\left(\frac{x}{y}\right)\approx&\frac{1}{KP_2^2}\big\{\Tr\left[O^2\mathcal{E}^{(2)}(\rho)\right]\\
-&2\Tr\left[O\mathcal{E}^{(2)}(\rho)\right]\Tr\left[O\mathcal{E}(\rho)\right]+\Tr\left[O\mathcal{E}^{(2)}(\rho)\right]^2\big\}.
\end{aligned}
\end{equation}
This result is easy to be generalised to any $M$ by changing $P_2$ and $\mathcal{E}^{(2)}$ to $P_M$ and $\mathcal{E}^{(M)}$, respectively.
Considering that the observable has a bounded spectral norm, $\norm{O}_{\infty}\le \mathrm{const}$, then the terms after $\frac{1}{KP_M^2}$ can also be bounded by some constant.
Thus, to suppress the variance to $\epsilon^2$, the experiment times should scale as $K\sim \mathcal{O}\left(\frac{1}{\epsilon^2P_M^2}\right)$.

\subsection{Implementation of maximally mixed states}\label{sec:impl_max_mix}
\begin{figure}
\centering
\includegraphics[width=1.0\linewidth]{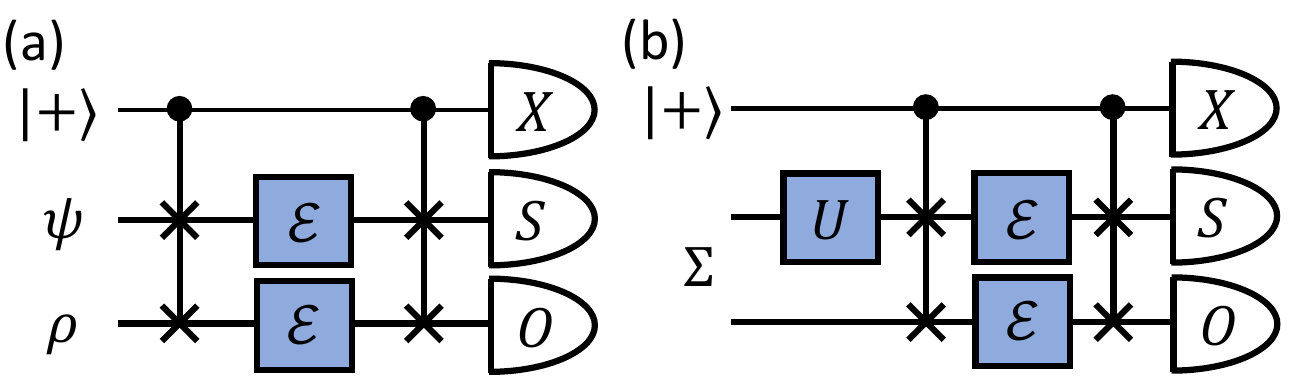}
\caption{
Two ways to implement the maximally mixed state. 
(a) One can initialise the ancillary register to be a random pure state, which varies in different experiment runs.
The only requirement for these random pure states is that the expectation of them should be maximally mixed state, $\mathbb{E}_\psi\ketbra{\psi}{\psi}=\frac{\mathbb{I}_{2^N}}{2^N}$.
(b) When the VCP is executed in the middle of the noisy circuit, as shown in Fig.~\ref{fig:data}(a), where the main and ancillary registers can be entangled and form a joint state labelled by $\Sigma$.
In this case, one can apply a random unitary on the ancillary state and effectively decouple the state from the state of the main register and also randomise the ancillary input state. This random unitary needs to be at least a unitary one-design, which can be achieved by the tensor product of single-qubit Pauli unitaries.
}
\label{fig:fake_mms}
\end{figure}

As shown in Fig.~\ref{fig:fake_mms}, we can use the random pure state or the random unitary evolution to replace the maximally mixed state, which will not cause an increase in sample complexity. 
Here we give a brief proof of it.
Let us take Eq.~\eqref{eq:var_3} as an example, when we change the maximally mixed state into a random pure state $\psi=\ketbra{\psi}{\psi}$, the variance becomes
\begin{equation}
\mathrm{Var}(x)=\frac{1}{K}\left[\mathbb{E}_\psi\expval{\mathbb{I}_2\otimes O^2}_\psi-\left(\mathbb{E}_\psi\expval{X\otimes O}_\psi\right)^2\right],
\end{equation}
where $\expval{\cdot}_\psi$ denotes the expectation value estimated when initialising the register as the state $\psi$.
Take the first term as an example, it is 
\begin{equation}
\begin{aligned}
&\mathbb{E}_\psi\expval{\mathbb{I}_2\otimes O^2}_\psi\\
=&\frac{1}{2}\mathbb{E}_\psi\sum_{i,j=0}^{4^N-1}p_ip_j\left[\Tr\left(O^2E_i\rho E_i^\dagger\right)\Tr\left(E_j\psi E_j^\dagger\right)+h.c.\right]\\
=&\frac{1}{2}\sum_{i,j=0}^{4^N-1}p_ip_j\left[\Tr\left(O^2E_i\rho E_i^\dagger\right)\Tr\left(E_j\mathbb{E}_\psi\psi E_j^\dagger\right)+h.c.\right]\\
=&\frac{1}{2}\sum_{i,j=0}^{4^N-1}\frac{p_ip_j}{2^N}\left[\Tr\left(O^2E_i\rho E_i^\dagger\right)\Tr\left(E_j\mathbb{I}_{2^N} E_j^\dagger\right)+h.c.\right]\\
=&\Tr\left[O^2\mathcal{E}(\rho)\right],
\end{aligned}
\end{equation}
which is equivalent to the case when the ancillary register is initialised as the maximally mixed state.
Following the same logic, $\mathbb{E}_\psi\expval{X\otimes O}$, together with all the variance and covariance terms are all equivalent with the maximally mixed state case, only requiring that $\mathbb{E}_\psi\ketbra{\psi}{\psi}=\frac{\mathbb{I}_{2^N}}{2^N}$.
This requirement can be easily achieved by randomly setting each qubit to $\ket{0}$ or $\ket{1}$ with equal probabilities.

Now we consider the second case, as shown in Fig.~\ref{fig:fake_mms}, where the ancillary and the main registers are entangled and together be described by a joint state $\Sigma$, with $\Tr_1(\Sigma)=\rho$ being the target input state of the main register.
This corresponds to the case of implementing VCP in the middle of a deep noisy circuit.
Similarly, we here take Eq.~\eqref{eq:var_3} as an example, which now becomes
\begin{equation}
\mathrm{Var}(x)=\frac{1}{K}\left[\mathbb{E}_U\expval{\mathbb{I}_2\otimes O^2}_U-\left(\mathbb{E}_U\expval{X\otimes O}_U\right)^2\right].
\end{equation}
Here, the first term is 
\begin{equation}
\begin{aligned}
&\mathbb{E}_U\expval{\mathbb{I}_2\otimes O^2}_U\\
=&\frac{1}{2}\mathbb{E}_U\sum_{i,j=0}^{4^N-1}p_ip_j\{\Tr\left[(\mathbb{I}_{2^N}\otimes O^2)(E_iU\otimes E_j)\Sigma(E_iU\otimes E_j)^\dagger\right]\\
&+\Tr\left[(\mathbb{I}_{2^N}\otimes O^2)(E_jU\otimes E_i)\Sigma(E_jU\otimes E_i)^\dagger\right]\}.
\end{aligned}
\end{equation}
Using the unitary one design property, we have 
\begin{equation}
\mathbb{E}_U(U\otimes\mathbb{I}_{2^N})\Sigma(U\otimes\mathbb{I}_{2^N})^\dagger=\frac{\mathbb{I}_{2^N}}{2^N}\otimes\Tr_1(\Sigma)=\frac{\mathbb{I}_{2^N}}{2^N}\otimes\rho.
\end{equation}
Therefore, we can prove that
\begin{equation}
\begin{aligned}
&\mathbb{E}_U\expval{\mathbb{I}_2\otimes O^2}_U\\
=&\frac{1}{2}\sum_{i,j=0}^{4^N-1}\frac{p_ip_j}{2^N}\left[\Tr\left(O^2E_i\rho E_i^\dagger\right)\Tr\left(E_j\mathbb{I}_{2^N} E_j^\dagger\right)+h.c.\right]\\
=&\Tr\left[O^2\mathcal{E}(\rho)\right],
\end{aligned}
\end{equation}
which is also equivalent to the case when the ancillary register is initialised as the maximally mixed state.
Similarly, the same logic can be utilised to prove that all the variance terms are unchanged in the case of Fig.~\ref{fig:fake_mms}(b).
It is worth noting that the unitary one design can be easily satisfied by setting the unitary to be the tensor product of single-qubit random Pauli unitaries.

\subsection{Error Suppression under Depolarising Noise} \label{sec:exp_err_suppr}
The $N$-qubit completely depolarising channel is just randomly applying any one of the $4^N$ Pauli operators $\{G_i\}$ with equal probability:
\begin{align}\label{eqn:comp_depol}
\mathcal{D}_{1}=\frac{1}{4^N}\sum_{i = 1}^{4^N} {\supop{G}}_i.
\end{align}
and it will turn any incoming state into the maximally mixed state:
\begin{align}\label{eqn:depol_output}
\mathcal{D}_{1}(\rho)= \frac{\mathbb{I}_{2^N}}{2^N}.
\end{align}
We can then define a $N$-qubit depolarising channel with error probability $P$ to be:
\begin{align*}
\mathcal{D}_{P} &= (1-P) \mathcal{I} + P \mathcal{D}_{1}\\
& = (1-P)\mathcal{I}+\frac{P}{4^N}\sum_{i = 1}^{4^N} {\supop{G}}_i
\end{align*}
where each error component occurs with $P/4^N$ probabilities. 

When acting this channel on some pure state $\ketbra{\psi}$ and using \cref{eqn:depol_output}, the output state is
\begin{align*}
\mathcal{D}_{P}(\ketbra{\psi})&=(1-P)\ketbra{\psi} + P\frac{\mathbb{I}_{2^N}}{2^N}\\
& = (1-P)\ketbra{\psi} + \frac{P}{2^N} \sum_{i=1}^{2^N} \ketbra{\phi_i}
\end{align*}
where we have decomposed the maximally mixed state into its $2^N$ error components $\ketbra{\phi_i}$, each occur with error probability $P/2^N$. 

In the following arguments, we will ignore the fact that one of ${\supop{G}}_i$ is the error-free component $\mathcal{I}$, similarly we will ignore the fact that one of $\ketbra{\phi_i}$ is $\ketbra{\psi}$. Due to their low probability compared to the main error-free component of the probability $1-P$, especially at large $N$, this will not affect any of our later arguments much. We are doing this to simplify our expression.

Applying VSP, the coefficient of every components in $\mathcal{D}_{P}(\ketbra{\psi})$ will be raised to its $M$th power along with a normalisation factor:
\begin{align*}
\rho^{(M)} &= \frac{(1-P)^M \ketbra{\psi}+(P/2^N)^M \sum_{i = 1}^{2^N} \ketbra{\phi_i} }{(1-P)^M + \sum_{i = 1}^{2^N} (P/2^N)^M } \\
&= \frac{(1-P)^M \ketbra{\psi}+2^N(P/2^N)^M (\mathbb{I}_{2^N}/2^N) }{(1-P)^M + 2^N (P/2^N)^M } \\
& \approx \left(\ketbra{\psi}+2^N\left(\frac{P}{2^N(1-P)}\right)^M \frac{\mathbb{I}_{2^N}}{2^N}\right) \\
& \quad \times \left(1 - 2^N\left(\frac{P}{2^N(1-P)}\right)^M\right)\\
& = \left(1 - 2^N\left(\frac{P}{2^N(1-P)}\right)^M\right) \ketbra{\psi} \\
&\quad + 2^N\left(\frac{P}{2^N(1-P)}\right)^M \frac{\mathbb{I}_{2^N}}{2^N} + \order{2^{-2N(M-1)}}
\end{align*}
The infidelity w.r.t. $\ketbra{\psi}$ is given as
\begin{align}
    \epsilon_{S} = 1 - \bra{\psi}\rho^{(M)}\ket{\psi} \approx 2^{N}\left(\frac{P}{2^N(1-P)}\right)^M
\end{align}
where again we have ignored the error-free components in $\frac{\mathbb{I}_{2^N}}{2^N}$ since it is $\order{2^N}$ smaller than the main component.

Applying VCP, the coefficient of every components in $\mathcal{D}_{P}$ will be raised to its $M$th power along with a normalisation factor:
\begin{align*}
\mathcal{D}_{P}^{(M)} = \frac{(1-P)^M\mathcal{I}+(P/4^N)^M \sum_{i = 1}^{4^N} {\supop{G}}_i }{(1-P)^M + \sum_{i=1}^{4^N} (P/4^N)^M } 
\end{align*}
When acting on $\ketbra{\psi}$, we will get
\begin{align*}
\mathcal{D}_{P}^{(M)}(\ketbra{\psi}) &= \frac{(1-P)^M\ketbra{\psi}+4^N(P/4^N)^M( \mathbb{I}_{2^N}/2^N)}{(1-P)^M + \sum_{i=1}^{4^N} (P/4^N)^M } \\
& = \left(1 - 4^N\left(\frac{P}{4^N(1-P)}\right)^M\right) \ketbra{\psi} \\
&\quad + 4^N\left(\frac{P}{4^N(1-P)}\right)^M \frac{\mathbb{I}_{2^N}}{2^N} + \order{4^{-2N(M-1)}}
\end{align*}
Following similar arguments as before, the infidelity w.r.t. $\ketbra{\psi}$ is given as
\begin{align}
    \epsilon_{C} = 1 - \bra{\psi}\mathcal{D}_{P}^{(M)}(\ketbra{\psi})\ket{\psi} \approx  4^{N}\left(\frac{P}{4^N(1-P)}\right)^M.
\end{align}
we see that this is a factor of $2^{N(M-1)}$ smaller than $\epsilon_S$, which is the infidelity obtained through VSP.

\subsection{Circuit Generalisation}\label{sec:circuit_generalisation}
Till now, we only consider the case in which 
the ancillary register in \cref{fig:circuit}(c) is initialised as the maximally mixed state and discarded at the end of the whole process.
Now let us consider the circuit in \cref{fig:vec_circuit}, in which the ancillary system is initialised as some general state $\sigma$ and measured with the observable $S$ at the end of the circuit, the expectation value will be
\begin{equation}\label{eq:gen_second_order}
\begin{aligned}
&\expval{X\otimes S\otimes O}\\
=&\frac{1}{2}\sum_{i,j=0}^{4^N-1}p_ip_j[\bra{0}X\ket{0}\Tr\left(O E_i\rho E_i^\dagger\right)\Tr\left(SE_j\sigma E_j^\dagger\right)\\
&+\bra{1}X\ket{1}\Tr\left(O E_j\rho E_j^\dagger\right)\Tr\left(SE_i \sigma E_i^\dagger\right)\\
&+\bra{0}X\ket{1}\Tr\left(O E_i\rho E_j^\dagger\right)\Tr\left(SE_j \sigma E_i^\dagger\right)\\
&+\bra{1}X\ket{0}\Tr\left(O E_j\rho E_i^\dagger\right)\Tr\left(SE_i \sigma E_j^\dagger\right)]\\
=&\sum_{i,j=0}^{4^N-1}p_ip_j\bigg[\Tr\left(O E_i\rho E_j^\dagger\right)\Tr\left(SE_j \sigma E_i^\dagger\right)\bigg].
\end{aligned}
\end{equation}
Therefore, any choices of $\sigma$ and $S$ that satisfies
\begin{equation}\label{eqn:purify_cond}
\Tr\left(SE_i \sigma E_j^\dagger\right)=\delta_{ij}
\end{equation}
can be used to perform VCP. If one has enough knowledge of the noise, by suitably choosing the input state and/or the observable $S$, one can completely remove the noise.
This is because if the state $\sigma$ and observable $S$ satisfy $\Tr(SE_j\sigma E_i^\dagger)=\delta_{i0}\delta_{j0}$, the expectation value
\begin{equation}
\frac{\expval{X\otimes S\otimes O}}{\expval{X\otimes S \otimes \mathbb{I}_{2^N}}}=\frac{p_0^2\Tr\left(OE_0\rho E_0^\dagger\right)}{p_0^2}=\Tr\left(O\rho\right)
\end{equation}
becomes the exact noiseless value.

Note that what we have discussed here can also be generalised to $M-1$ ancillary registers, each can come with a different input state $\sigma_m$ and measurement $S_m$ with the corresponding expectation value being:
\begin{equation}\label{eq:gen_higher_order}
\begin{aligned}
&\expval{X\otimes \bigotimes_{m=1}^{M-1} S_m \otimes O}=\sum_{i_1,\cdots,i_M=0}^{4^N-1}\left(\prod_{m=1}^M p_{i_m}\right)\\
 &\quad\times \Tr\left(O E_{i_M}\rho E_{i_1}^\dagger\right) \prod_{m=1}^{M-1}\Tr\left(S_mE_{i_m}\sigma_m E_{i_{m+1}}^\dagger\right)
\end{aligned}
\end{equation}

\section{Combining with Quantum Error Correction}\label{sec:vir_with_qec}

\subsection{Mixing unencoded state and code state}\label{sec:mix_unencode_code}
Following the notations and arguments from \cref{sec:circuit_generalisation}, but focusing on only the $M=2$ case and allowing for different noise channels $\mathcal{E}_1(\rho) = \sum_{i} p_i E_i \rho E_i^\dagger$ and $\mathcal{E}_2(\rho) = \sum_{j} q_j E_j \rho E_j^\dagger$ acting on the two registers, then \cref{eq:gen_second_order} will becomes:
\begin{equation}
\begin{aligned}
&\expval{X\otimes S\otimes O}
=&\sum_{i,j=0}^{4^N-1}p_iq_j\bigg[\Tr\left(SE_j \sigma E_i^\dagger\right)\Tr\left(O E_i\rho E_j^\dagger\right)\bigg].
\end{aligned}
\end{equation}
This essentially means that after measuring $X$ on the control qubit, the above expectation value can be effectively viewed as performing $O \otimes S$ on the state:
\begin{equation}\label{eq:exp_val_QEC}
\begin{aligned}
&\sum_{i,j} p_{i}q_{j}\underbrace{\left(E_{j}\sigma E_{i}^\dagger\right)}_{\text{ancillary}} \otimes \underbrace{\left(E_{i}\rho E_{j}^\dagger \right)}_{\text{main}}.
\end{aligned}
\end{equation}
which is the effective output ``state'' of the main and ancillary registers after post-processing using the $X$ measurement on the control qubit (i.e. a state obtained by measuring $X$ on the control qubit and attaching the measurement result of $\pm 1$ as a factor to the output states).

By having the ancilla input being the code state $\sigma = \sigma_E$ and performing the corresponding stabiliser measurements on the ancillary register, we will project it into the $k$th syndrome subspace defined by $E_k \Pi_E E_k^\dagger$, and the resultant ancillary state will become
\begin{align*}
    &(E_k \Pi_E E_k^\dagger) \left(E_{i}\sigma_E E_{j}^\dagger\right)(E_k \Pi_E E_k^\dagger) \\
    = & E_k (\Pi_E E_k^\dagger E_{i} \Pi_E) \sigma_E (\Pi_E E_{j}^\dagger E_k \Pi_E) E_k^\dagger \\
    = & \delta_{i, k} \delta_{j, k} E_k \sigma_E E_k^\dagger.
\end{align*}
Here we have made use of the Knill-Laflamme condition in \cref{eqn:kl_ortho}. Substituting into \cref{eq:exp_val_QEC}, the resultant state after we perform stabiliser measurement on the ancillary state is 
\begin{align*}
    p_kq_k\left(E_k\sigma_E E_k^\dagger\right) \otimes \left(E_k\rho E_k^\dagger \right).
\end{align*}
Since we are considering a non-degenerate code that satisfies \cref{eqn:kl_ortho}, the stabiliser measurement result will unambiguously tell us that the error inflicted is $E_k$. Hence, we can apply the corresponding correction $E_k$ to the main register and trace out the ancilla QEC state, giving us the state  $p_kq_k \rho$. The additional $p_kq_k$ factor in front is due to the fact that we are measuring $X$ on the controlled qubit and thus the state is actually virtually obtained from post-processing. Averaging over all possible measurement outcome of the ancilla qubit, the effective output state is then 
\begin{align*}
    \left(\sum_k p_kq_k\right) \rho
\end{align*}
The factor in front is the same as the normalising factor of 2nd-order purification. It is also the expectation value of the control qubit $X$ measurement, as discussed at the start of
\cref{sec:comb_with_qec} where both registers are unmeasured. The inverse of this factor, $(\sum_k p_kq_k)^{-1}$, is then the increase of the range of the random variable we measure and thus $(\sum_k p_kq_k)^{-2}$ will be the sampling cost, which is the same as the usual 2nd-order purification.

\subsection{Application of VEC}\label{sec:apply_vec}
Looking at the VEC circuit in \cref{fig:vec_circuit}, if $\rho$ is an $N$-qubit state, the number of logical qubits we need to put in $\sigma_E$ is $\lceil N/K \rceil$. Hence, there need to be $\lceil N/K \rceil K$ qubits in the code register. Since $N \leq \lceil N/K \rceil K$, we will have a VEC circuit whose main register is smaller than the ancilla register. This can work by simply only applying \textsc{cswap}s to any $N$ qubits within the $\lceil N/K \rceil K$ qubits in the code register. It does not matter which $N$ qubits we choose since all of these qubits will have the same error tolerance we need. Note that this is the same as virtually having some additional qubits in the main register such that the main register is the same size as the code register. We will apply the VEC circuit in the same way as before and any errors on these additional main register qubits will also be corrected. However, since we do not care about the information stored in this qubit, we can simply remove these qubits and also remove any \textsc{cswap}s and corrections that we have performed on them. In this way, the number of qubits needed for VEC is
\begin{align*}
    N + \lceil N/K \rceil K + 1 \leq N + (N/K + 1) K + 1 = 2N + K + 1
\end{align*}
The number of qubits needed when using QEC to encode $\rho$ is $NK$. Hence, making the practical assumption of $K \gg 1$, the factor of qubit overhead saving brought by VEC will be more than:
\begin{align*}
    \frac{NK}{2N + K + 1} = \begin{cases}
        N \quad &N \ll K\\
        K/3\quad &N \sim K\\
        K/2 \quad & N \gg K
    \end{cases}
\end{align*}
i.e. we have a factor of $\order{K}$ qubit saving when $N \gtrsim K$, and a factor of $N$ saving when $N \ll K$.

\subsection{Mixing different code states} \label{sec:mixing_code_state}
Now let us suppose we have two code states, with $\sigma_E$ from code $E$ can correct the set of errors $\mathbb{E} = \{E_i\}$ and $\sigma_F$ from code $F$ can correct the set of errors $\mathbb{F} = \{F_j\}$. Here we will focus on the case in which code $E$ is unable to detect any errors in $\mathbb{F}$ and similarly, code $F$ is unable to detect any errors in $\mathbb{E}$. Denoting the code space projector of code $E$ and $F$ as $\Pi_E$ and $\Pi_F$, respectively, this implies that $\Pi_E$ commutes any elements in $\mathbb{F}$ and similarly $\Pi_F$ commutes any elements in $\mathbb{E}$. Note that this implies that two codes do not share any correctable errors other than $I$, i.e. $\mathbb{E} \cap \mathbb{F} = \{I\}$ (but the inverse may not be true). In particular, this covers the important scenario in which one of the codes can only correct $X$ errors while the other can only correct $Z$ errors. 

The error channel that our states suffer will be a combination of error elements in $\mathbb{E}$ and $\mathbb{F}$: $\mathcal{G}(\rho) = \sum_{i,j}p_{i,j} E_iF_j \rho F_j^\dagger E_i^\dagger$ (note that for simplicity we are considering Pauli errors here thus the ordering between $E_i$ and $F_j$ is not important). Using the circuit in \cref{fig:vec_circuit_gen}(c), after measuring $X$ on the control qubit, we will virtually obtain the following state
\begin{equation}\label{eq:exp_val_QEC_2}
\begin{aligned}
&\sum_{i_1,i_2, j_1, j_2} p_{i_1, j_1}p_{i_2, j_2}\left(E_{i_2}F_{j_2} \sigma_{E} F_{j_1}^\dagger E_{i_1}^\dagger \right)\otimes \left(E_{i_1}F_{j_1} \sigma_{F} F_{j_2}^\dagger E_{2}^\dagger \right).
\end{aligned}
\end{equation}

By performing stabiliser measurement on the two registers, we will project the first code register into the $i$th syndrome subspace defined by $E_i \Pi_E E_i^\dagger$ with the resultant output state becomes
\begin{align*}
    &(E_i \Pi_E E_i^\dagger) \left(E_{i_2}F_{j_2} \sigma_{E} F_{j_1}^\dagger E_{i_1}^\dagger\right)(E_i \Pi_E E_i^\dagger) \\
    = & E_i \Pi_E E_i^\dagger E_{i_2}F_{j_2}  (\Pi_E \sigma_{E}\Pi_E) F_{j_1}^\dagger E_{i_1}^\dagger E_i \Pi_E E_i^\dagger \\
    = & E_i (\Pi_E E_i^\dagger E_{i_1} \Pi_E) F_{j_2} \sigma_E F_{j_1}^\dagger  (\Pi_E E_{i_2}^\dagger E_i \Pi_E) E_j^\dagger \\
    = & \delta_{i_1, i} \delta_{i_2, i} E_i F_{j_2} \sigma_E F_{j_1}^\dagger E_i^\dagger.
\end{align*}
where we have make use of $\Pi_E$ commuting with $F_j$ and also the Knill-Laflamme condition in \cref{eqn:kl_ortho}. Similarly, after performing stabiliser measurement on the second code register and projecting into the $j$th syndrome subspace defined by $F_j \Pi_F F_j^\dagger$, the output states become 
\begin{align*}
    &(F_j \Pi_F F_j^\dagger) \left(E_{i_2}F_{j_2} \sigma_{F} F_{j_1}^\dagger E_{i_1}^\dagger\right)(F_j \Pi_F F_j^\dagger)\\ 
    = & \delta_{j_1, j} \delta_{j_2, j} F_j E_{i_2} \sigma_F E_{i_1}^\dagger F_j^\dagger.
\end{align*}
Substituting into \cref{eq:exp_val_QEC_2} we get
\begin{equation}
\begin{aligned}
&\sum_{i_1,i_2, j_1, j_2} p_{i_1, j_1}p_{i_2, j_2} \left(\delta_{i_1, i} \delta_{i_2, i} E_i F_{j_2} \sigma_E F_{j_1}^\dagger E_i^\dagger\right) \\
 &\times  \left(\delta_{j_1, j} \delta_{j_2, j} F_j E_{i_2} \sigma_F E_{i_1}^\dagger F_j^\dagger\right)\\
 = & p_{i, j}^2  \left(E_i F_{j} \sigma_E F_{j}^\dagger E_i^\dagger\right)  \left(F_j E_{i} \sigma_F E_{i}^\dagger F_j^\dagger\right)
\end{aligned}
\end{equation}
Since we are considering a non-degenerate code that satisfies \cref{eqn:kl_ortho} for both code $E$ and $F$, the stabiliser measurement result will unambiguously tell us that the error inflicted is $E_i$ and $F_j$. Hence, we can apply the corresponding correction $E_i$ and $F_j$ to \emph{both registers}, giving us the state  $p_{i, j}^2 \sigma_E \otimes \sigma_F$. The additional $p_{i,j}^2$ factor in front is due to the fact that we are measuring $X$ on the controlled qubit and thus the state is actually virtually obtained from post-processing. Averaging over all possible measurement outcome of the ancilla qubit, the effective output state is then 
\begin{align*}
    (\sum_{i,j} p_{i, j}^2) \rho
\end{align*}
The factor in front is the same as the normalising factor of 2nd-order purification. It is also the expectation value of the control qubit $X$ measurement, as discussed at the start of
\cref{sec:comb_with_qec} where both registers are unmeasured. The inverse of this factor, $(\sum_{i,j} p_{i,j}^2)^{-1}$, is then the increase of the range of the random variable we measure and thus $(\sum_{i,j} p_{i,j}^2)^{-2}$ will be the sampling cost, which is the same as the usual 2nd-order purification. The case in \cref{sec:mix_unencode_code} is simply the special case of $\mathbb{F} = \{I\}$. 

\subsection{Mitigating \textsc{cswap} noise in VEC}\label{sec:cswap_noise_vec}
The arguments in \cref{sec:qem_before_qec} on how to apply QEM before QEC can be directly applied to the VEC circuit. We will focus on \cref{fig:vec_circuit}(b) in this section and the arguments can be easily generalised to the case in \cref{fig:vec_circuit_gen}(c). In the VEC circuit in \cref{fig:vec_circuit}(b), we are performing syndrome check on one register and correction on another, this simply means that our parity check is on a sub-register ${\supop{\Pi}}_s \Rightarrow {\supop{\Pi}}_s \otimes \mathcal{I}$ and the correction is on another 
$\mathcal{C}_s \Rightarrow \mathcal{I} \otimes \mathcal{C}$. Thus, the overall error correction process for VEC is in the form of
\begin{align*}
    \mathcal{R}_{\mathrm{qec}} = \sum_s\mathcal{C}_s{\supop{\Pi}}_s \Rightarrow \mathcal{R}_{\mathrm{qec}} = \sum_s {\supop{\Pi}}_s \otimes \mathcal{C}_s.
\end{align*}
Let us denote the core part of the VEC circuit, which is the noise process along with the two $\textsc{cswap}$s, as $\mathcal{V} = \textsc{cswap} (\mathcal{I} \otimes \mathcal{E}_1 \otimes \mathcal{E}_2) \textsc{cswap}$, then the overall VEC circuit can be written as
\begin{align*}
    \expval{O}_{\mathrm{vec}} = \pbra{X \otimes I \otimes O}\sum_s(\mathcal{I} \otimes {\supop{\Pi}}_s \otimes \mathcal{C}_s)\mathcal{V}\pket{+ \otimes \rho \otimes \sigma}
\end{align*}
Now if there is noise associated with the \textsc{cswap}s, we can commute the noise of the first \textsc{cswap} to the front and denote it as $\mathcal{F}$, while the noise of the second \textsc{cswap} will be commuted to the back and denoted as $\mathcal{K}$. The resultant noisy expectation value will be
\begin{align*}
    \expval{O}_{\mathrm{noi}} = \pbra{X \otimes I \otimes O}\sum_s(\mathcal{I} \otimes {\supop{\Pi}}_s \otimes \mathcal{C}_s)\mathcal{K}\mathcal{V}\mathcal{F}\pket{+ \otimes \rho \otimes \sigma}
\end{align*}
Note that this whole expectation value is still linear in $\mathcal{K}$ and $\mathcal{F}$, so we should be able to use some linear combination of different circuit configurations to cancel out these noises. For example, we can use PEC to virtually implement $\mathcal{F}^{-1} = \sum_{i} \alpha_i \mathcal{A}_i$ and $\mathcal{K}^{-1} = \sum_{j} \beta_j \mathcal{B}_j$, where $\{A_i\}$ and $\{B_j\}$ are some implementable basis operations, to remove the noise:
\begin{align*}
    &\expval{O}_{\mathrm{mit}} = \sum_{i,j} \alpha_i \beta_j \\
    &\quad \times\pbra{X \otimes I \otimes O}\sum_s(\mathcal{I} \otimes {\supop{\Pi}}_s \otimes \mathcal{C}_s)\mathcal{B}_j\mathcal{K}\mathcal{V}\mathcal{F}\mathcal{A}_i\pket{+ \otimes \rho \otimes \sigma}\\
    & = \pbra{X \otimes I \otimes O}\sum_s(\mathcal{I} \otimes {\supop{\Pi}}_s \otimes \mathcal{C}_s)\mathcal{K}^{-1}\mathcal{K}\mathcal{V}\mathcal{F}\mathcal{F}^{-1}\pket{+ \otimes \rho \otimes \sigma}\\
    & = \expval{O}_{\mathrm{vec}}.
\end{align*}

\section{Optimal Number of Layers in VCP}\label{sec:vcp_err}
\subsection{Overall derivation}\label{sec:vcp_err_overall}
We will be considering a quantum circuit with $N$ qubits, depth $D$ and gate error rate $p$. The total number of gates in the whole circuit is approximately $ND$ with the circuit fault rate (average number of faults per circuit run) being $\lambda = NDp$. For the VCP considered in this section, the order of purification is denoted as $M$ and the number of VCP layers is denoted as $L$. For a large enough circuit, we can assume the errors in the circuit follow a Poisson distribution and thus the probability that the circuit is error-free is around $e^{-\lambda}$~\cite{caiPracticalFrameworkQuantum2021}. Following similar arguments, for each VCP layer, the probability that the VCP layer is error-free is around $e^{-\lambda/L}$. 

With noiseless \textsc{cswap}s, the error rate of the circuit after VCP is:
\begin{align}\label{eqn:P_c_cir}
    P_{\mathrm{c,cir}} &\approx 1 - \left[1-n_c^{1-M}\left(1 - e^{-\lambda/L}\right)^M\right]^L
\end{align}
Here $n_c$ is the number of (dominating) error components in the channel. The discussion in \cref{sec:stronger_suppresion} focuses on the region where $\lambda \ll L$, in which $P_{\mathrm{c,cir}} \approx n_cL\left(\lambda/(n_cL)\right)^M$. For the expression in \cref{eqn:P_c_cir} to be accurate, we need the noiseless component to be dominating, which means $e^{-\lambda/L} \gg n_c^{-1}(1-e^{-\lambda/L})$ and thus $n_c +1 \gg e^{\lambda/L}$. For global depolarising noise where $n_c = 4^N$, we have $N \gg \lambda/(\ln(4)L)$ as a requirement. A similar but stronger restriction also applies to the corresponding expression for VSP in \cref{eqn:pscir}.

The number of \textsc{cswap}s in VCP is $2NML$ and each of them will have a gate error rate $\alpha p$, i.e. the \textsc{cswap} gate errors is $\alpha$ times stronger. As mentioned in \cref{sec:practical_imple}, the noise in half of these \textsc{cswap}s can be mitigated by VCP, thus the remaining average fault rate due to \textsc{cswap} after VCP is approximately $NML\alpha p = \alpha ML\lambda/D$. The corresponding error rate due to \textsc{cswap}s is then:
\begin{align}\label{eqn:P_c_sw}
    P_{\mathrm{c,sw}} &\approx 1 - e^{-\alpha ML\lambda/D}
\end{align}
At $L  = 1$ and small $M$, we usually expect $P_{\mathrm{c,cir}}(L = 1) > P_{\mathrm{c,sw}}(L=1)$ since the main circuit is in general much deeper than the layers of additional \textsc{cswap}s in single-layer VCP ($D \gg M$). As we increase $L$, $P_{\mathrm{c,cir}}$ will decrease exponentially (we will discuss the case where $P_{\mathrm{c,cir}}$ is not always decreasing in \cref{sec:vcp_err_furth_consider}) while $P_{\mathrm{c,sw}}$ will increase (at first linearly). We will increase the number of layers until $P_{\mathrm{c,cir}}(L^*) \sim P_{\mathrm{c,sw}}(L^*)$ because beyond this point $P_{\mathrm{c,sw}}$ will dominate and the overall errors will increase with further increase in $L$. At this optimal $L^*$, the overall error rate after VCP
\begin{equation}
    \begin{split}
         P_{\mathrm{c,tot}}(L) & = P_{\mathrm{c,cir}}(L) + P_{\mathrm{c,sw}}(L)
    \end{split}
\end{equation}
will be minimised. The exact number of optimal layers should be obtained by taking the derivative of the total error rate and setting it to zero, which can be done numerically. In order to obtain an analytical expression, we will obtain an approximate optimal number of layers based on the arguments above by solving:
\begin{align}
    P_{\mathrm{c,sw}}(L^*) &= P_{\mathrm{c,cir}}(L^*) \nonumber\\
    e^{-\alpha ML^*\lambda/D}&= \left[1-n_c^{1-M}\left(1 - e^{-\lambda/L^*}\right)^M\right]^{L^*} \nonumber\\
     L^* &= -\lambda/\ln(1 - n_c^{\frac{M-1}{M}}\left(1 - e^{-\alpha M\lambda/D}\right)^{\frac{1}{M}})\label{eqn:L_opt_full_expr}
\end{align}
Following this assumption of $P_{\mathrm{c,sw}}(L^*) = P_{\mathrm{c,cir}}(L^*)$, the overall error after VCP at the optimal layer number is then given as:
\begin{equation}\label{eqn:P_c_tot}
    \begin{split}
         P_{\mathrm{c,tot}}(L^*) & = P_{\mathrm{c,cir}}(L^*) + P_{\mathrm{c,sw}}(L^*)\\
         & = 2 P_{\mathrm{c,sw}}(L^*)= 2 P_{\mathrm{c,cir}}(L^*)
    \end{split}
\end{equation}

which can be expressed into more explicit form by substituting \cref{eqn:L_opt_full_expr} into \cref{eqn:P_c_cir,eqn:P_c_sw,eqn:P_c_tot}.
\begin{figure}
\centering
\includegraphics[width=0.48\textwidth]{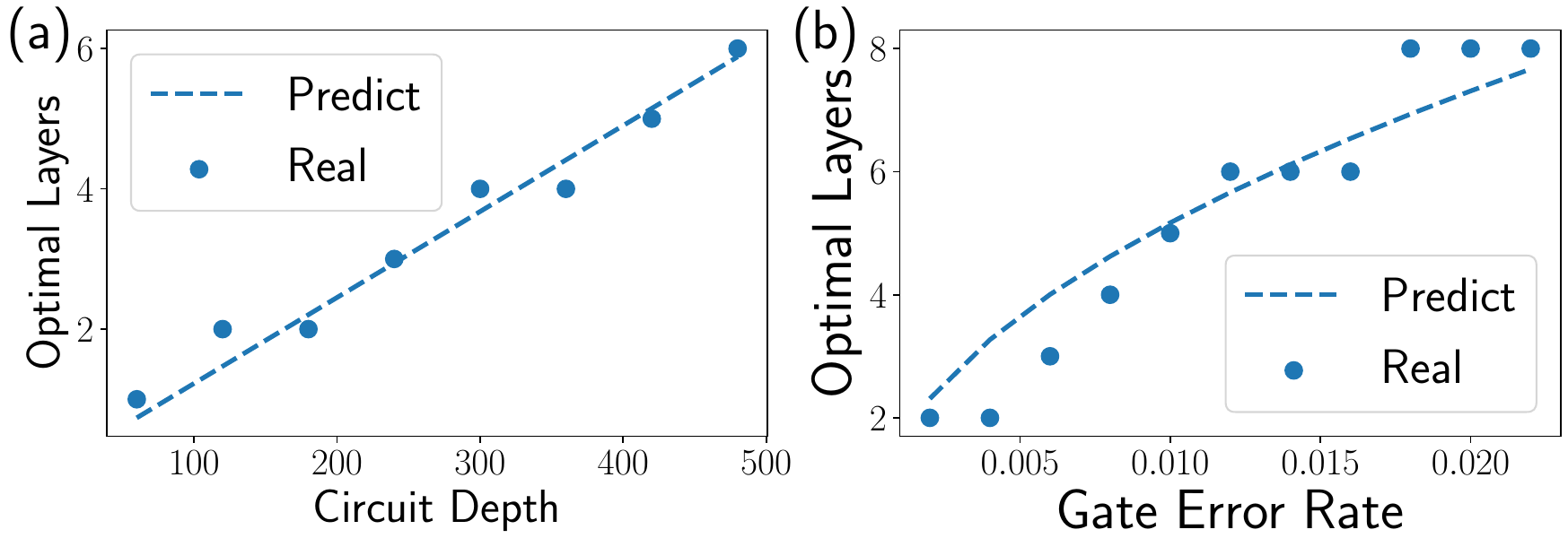}
\caption{ (a) shows the scaling of optimal layers with circuit depth, where we consider the same four-qubit random unitary circuit as in the main text with gate error rate $p=0.005$.
(b) shows the scaling of optimal layers with gate error rate, with circuit depth $D=240$. 
}
\label{fig:data_figure_app}
\end{figure}

\subsection{Small noise limit}\label{sec:vcp_err_small_noise}

An interesting limit to consider is when $n_c^{M-1}\alpha M\lambda/D \ll 1$ which is true for many practical cases. In this case, we have
\begin{align*}
     L^* \approx \lambda \left(\frac{D}{\alpha Mn_c^{M-1}\lambda}\right)^{1/M} 
      = D \left(\alpha M\right)^{-\frac{1}{M}} \left(\frac{Np}{n_c}\right)^{\frac{M-1}{M}} 
\end{align*}
For a fixed gate layer error rate $Np = \lambda/D$, we have $L^*$ grows linearly with $D$. On the other hand, at fixed $D$, we have $L^* \propto (Np)^{1/2}$ at $M =2$. We indeed see such trends in the numerical experiments in \cref{fig:data_figure_app}. The mismatch is mainly due to rounding $L^*$ to the nearest integer.

The optimal depth per VCP layer is given as:
\begin{align}\label{eqn:d_opt_1}
    d^*  = D / L^* \approx (\alpha M)^{\frac{1}{M}}\left(\frac{n_c}{Np}\right)^{\frac{M-1}{M}}. 
\end{align}
which is independent of $D$ and only determined by the error rate in each gate layer $Np$.

By noting that $P_{\mathrm{c,sw}}(L^*) = P_{\mathrm{c,cir}}(L^*)$ implies:
\begin{align*}
    n_c^{1-M}\left(1 - e^{-\lambda/L^*}\right)^M = 1 - e^{-\alpha M\lambda/D}
\end{align*}
Further using the fact that $n_c^{M-1}\alpha M\lambda/D \ll 1$ implies $\alpha M\lambda/D \ll 1$, we can then obtain
\begin{align*}
    n_c^{1-M}\left(1 - e^{-\lambda/L^*}\right)^M = \alpha M\lambda/D \ll 1
\end{align*}
This enable us to further simplify the expression of \cref{eqn:P_c_cir} and \cref{eqn:P_c_sw} at $L = L^*$ to
\begin{align*}
    P_{\mathrm{c,cir}}(L^*) &\approx L^*n_c^{1-M}\left(1 - e^{-\lambda/L^*}\right)^M\\
    P_{\mathrm{c,sw}}(L^*) &\approx \alpha ML^*\lambda/D
\end{align*}
Hence, the total error rate is
\begin{align}
    P_{\mathrm{c,tot}}(L^*) & = P_{\mathrm{c,cir}}(L^*) + P_{\mathrm{c,sw}}(L^*) \nonumber \\
    \approx 2 P_{\mathrm{c,cir}}(L^*)&= 2 L^*n_c^{1-M}\left(1 - e^{-\lambda/L^*}\right)^M \label{eqn:vcp_tot_err_circ}\\
    \approx 2 P_{\mathrm{c,sw}}(L^*) & = 2\alpha ML^*\lambda/D\label{eqn:vcp_tot_err_sw}
\end{align}
Using the expression of $L^*$, we can simplify this to
\begin{align*}
    P_{\mathrm{c,tot}}(L^*) &= 2\alpha ML^*\lambda/D\\
    & = 2 D \left(\frac{\alpha M}{n_c}\right)^{\frac{M-1}{M}}  \left(Np\right)^{\frac{2M-1}{M}} 
\end{align*}

\subsection{Alternative Derivation}
From \cref{eqn:d_opt_1}, we see that the optimal depth of each VCP layer is solely determined by the error rate of each gate layer and is independent of the overall depth of the circuit. Deeper circuits simply mean more repetition of these VCP layers of optimal depth. Hence, unsurprisingly the optimal depth of the circuit can be derived by looking at only individual VCP layers without considering the whole circuit. Denoting the depth of the VCP layer as $d$, then the unmitigated error rate of each VCP layer is around $dNp$ in the small noise regime of $dNp < 1$. In each VCP layer, there is $2MN$ \textsc{cswap}s with error rate $\alpha p$ and as mentioned before, only half of them will contribute to the \textsc{cswap} errors, thus the \textsc{cswap} error rate in each VCP layer is around $MN\alpha p$ in the small noise regime. In such a way, the total error rate per unit depth after applying VCP to the layer is
\begin{align}\label{eqn:tot_err_d}
    P_{c,\mathrm{tot}}/D = d^{-1}\left(n_c^{1-M}(dNp)^M + MN\alpha p\right)
\end{align}
Taking the derivative against $d$ and setting it to $0$ we can obtain 
\begin{align*}
    {d^*}^{-2} (MN\alpha p) &= (M-1){d^*}^{M-2}n_c^{1-M} (Np)^M \\
    d^* & = \left(M-1\right)^{-\frac{1}{M}}(\alpha M)^{\frac{1}{M}}\left(\frac{n_c}{Np}\right)^{\frac{M-1}{M}}
\end{align*}
We see that this is the same as \cref{eqn:d_opt_1} with an additional factor of $(M-1)^{-1/M}$ since we are taking the derivative rather than simply setting the two noise terms to be the same. Note that $(M-1)^{-1/M} = 1$ for the most practical $M =2$ case.

\subsection{Further considerations}\label{sec:vcp_err_furth_consider}
Note that there are still some factors we have not taken into account in the above analysis. The number of channel noise components per VCP layer $n_c$ should be dependent on the depth per VCP layer $d$. Furthermore, we have not considered the simultaneous occurrence of errors in both the circuit and \textsc{cswap}s such that the total error rate for VCP is $P_{\mathrm{c,cir}}(L^*) + P_{\mathrm{c,sw}}(L^*) - P_{\mathrm{c,cir}}(L^*) P_{\mathrm{c,sw}}(L^*)$ and similarly for VSP. Moreover, the Poisson approximation we made for noise rate may not be fully suitable for some extreme cases. For the \textsc{cswap} noise, only the noise acting on the register that we are going to measure will have the strongest effect, at least for \textsc{cswap}s at the end of the circuit. For the parameter regime that we consider, these should not change any qualitative conclusion we draw in any significant way. 

There is another assumption that we have not mentioned in the analysis above, that is, $P_{\mathrm{c,cir}}$ always decreases as $L$ increases. Or equivalently, with noiseless \textsc{cswap}s, the error rate per unit depth decreases as $d$ decreases. This is actually only true in specific parameter regimes. Let us look at the error rate per unit depth with noiseless \textsc{cswap}s without taking the small noise limit:
\begin{align*}
    P_{\mathrm{c,cir}}/D &\approx d^{-1}n_c^{1-M}\left(1 - e^{-dNp}\right)^M
\end{align*}
At small $d$, we have $P_{\mathrm{c,cir}}/D \approx \left(\frac{dNp}{n_c}\right)^{M-1}Np \propto d^{M-1}$, which increases as $d$ increase. This is the region that we have considered before. However, if we look at large $d$, we have $P_{\mathrm{c,cir}}/D \approx n_c^{1-M} d^{-1}$ which decreases as $d$ increase.

To find the transition point, we can differentiate $P_{\mathrm{c,cir}}/D$ against $d$:
\begin{align*}
    \pdv{d}(P_{\mathrm{c,cir}}/D) &\approx n_c^{1-M}d^{-2}\left(1 - e^{-dNp}\right)^{M-1} \\
    & \quad \times \left((1+MdNp)e^{-dNp}-1\right)
\end{align*}
This will be $0$ at a threshold value of $d_{th}Np < 1.256$ for $M=2$. When $dNp < 1.256$, we have $1+MdNp > e^{dNp}$ and thus $\pdv{d}(P_{\mathrm{c,cir}}/D) > 0$ and $P_{\mathrm{c,cir}}/D$ increases as $d$ increases as expected. However,  when $dNp > 1.256$, we have $\pdv{d}(P_{\mathrm{c,cir}}/D) < 0$ and $P_{\mathrm{c,cir}}/D$ decreases as $d$ increases. 

Hence, when the circuit fault rate $\lambda = DNp < 1.256$, then we naturally have $dNp < 1.256$ and all of our arguments above are valid. On the other hand, if we have $\lambda = DNp > 1.256$, then the optimal $d^*$ we found through setting the derivative of the total error to $0$ will always be on the side of $d^*Np < 1.256$ (since only on this side we have a positive gradient cancel with a negative gradient of the \textsc{cswap} noise). On the side of $DNp < 1.256$, the smallest error is always given by $d = D$ since the error decreases with the increase of $d$. Hence, all we need to do is to compare $d^*$ with $d = D$ to see which one is a better solution. 
As $M$ increases, the threshold value for $dNp$ will increase and thus the threshold value for $DNp$ will decrease.

\section{Comparison between VSP and VCP}\label{sec:vsp_vcp_compare}
\subsection{Applicability of VSP and VCP}
Here we follow the same notations outlined in \cref{sec:vcp_err_overall} to consider when is VSP and VCP applicable, respectively. We are applying VCP to individual VCP layers, thus we only need to look at the requirement on the individual layers. While we want to increase the layer number so that the dominant component per layer becomes the noiseless component, we also want to restrict the layer number so that the number of gates per layer is large compared to the number of \textsc{cswap}s per layer $2MN$, to minimise the damages due to \textsc{cswap} noise. We will use $n_c$ to denote the number of (leading) error components in the noise channel. In order for VCP to work, we need the identity component to be leading, i.e. $e^{-\lambda/L} > (1-e^{-\lambda/L})/n_c$, which means $\ln (n_c + 1) > \lambda/L$. For global depolarising noise where $n_c + 1 = 4^N$, we have $2N\ln(2)> \lambda/L$ as requirement. Using $\lambda = DNp$ we have
\begin{align*}
\frac{D}{L} = d < \frac{2\ln(2)}{p}
\end{align*}
as the requirement, where $d = D/L$ is the depth per VCP layer. If $p = 0.01$, then we can have $D/L$ up to $60$, which is much more than the depth of the \textsc{cswap}s $2M$ for $M = 2$ and thus the number of gates per layer is much larger than the numbers of \textsc{cswap}s. The restriction on VSP is essentially the same with $L = 1$ and $n_c = 2^N$, i.e. we have
\begin{align*}
D_{\mathrm{vsp}} < \frac{\ln(2)}{p}
\end{align*}
This places a hard bound on the depth of the circuit for a given gate error rate for VSP, while such depth bound only applies to the individual VCP layers in VCP and we can apply it to circuits of arbitrary depth by simply increasing the number of VCP layers.

\subsection{Comparing Error Suppression of VSP and optimal VCP}
Here we will consider the region where that noise is not so large such that VSP is still applicable, which also implies that VCP is also applicable. Following the same argument as \cref{sec:vcp_err_overall}, the error rate of the circuit after $M$th order VSP without \textsc{cswap} noise and the error rate due to the \textsc{cswap}s are:
\begin{align}
    P_{\mathrm{s,cir}} &\approx n_s^{1-M}\left(1 - e^{-\lambda}\right)^M\label{eqn:pscir}\\
    P_{\mathrm{s,sw}} &\approx 1 - e^{-\alpha M\lambda/D}
\end{align}
respectively. Here $n_s$ is the number of (dominating) error components in the state. Note that $P_{\mathrm{s,cir/sw}}$ is the same as $P_{\mathrm{c,cir/sw}}$ with $L = 1$ and $n_c = n_s$. In general we have $n_c \geq n_s$ as discussed in the main text, and thus $P_{\mathrm{c,cir}} < P_{\mathrm{s,cir}}$. 

The ratio between the error rate after VSP and optimal-layer VCP is:
\begin{align}\label{eqn:error_ratio}
    R&= \frac{P_{\mathrm{s,cir}} + P_{\mathrm{s,sw}}}{P_{\mathrm{c,cir}}(L^*) + P_{\mathrm{c,sw}}(L^*)} \approx \frac{P_{\mathrm{s,cir}}}{2P_{\mathrm{c,cir}}(L^*)} + \frac{P_{\mathrm{s,sw}}}{2P_{\mathrm{c,sw}}(L^*)} \nonumber\\
     &\geq \frac{P_{\mathrm{s,cir}}}{2P_{\mathrm{c,cir}}(L^*)} 
\end{align}
In other words, this is lower-bounded by the improvement of optimal-layer VCP over VSP with noiseless \textsc{cswap}s divided by $2$. 

Starting from the small noise limit in \cref{sec:vcp_err_small_noise} where $\alpha M\lambda/D \ll 1$, this implies that $P_{\mathrm{s,sw}} \approx M\lambda/D \ll 1$. Hence, the resultant error ratio between VSP and VCP using \cref{eqn:error_ratio} is then
\begin{align*}
 \frac{P_{\mathrm{s,cir}}}{2P_{\mathrm{c,cir}}(L^*)} = \frac{1}{2L^*} \left(\frac{n_{c}}{n_{s}}\right)^{M-1} \left(\frac{1-e^{-\lambda}}{1 - e^{-\lambda/L^*}}\right)^M
\end{align*}
For small $\lambda$ such that $\lambda \ll L^*$ (i.e. $d^* Np \ll 1$), we have:
\begin{align*}
     R \approx \frac{1}{2}{L^*}^{M-1} \left(\frac{n_{c}}{n_{s}}\right)^{M-1}
\end{align*}
which is always larger than $1$ if $n_c \geq n_s$ and $L \geq 2$ (given that $M \geq 2$). 

If we assume weak dependence of $n_c$ on $L$, then $(n_{c}/n_{s})^{M-1}$ is approximately the improvement of single-layer VCP over VSP with noiseless \textsc{cswap}s using \cref{eqn:P_c_cir,eqn:pscir}
\begin{align*}
    \left(\frac{n_{c}}{n_{s}}\right)^{M-1} \approx \frac{P_{\mathrm{s,cir}}}{P_{\mathrm{c,cir}}(L=1)}.
\end{align*}
Hence, the $R$ can be rewritten as:
\begin{align*}
    R \approx \frac{1}{2}{L^*}^{M-1}\frac{P_{\mathrm{s,cir}}}{P_{\mathrm{c,cir}}(L=1)}.
\end{align*}

\section{Practical Considerations}

\subsection{Circuit Variants with Separate Control Qubits}\label{sec:circ_variant}
\begin{figure}[htbp]
\centering
\includegraphics[width = 0.27\textwidth]{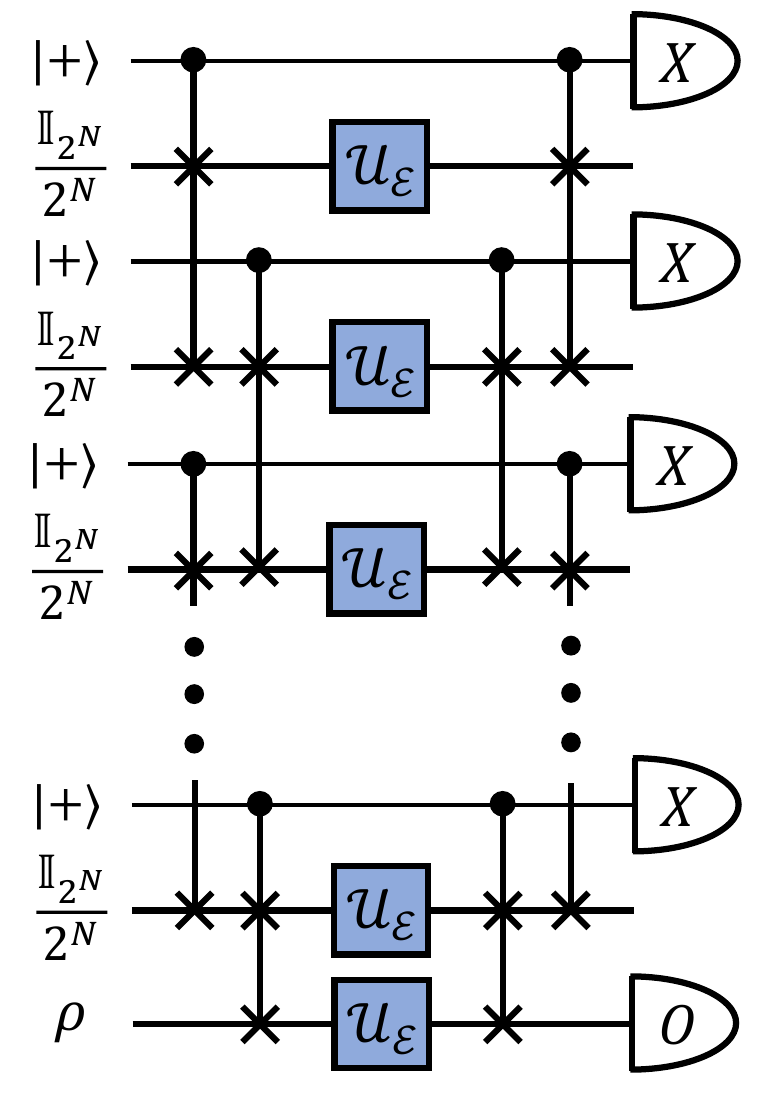}
\caption{Circuit for virtual channel purification with separate control qubits for each \textsc{cswap}.}
\label{fig:vir_dis_circ_sep_anc}
\end{figure}
Instead of having one control qubit for all controlled register-swaps, we can also have a control qubit for each pair of controlled register-swaps acting between a given pair of registers. In \cref{eq:exp_val_higher_order} from \cref{fig:higher_order}(a), the right Kraus operator is a cyclic permutation of the left Kraus operator. 
The new output from \cref{fig:vir_dis_circ_sep_anc} will be a superposition of different terms with the right Kraus operator being a different derangement of the left Kraus operator. The end result when we remove all the cross terms in virtual channel purification will be the same as \cref{eq:exp_val_higher_order}. 
The benefit of using separable control qubits is that one can choose an optimal permutation operator $D_M$ in \cref{fig:higher_order}(a) to achieve a constant circuit depth.

The validation of \cref{fig:vir_dis_circ_sep_anc} requires the fact that for a circuit that applies the whole set of linear nearest neighbour swaps in a random order, the whole circuit is always a derangement operation. This can be shown by realising that the qubit at position $i$ is touched by only two swaps in the entire circuit. One takes it to $i+1$ and the other takes it to $i-1$. If the first swap it meets takes it to $i+1$, then the only next possible swap that it can meet at $i+1$ is the one that goes from $i+1$ to $i+2$ since the other swap that goes from $i+1$ to $i$ is already used. Hence, in this way, the qubit can only remain at $i+1$  or increase its index to $i+2$. If the first swap it meets takes it to $i-1$, then it can only remain at $i-1$  or decrease its index to $i-2$. The qubit will never end up back in its original position $i$ and thus the circuit we show can ``derange'' the qubit indices. 

\begin{figure}[htbp]
    \centering
    \includegraphics[width = 0.5\textwidth]{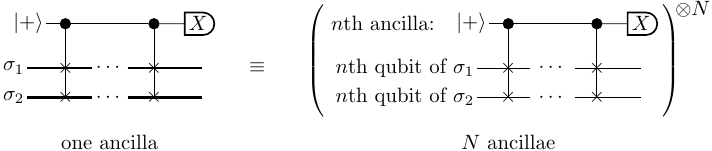}
    \caption{Circuit implementing controlled register-swaps using one control qubit can be replaced by a circuit using one control qubit for each pair of corresponding qubits. The overall control qubit measurement outcome is simply the product of the measurement of all control qubits.}
    \label{fig:cswap_decomp}
\end{figure}

For a given controlled register-swap, instead of just using one control qubit and performing \textsc{cswap} between all of the corresponding qubit pairs within different registers, we can assign one control qubit for each pair of the corresponding qubit and perform the \textsc{cswap} using that dedicated control qubit as shown in \cref{fig:cswap_decomp}. The overall control qubit measurement outcome is simply the product of the measurement of all control qubits. This can be done because the swap operations between different qubit pairs commute with each other, and thus it does not matter whether we perform it on the left or right Kraus operators since they are all the same when we take the trace. 

\subsection{Deviation from Pauli Noise Model}\label{sec:deviation_from_Pauli}
A given noise channel $\mathcal{E}$ acting on a $2^N$ dimensional Hilbert space can always be written in the normalised Kraus representation:
\begin{align}\label{eqn:norm_noise_channel}
\mathcal{E}(\rho) = \sum_{i = 0}^{4^N-1} p_i E_{i}\rho E_{i}^\dagger = \sum_{i = 0}^{4^N-1} p_i {\supop{E}}_{i}(\rho)
\end{align}
with non-negative $p_i$ and normalised Kraus operators $\Tr(E_{i}^\dagger E_{i}) = 2^N$. The trace-preserving condition of the Kraus operators implies that $\sum_i p_i = 1$. We have used the overbracket ${\supop{E}}_{i}$ to denote the channel corresponding to the operator $E_{i}$. Without loss of generality, we will assume all Kraus operators are ordered in descending order of $p_i$, i.e. $p_{0} \geq p_{1} \cdots p_{4^N - 1} \geq 0$. 

Just like regular Kraus representation, the normalised version in \cref{eqn:norm_noise_channel} is not unique for a given quantum channel. We will look at noise channels that, in at least one of their normalised Kraus representation, have their dominant Kraus operator $E_0$ being the identity operator $\mathbb{I}_{2^N}$, i.e. $p_0 > p_1$ and $E_0 = \mathbb{I}_{2^N}$:
\begin{align}\label{eqn:id_noise_channel_gen}
\mathcal{E} = p_0 \mathcal{I} + \sum_{i = 1}^{4^N-1} p_i {\supop{E}}_{i}
\end{align}
We will call such a channel the \emph{identity-dominant channel}, with the corresponding normalised Kraus representation being the \emph{identity-dominant Kraus representation}. Note that the identity-dominant Kraus representation is not unique.

Identity-dominant channels become \emph{stochastic noise}~\cite{graydonDesigningStochasticChannels2022} when one of the {identity-dominant Kraus representation} coincide with the canonical Kraus representation of the channel, such that the Kraus operators in \cref{eqn:id_noise_channel} are orthogonal to each other
\begin{align}\label{eqn:noise_ortho_gen}
    \Tr(E_{i} E_{j}^\dagger) = \delta_{ij} 2^N.
\end{align}
Stochastic noise contains \emph{Pauli noise}~\cite{carignan-dugasPolarDecompositionQuantum2019} (including dephasing and depolarising noise) that we have discussed and also more general mixed-unitary noise model that contains identity component.

Our purification scheme can be easily applied to stochastic noise channels beyond Pauli noise. For stochastic noise that satisfies \cref{eqn:noise_ortho_gen}, we can use the maximally mixed ancillary state $\sigma = \mathbb{I}_{2^N}/2^N$ to satisfy the purification condition in \cref{eqn:purify_cond}. However, in general the purified channel in the form of \cref{eqn:purified_channel} is not properly normalised because in general $\sum_{i}p_iE_i^\dagger E_i = \mathbb{I}_{2^N}$ does not imply $P_M^{-1}\sum_{i} p_i^M E_i^\dagger E_i = \mathbb{I}_{2^N}$. In order to normalised it, we need to measure a normalising constant for a given input $\rho$ by simply setting $O=\mathbb{I}_{2^N}$. However, it is possible that even the unnormalised version will already give us a better output than the noisy version. This can be left for future investigation. 

Similar to VSP, our method will also suffer from coherent mismatch when there is coherent noise in the target noise channel~\cite{hugginsVirtualDistillationQuantum2021,koczorExponentialErrorSuppression2021}. Though such coherent noise is expected to be small for circuits of practical sizes~\cite{koczorDominantEigenvectorNoisy2021}.

\subsection{Tensor Product of Channels}\label{sec:channel_tensor_prod}
Given a product of two noise channels $\mathcal{E}$ and $\mathcal{F}$, their tensor product is simply
\begin{align*}
    \mathcal{E} \otimes \mathcal{F} &=\left(\sum_{i=0}^{4^N-1}p_i{\supop{E}}_i\right) \otimes \left(\sum_{j=0}^{4^N-1}q_j{\supop{F}}_j\right) \\
    & = \sum_{i,j=0}^{4^N-1}p_iq_j{\supop{E}}_i\otimes{\supop{F}}_j.
\end{align*}
If we perform purification on $\mathcal{E} \otimes \mathcal{F}$, we have
\begin{align*}
    &\quad (\mathcal{E} \otimes \mathcal{F})^{(M)}\\
    & = \frac{1}{\sum_{i,j=0}^{4^N-1}(p_iq_j)^M} \sum_{i,j=0}^{4^N-1}(p_iq_j)^M{\supop{E}}_i \otimes {\supop{F}}_j\\
    &=\frac{1}{\sum_{i=0}^{4^N-1}p_i^M}\left(\sum_{i=0}^{4^N-1}p_i^M{\supop{E}}_i\right) \otimes \frac{1}{\sum_{i=0}^{4^N-1}q_j^M}\left(\sum_{j=0}^{4^N-1}q_j^M{\supop{F}}_i\right) \\
& = \mathcal{E}^{(M)} \otimes \mathcal{F}^{(M)}
\end{align*}
which is the same as performing purification on $\mathcal{E}$ and $\mathcal{F}$ separately. More generally, performing purification on a tensor product of channels will obtain a tensor product of purified channels. 

Note that we may be tempted to say that the same is true for  $\mathcal{E}\mathcal{F}$ instead of $\mathcal{E} \otimes \mathcal{F}$. This will be true if all $E_iF_j$ are different. However, actually some $E_iF_j$ for different $i$ and $j$ will be the same, causing the merging of terms such that the argument above is not valid. This will be obvious when considering $\mathcal{E}\mathcal{E}$ versus $\mathcal{E}\otimes \mathcal{E}$. This is why we have the advantages for layer-by-layer VCP.

The leading error terms of a tensor product of channels is approximately the sum of the leading error terms of the individual local channel. Hence, if we perform purification to a tensor product of channels, the factor of error suppression is roughly the same as the error suppression factor of the local channels. Similarly, the factor of gain of VCP over VSP (the ratio between their error suppression factors) for the tensor product of local channels is roughly the same as that of the component local channels.

\section{Connection to VSP}\label{sec:vsp_connection}

\subsection{State purification}
For the given input state $\rho = \ketbra{\psi}$, we can always choose a Kraus representation such that
\begin{align}\label{eqn:kl_ortho_state}
    \bra{\psi} E^\dagger_j E_i \ket{\psi} = \delta_{ij}
\end{align}
If under this Kraus representation, the leading Kraus operator is still $E_0 = \mathbb{I}_{2^N}$, which simply means that the non-identity noise elements will take the incoming state into an orthogonal state, then choosing the ancillary state to be the same as the input state $\sigma = \rho = \ketbra{\psi}$ will allow us to satisfy \cref{eqn:purify_cond}, which then allow us to perform channel purification. The reason why the noise assumption in this section is likely to be valid in practice is discussed in Ref.~\cite{koczorDominantEigenvectorNoisy2021}. 

\subsection{State verification}
As derived in \cref{sec:circuit_generalisation}, if we add an $S$ measurement to all ancillary states, then the output of the circuit is
\begin{align}\label{eqn:circ_output_with_meas}
    & \expval{X\otimes S^{\otimes M-1} \otimes O}  = \sum_{i_1,\cdots,i_M=0}^{4^N-1}\left(\prod_{m=1}^M p_{i_m}\right) \nonumber\\
    & \quad\quad\quad \times \Tr(E_{i_{M}}U\rho U^\dagger E^\dagger_{i_{1}} O) \prod_{m = 1}^{M-1}\Tr(E_{i_{m}} U\sigma U^\dagger E^\dagger_{i_{m+1}}S)
\end{align}
If we are performing state purification using input state $\ket{\psi}$ and the same ancillary states, we can further suppress the noise using the state verification protocol introduced in \cite{obrienErrorMitigationVerified2021}. It is performed by making noisy projective measurements of $S = \mathcal{U}_{\epsilon}(\ketbra{\psi})$, which can be approximately carried out by applying the noisy inverse channel and then performing projective measurement of $\ketbra{\psi}$~\cite{obrienErrorMitigationVerified2021,huoDualstatePurificationPractical2022,caiResourceefficientPurificationbasedQuantum2021}. In such a case, we have:
\begin{align}\label{eqn:purified_cond_state_ver}
    &\quad \Tr(E_iU\sigma U^\dagger E^\dagger_j S) \nonumber \\
    &= \sum_a p_a \Tr(E_iU \ketbra{\psi} U^\dagger E^\dagger_j E_a U \ketbra{\psi} U^\dagger E^\dagger_a) \nonumber \\
    & = \sum_{a} p_a \bra{\psi}U^\dagger E^\dagger_j E_a U \ket{\psi}  \bra{\psi}U^\dagger E^\dagger_a E_iU \ket{\psi} = p_i \delta_{ij}
\end{align}
where we have used \cref{eqn:kl_ortho_state} in the last step. Substituting into \cref{eqn:circ_output_with_meas} we see that this can further suppress the noise by an additional $M-1$ order.

\section{Coherent and incoherent detector}\label{sec:detector}

\subsection{Spacetime check}
In the recently proposed spacetime picture of viewing QEC~\cite{mcewenRelaxingHardwareRequirements2023,gottesmanOpportunitiesChallengesFaultTolerant2022,delfosseSpacetimeCodesClifford2023}, correction of circuit faults in conventional $D$ dimensional space codes can be viewed as correction of Pauli errors in $D+1$ dimension spacetime code. The building blocks of these spacetime codes are called ``detectors'', which are a set of stabiliser checks of the corresponding $D$ dimensional space code that has a deterministic collective parity under noiseless execution, i.e. these set of checks should output results that are consistent with one another. The simplest example of a detector is a pair of consecutive checks of the same stabiliser. In the noiseless case, the result of these two checks will be the same, i.e. their collective parity should be even. In the region between these two checks, any faults that anti-commute with the check will flip the collective check parity and be detected, thus this region is called the detection region of the detector.

\begin{figure}
    \centering
    \includegraphics[width=0.45\textwidth]{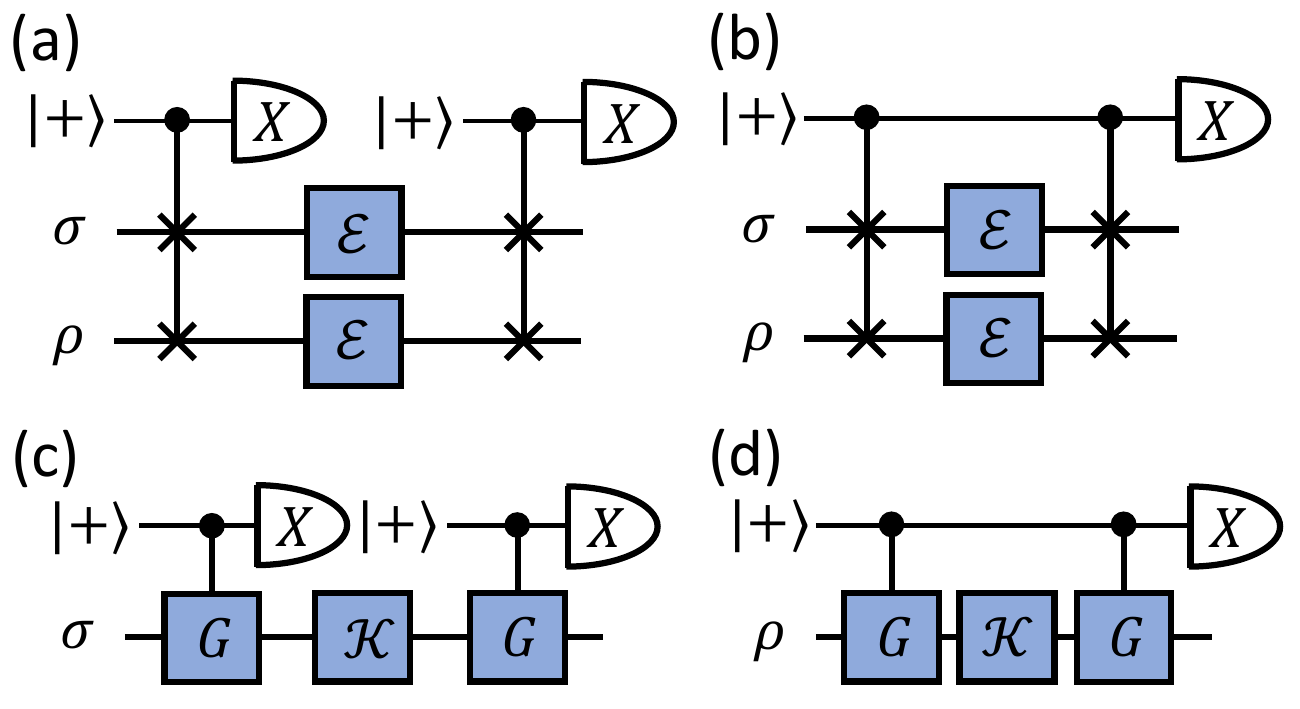}
    \caption{Different circuits for measuring the detector.}
    \label{fig:detector_circ_incoh}
\end{figure}

For the case in which our check is the \textsc{swap} operator, the circuit is simply shown in \cref{fig:detector_circ_incoh}(a). 
By post-selecting on the cases in which the measurement results of the two control qubits are the same (the collective parity of the two checks are even), we can project the incoming state and the outgoing state both into the $+1$ or both into the $-1$ eigenspace of the \textsc{swap} operator. 
The detection region in this case is simply the region between the two \textsc{swap}s where the noise channel $\mathcal{E} \otimes \mathcal{E}$ happens. 
As we will show later, our detector can remove any components of $\mathcal{E} \otimes \mathcal{E}$ that are not invariant under the conjugation of \textsc{swap}.

Instead of measuring the two checks one by one and computing their parity in post-processing, we can directly measure the collective parity of these two checks using one control qubit as shown in \cref{fig:detector_circ_incoh}(b). In this way, we essentially directly measure the stabiliser checks of the space-time code~\cite{delfosseSpacetimeCodesClifford2023} instead of trying to obtain it via post-processing the check results of the corresponding spatial code. 
This is also the type of circuit that is used for flag fault-tolerance~\cite{chaoFlagFaultTolerantError2020}. 
We will refer to this type of circuit implementation as the \emph{coherent detector check} and the one in \cref{fig:detector_circ_incoh}(a) as the \emph{incoherent detector check}. 
As we will show later, the coherent detector checks in \cref{fig:detector_circ_incoh}(b) can remove the noise components in the detection region that are not invariant under \textsc{swap}, just like the incoherent detector checks. 
However, unlike the incoherent detector check, the coherent detector check will \emph{not} project the incoming state and the outgoing state into the eigensubspace of \textsc{swap}, i.e. the coherent detector check acts only on the channel in the detection region. 

Since our channel purification protocol also only focuses on improving the channel rather than placing restrictions on the incoming and output state, it is natural for us to use the coherent detector circuit to perform channel purification as we have outlined before. If we indeed try to use the incoherent detector circuit for channel purification, the symmetrisation of the input state means that we have $\frac{1}{2} \left(\rho \otimes \sigma + \sigma \otimes \rho\right)$ as output state instead. Hence, when it comes to the measurement stage, the only way to perform $O$ measurement on $\rho$ is by performing $O$ measurement on both of the registers since they both have $\rho$ components, or by having $\sigma = \rho$. Both of these cases can be challenging in choosing the ancillary state and measurement $\sigma$ and $S$ in order to fulfil the channel purification condition in \cref{eqn:purify_cond} as will be explained later. This is why we focus on using the coherent detector to perform channel purification in this paper.

\subsection{Property of involution operators}
For a given involution operator $G$ (square to $I$), we can decompose any operator $A$ into components that commute and anti-commute with $G$:
\begin{align}\label{eqn:involution_decomp}
    A = A_+ + A_-
\end{align}
with
\begin{align*}
    A_\pm = \frac{A\pm GAG}{2}.
\end{align*}
It is easy to see that $GA_\pm = \pm A_\pm G$. 

Further defining the projection operators
\begin{align*}
    \Pi_{\pm} = \frac{\mathbb{I} \pm G}{2},
\end{align*}
in this way, we can obtain the components of $A$ that are in the $\pm 1$ eigensubspace of $G$ via:
\begin{equation}\label{eqn:even_parity_swap_id}
    \begin{split}
        \Pi_\pm A \Pi_{\pm}  = \Pi_{\pm} A_{+} \Pi_{\pm}
    = \Pi_{\pm} A_{+}  =  A_{+} \Pi_{\pm}
    \end{split}
\end{equation}
Note that $\Pi_{\pm} A \Pi_{\pm} \neq A_\pm$. $A_\pm$ are components in $A$ that gain a $\pm 1$ factor when \emph{conjugated} with $G$, while $\Pi_\pm A \Pi_{\pm}$ are components that get an $\pm1$ factor when \emph{acted by $G$ on either side}.

For the components of $A$ that connect between the $\pm 1$ eigensubspaces of $G$, we have:
\begin{equation}\label{eqn:odd_parity_swap_id}
    \begin{split}
        \Pi_\pm A \Pi_\mp  = \Pi_\pm A_{-} \Pi_\mp  
        =  \Pi_{\pm}A_{-} = A_{-}\Pi_{\mp}
    \end{split}
\end{equation}

We can rewrite $A$ as
\begin{align*}
    A &= \underbrace{\Pi_{+} A \Pi_{+} + \Pi_{-} A \Pi_{-}}_{A_+} + \underbrace{\Pi_{+} A \Pi_{-} + \Pi_{-} A \Pi_{+}}_{A_-}\\
    &= \underbrace{\Pi_{+} A_+ \Pi_{+} + \Pi_{-} A_+ \Pi_{-}}_{A_+} + \underbrace{\Pi_{+} A_- \Pi_{-} + \Pi_{-} A_- \Pi_{+}}_{A_-}
\end{align*}
where we see that $A_+$ contains the components of $A$ in \emph{both} of the $\pm 1$ eigenspace, while $A_+$ contains the components of $A$ that represent the transition between the two eigensubspaces.

\subsection{Coherent detector}
For the noise channel $\mathcal{K}(\rho) = \sum_{i} K_{i} \rho K_{i}^\dagger$, each Kraus element can be decomposed into $K_i = K_{i,+} + K_{i,-}$ as discussed in \cref{eqn:involution_decomp}. 

When post-selecting on $\ket{\pm}$ output on the control qubit of \cref{fig:detector_circ_incoh}(d), the output state is given as:
\begin{align}\label{eqn:coh_detector_output}
 \mathcal{K}_{\mathrm{coh}}^{\pm}(\rho) 
    &= \frac{1}{4}\sum_{i} \bigg[\left(K_{i,+} + K_{i,-}\right) \rho \left(K_{i,+}^{\dagger} + K_{i,-}^{\dagger}\right) \nonumber\\
    &\quad +  \left(K_{i,+} - K_{i,-}\right) \rho \left(K_{i,+}^{\dagger} - K_{i,-}^{\dagger}\right) \nonumber\\
    &\quad \pm \left(K_{i,+} + K_{i,-}\right) \rho \left(K_{i,+}^{\dagger} - K_{i,-}^{\dagger}\right) \nonumber\\
    &\quad \pm \left(K_{i,+} - K_{i,-}\right) \rho \left(K_{i,+}^{\dagger} + K_{i,-}^{\dagger}\right) \bigg]\nonumber\\
    &= \sum_{i} K_{i,\pm} \rho  K_{i,\pm}^{\dagger}
\end{align}
i.e. when for $+ 1$ outcome, we have removed the noise components that anticommute with $G$, and for $-1$ outcome, we have removed the noise components that commute with $G$. Note that $\mathbb{I}$ commutes with $G$, thus if we want to keep the identity component, we need to keep the $+1$ outcome.

Note that the coherent detector only removes the noise components of the wrong symmetry in $\mathcal{K}$ (make it invariant under conjugation of $G$). It does not enforce symmetry on the incoming state $\rho$ and also does not enforce symmetry on the output state.

\subsection{Incoherent detector}
Instead of performing the detector check coherently using one control qubit, we can also perform it incoherently using \cref{fig:detector_circ_incoh}(c) and keep only the output for which the two consecutive control qubit measurement is consistent (i.e. their collective parity is even). In this case, using \cref{eqn:even_parity_swap_id}, the output state is given as:
\begin{align*}
\mathcal{K}_{\mathrm{inc}}^{\mathrm{even},\pm}(\rho)& =  \sum_{i} \Pi_{\pm}  K_{i} \Pi_{\pm}\ \rho\ \Pi_{\pm}  K_{i}^\dagger \Pi_{\pm}  \\
&= \sum_{i} \Pi_{\pm}  K_{i,+} \left(\Pi_{\pm}\rho\Pi_{\pm}\right)   K_{i,+}^\dagger \Pi_{\pm}
\end{align*}
In this case, the anticommuting components in the noise operators $K_{i}$ are removed just like in the coherent detector case. \emph{However, on top of that, the input state and the output state are restricted to the $\pm 1$ eigensubspace of $G$. }

It can be further written as:
\begin{align}\label{eqn:incoh_detector_output_even}
    &\quad \mathcal{K}_{\mathrm{inc}}^{\mathrm{even},\pm}(\rho) = \sum_{i}  K_{i,+}\ \rho_{+}\  K_{i, +}^\dagger \Pi_{\pm}\nonumber\\
    & = \frac{1}{4} \big[\mathcal{K}_{\mathrm{coh}}^{+}(\rho) + \mathcal{K}_{\mathrm{coh}}^{+}(G\rho G)  \pm \left(\mathcal{K}_{\mathrm{coh}}^{+}(\rho)  + \mathcal{K}_{\mathrm{coh}}^{+}(G\rho G)\right)G\big] 
\end{align}
which we can explicitly see that the coherent detector channel is just one of its components.

Similarly, the output of a circuit in \cref{fig:detector_circ_incoh}, with the collective parity of the two measurements being odd, is given as:
\begin{align}\label{eqn:incoh_detector_output_odd}
    &\mathcal{K}_{\mathrm{inc}}^{\mathrm{odd},\pm}(\rho)\nonumber =  \sum_{i} \Pi_{\pm}  K_{i} \Pi_{\mp}\ \rho\ \Pi_{\mp}  K_{i}^\dagger \Pi_{\pm} \nonumber\\
    & =  \sum_{i} \Pi_{\pm}  K_{i,-} \left(\Pi_{\mp}\rho\Pi_{\mp}\right) K_{i,-}^\dagger \Pi_{\pm}\nonumber\\
    & = \frac{1}{4} \big[\mathcal{K}_{\mathrm{coh}}^{-}(\rho) + \mathcal{K}_{\mathrm{coh}}^{-}(G\rho G)  \pm \left(\mathcal{K}_{\mathrm{coh}}^{-}(\rho)  + \mathcal{K}_{\mathrm{coh}}^{-}(G\rho G)\right)G\big] 
\end{align}

\subsection{Implication for virtual channel purification}
In the case of virtual channel purification, we have:
\begin{align*}
    G &\Rightarrow \textsc{swap}\\
    \rho &\Rightarrow \sigma \otimes \rho\\
    \mathcal{K} &\Rightarrow \mathcal{E} \otimes \mathcal{E}
\end{align*}
and the output state using the coherent detector using \cref{eqn:coh_detector_output} is:
\begin{align*}
    \mathcal{K}_{\mathrm{coh}}^{\pm}(\sigma \otimes \rho) & = \frac{1}{2}\sum_{i,j}  \bigg[ \left(K_{i} \otimes K_{j}\right)  \left(\sigma \otimes \rho\right) \left(K_{i}^\dagger \otimes K_{j}^\dagger\right)\\
    &\quad \pm \left(K_{i} \otimes K_{j}\right) \left(\sigma \otimes \rho\right)  \left(K_{j}^\dagger \otimes K_{i}^\dagger\right)\bigg]
\end{align*}

Using the coherent detector circuit, we can obtain the state for virtual channel purification by combining the result of the $\pm$ outcome:
\begin{align*}
    \mathcal{K}_{\mathrm{dis}}(\sigma \otimes \rho) &= \mathcal{K}_{\mathrm{coh}}^{+}(\sigma \otimes \rho) - \mathcal{K}_{\mathrm{coh}}^{-}(\sigma \otimes \rho)\\
    &= \sum_{i,j}\left(K_{i} \otimes K_{j}\right) \left(\sigma \otimes \rho\right)  \left(K_{j}^\dagger \otimes K_{i}^\dagger\right)
\end{align*}

The output state using incoherent detector using \cref{eqn:incoh_detector_output_even,eqn:incoh_detector_output_odd} is:
\begin{align*}
    \mathcal{K}_{\mathrm{inc}}^{\mathrm{even},\pm}(\sigma \otimes \rho)
    & = \frac{1}{4} \big(\mathcal{K}_{\mathrm{coh}}^{+}(\sigma \otimes \rho) + \mathcal{K}_{\mathrm{coh}}^{+}(\rho \otimes \sigma) \\
    &\quad  \pm \left(\mathcal{K}_{\mathrm{coh}}^{+}(\sigma \otimes \rho)  + \mathcal{K}_{\mathrm{coh}}^{+}(\rho \otimes \sigma)\right)\textsc{swap}\big) \\
    \mathcal{K}_{\mathrm{inc}}^{\mathrm{odd},\pm}(\sigma \otimes \rho)
    & = \frac{1}{4} \big(\mathcal{K}_{\mathrm{coh}}^{-}(\sigma \otimes \rho) + \mathcal{K}_{\mathrm{coh}}^{-}(\rho \otimes \sigma) \\
    &\quad  \pm \left(\mathcal{K}_{\mathrm{coh}}^{-}(\sigma \otimes \rho)  + \mathcal{K}_{\mathrm{coh}}^{-}(\rho \otimes \sigma)\right)\textsc{swap}\big) 
\end{align*}

For the incoherent detector circuit, by adding the results with the same collective parity and using the even parity result to subtract the odd parity result, we have:
\begin{align*}
    \mathcal{K}_{\mathrm{nodis}}(\sigma \otimes \rho) &=  \left(\mathcal{K}_{\mathrm{inc}}^{\mathrm{even},+}(\sigma \otimes \rho) + \mathcal{K}_{\mathrm{inc}}^{\mathrm{even},-}(\sigma \otimes \rho)\right)\\
    &\quad - \left(\mathcal{K}_{\mathrm{inc}}^{\mathrm{odd},-}(\sigma \otimes \rho) + \mathcal{K}_{\mathrm{inc}}^{\mathrm{odd},-}(\sigma \otimes \rho)
    \right)\\
    & = \frac{1}{2} \bigg(\mathcal{K}_{\mathrm{coh}}^{+}(\sigma \otimes \rho) + \mathcal{K}_{\mathrm{coh}}^{+}( \rho \otimes \sigma)\\
    &\quad - \mathcal{K}_{\mathrm{coh}}^{-}(\sigma \otimes \rho) - \mathcal{K}_{\mathrm{coh}}^{-}(\rho \otimes \sigma) \bigg)\\
    &= \frac{1}{2} \left(\mathcal{K}_{\mathrm{dis}}(\sigma \otimes \rho) +  \mathcal{K}_{\mathrm{dis}}(\rho \otimes \sigma)\right)
\end{align*}
This seems to be able to be used to purify the channel. However, it is a mixture of output states with $\rho$ in the main register and also $\rho$ in the ancillary register, and this mixture cannot be separated by the control qubit measurement results. Hence, if we measure $O$ on one of the qubits, part of the time, it will measure on the $\sigma$ ancillary state, which is not our targeted result. 

In order to perform channel purification using the incoherent detector circuit. There are two possible solutions:
\begin{itemize}
    \item Choose the ancillary state to be $\rho$ and try to identify an ancillary measurement that satisfy \cref{eqn:purify_cond}.
    \item Fixed the ancillary measurement to be $O$ and try to identify an ancillary input state that satisfy \cref{eqn:purify_cond}.
\end{itemize}
Both of these are not trivial to perform and we will leave it for future investigation.

\section{Connection and Comparison to other Techniques}\label{sec:generalisation}
\subsection{Connection to Existing Methods}
We have seen in \cref{sec:comb_qec}, we can have other choices of ancillary input states beyond the maximally mixed state in order to achieve error mitigation. 
In fact, if the input state is a pure state $\rho = \ketbra{\psi}$, we can choose the ancillary input state to be the same as the main input state such that the first controlled-permutation operator acts trivially and can be removed. In this way, we have recovered the VSP circuit \cite{koczorExponentialErrorSuppression2021}. We have 
discussed in more detail how VSP fit into the general framework of VCP in \cref{sec:vsp_connection}. In the context of state purification, noise can be further suppressed by measuring noisy projectors of the target state on the ancillary systems~\cite{obrienErrorMitigationVerified2021,huoDualstatePurificationPractical2022,caiResourceefficientPurificationbasedQuantum2021}. Adding ancillary measurements is a way to further generalise the VCP protocol as we have also seen in \cref{sec:comb_qec}.

\subsection{Connection to Flag Fault Tolerance} \label{sec:additional_symmetry}
\begin{figure}
    \centering
    \includegraphics[width=0.45\textwidth]{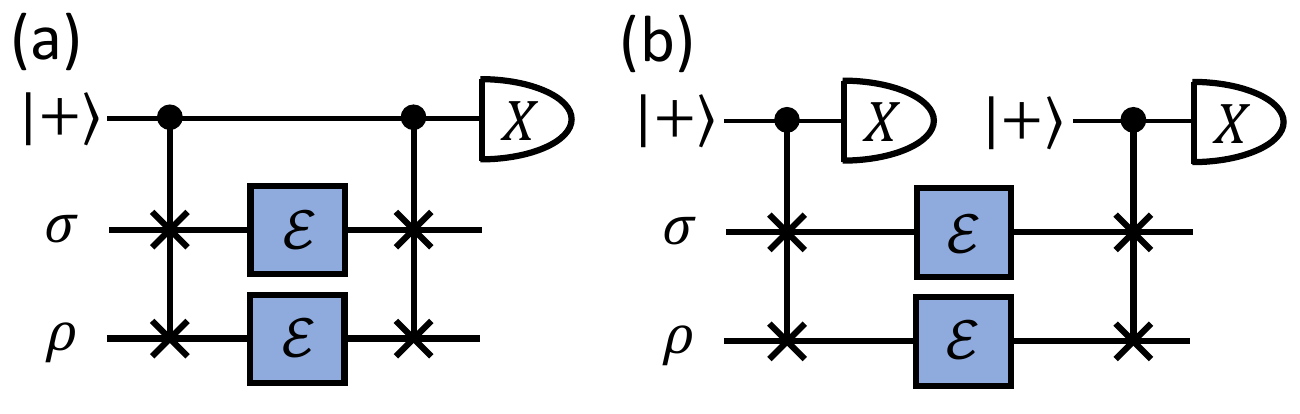}
    \caption{Circuits for measuring the \textsc{swap}-based detector. (a) Coherent \textsc{swap} detector. (b) Incoherent \textsc{swap} detector.}
    \label{fig:swap_detector_circ}
\end{figure}

As mentioned before, the circuit structure we use for VCP is similar to the circuit used for flag fault-tolerance~\cite{chaoQuantumErrorCorrection2018,chaoFlagFaultTolerantError2020} or spacetime checks in the recently proposed spacetime picture of viewing quantum error correction~\cite{mcewenRelaxingHardwareRequirements2023,gottesmanOpportunitiesChallengesFaultTolerant2022,delfosseSpacetimeCodesClifford2023}. 
In VCP, we probe the noisy state before and after the noise channel using the \textsc{cswap}s operators and we measure the collective parity of these two checks using one control qubit as shown in \cref{fig:swap_detector_circ}(a) to see if the two probes are consistent with each other or not. A natural question to ask is whether we can perform the two checks one by one as shown in \cref{fig:swap_detector_circ}(b) and simply compare the check results in post-processing, which is how the conventional purely spatial codes operate. We will call the circuit in \cref{fig:swap_detector_circ}(a) and \cref{fig:swap_detector_circ}(b) as coherent and incoherent detectors, respectively.

As we have shown in \cref{sec:detector}, both the coherent and incoherent detectors will remove the noise components in the \emph{detection region} sandwiched between the \textsc{cswap}s. However, the incoherent detector will project the incoming state and the outgoing state into the eigensubspace of \textsc{swap} while the coherent detector will not act on the incoming states at all. As a result, VCP can only be performed using the coherent detector circuit. 

It will also be very interesting to see how we can expand the application of such coherent detector circuits for symmetries beyond \textsc{swap} operations. 
For example, we can extend symmetry verification~\cite{mcardleErrorMitigatedDigitalQuantum2019,bonet-monroigLowcostErrorMitigation2018} of problem-specific symmetries from the level of quantum states to channels. 
We can also try to find ways to apply this to the symmetry expansion and generalised subspace expansion protocols~\cite{caiQuantumErrorMitigation2021,yoshiokaGeneralizedQuantumSubspace2022} which are frameworks that further generalise the virtual purification scheme.

\subsection{Comparison to other techniques}

We have discussed the main difference between VCP and other major QEM techniques in \cref{sec:intro} and also made extensive comparisons between VCP and VSP in \cref{sec:vcp_performance}. 
There are two other error suppression techniques in the name of error filtration (EF)~\cite{gisinErrorFiltrationEntanglement2005,leeErrorSuppressionArbitrarySize2023} and superposed quantum error mitigation (SQEM)~\cite{miguel-ramiroSuperposedQuantumError2023} that also use multiple copies of the noisy channel. While in VCP we use \emph{one single control qubit} to create interference between the original noise term and one other permuted noise term, EF and SQEM use a \emph{multi-qubit controlled register} to create interference among multiple noise terms that correspond to \emph{all possible swappings} between the main noise channel and the other noise channel copies. Then, they measure the control qubit and post-select the result, while in VCP, we keep all the results and perform post-processing. Due to these differences, EF and SQEM can only achieve suppression of noise \emph{linear} in the number of copies of noisy channel $M$ instead of \emph{exponential} in $M$ in VCP. Note that in SQEM~\cite{miguel-ramiroSuperposedQuantumError2023} they also proposed incorporating active corrections to increase the efficiency of the protocol. However, they did not make the connection to QEC. Instead, they resorted to performing black-box optimisation for the selection of the input state, the measurement basis and the correction unitaries altogether. It is challenging to have a rigorous performance guarantee for such an optimisation routine and it can be very costly to implement as we increase the system size, and thus it is only suitable for small noisy gates. 

Of course, rather than comparing VCP against these methods, it often makes more sense to combine the key ideas within them. We have mentioned in \cref{sec:vcp_performance} about the effectiveness of using other QEM methods like probabilistic error cancellation and zero-noise extrapolation to remove the noise in the \textsc{cswap}s and the ancilla in VCP. We can also bring in ideas from EF to explore the possibility of using fewer registers for higher $M$ (at the cost of more control qubits). We can also try to apply the idea of using maximally mixed states and QEC code states as the ancillary input state from VCP to EF and SQEM, especially when applying to communication. It will also be interesting to look at VCP in other application scenarios like QRAM mentioned in Ref.~\cite{leeErrorSuppressionArbitrarySize2023}.

\end{document}